\documentclass[]{article}
\usepackage{mybaltzer}
\usepackage[dvips]{graphicx}
\usepackage{pifont}
\usepackage{eepic}
\usepackage{epic}
\theoremstyle{plain} \theorembodyfont{\itshape} \newtheorem{Def}{Definition}
\theoremstyle{plain} \theorembodyfont{\rmfamily} \newtheorem{The}{Theorem}
\theoremstyle{plain} \theorembodyfont{\rmfamily} \newtheorem{Pro}{Proposition}
\theoremstyle{plain} \theorembodyfont{\rmfamily} \newtheorem{Exa}{Example}
\theoremstyle{plain} \theorembodyfont{\rmfamily} \newtheorem{Rem}{Remark}
\newcommand{\be}{\begin{equation}}
\newcommand{\ee}{\end{equation}}
\newcommand{\bea}{\begin{eqnarray}}
\newcommand{\eea}{\end{eqnarray}}
\newlength{\hoehe}
\newlength{\tiefe}
\newcommand{\ind}[1]{\ensuremath{\int \! d#1 \,}}
\newcommand{\ave}[1]{\ensuremath{\left\langle #1\right\rangle}}
\newcommand{\avi}[2]{\ensuremath{\left\langle #1\right\rangle}_{{}_{#2}}}
\newcommand{\scp}[2]{\ensuremath{\left\langle #1 | #2 \right\rangle}}

\newcommand{\mel}[3]{\ensuremath{
                 \settoheight{\hoehe}{\ensuremath{#1 #3}}
                 \settodepth{\tiefe} {\ensuremath{#1 #3}}
                 \left\langle#1\left| #2\rule[\tiefe]{0cm}{\hoehe}
                               \right|#3\right\rangle}}

\volume{}       %optional
\firstpage{1}  %optional
\lastpage{88}   %optional
\begin{document}
\begin{frontmatter}
%
%\pretitle{}                  %   e.g.: \pretitle{Guide}
\title{
How to Implement A Priori Information:\\ 
A Statistical Mechanics Approach\thanks{Report-no: MS-TP1-98-12}
}
%\thanks{} is optional
%\subtitle{}                  %   e.g.: \subtitle{Dedicated ...}
\author[A]{J\"org C. Lemm}%
%\thanks{...}
%}   \thanks{} is optional
%\author[B]{...}
\address[A]{Institute for Theoretical Physics I\\
            M\"unster University\\
            Wilhelm--Klemm--Str. 9\\
            D-48149 M\"unster, Germany\\
{\rm http://pauli.uni-muenster.de/{${}^\sim$}lemm}
\email{lemm@uni-muenster.de}}   
%   \email{} is optional
%\address[B]{...}
\runningauthor{Lemm, J.C.}
%\runningtitle{Prior Concepts}
\runningtitle{How to Implement A Priori Information}
% HISTORY:    %% HISTORY is optional
%\received{}
%\revised{}
%\accepted{}
%\dedicated{}
%\presented{}
%
\begin{abstract}  % abstract goes here
A new general framework is presented for implementing 
complex a priori knowledge,
having in mind especially situations where
the number of available training data is small compared to the
complexity of the learning task. 
A priori information is hereby decomposed into simple components
represented by quadratic building blocks (quadratic concepts)
which are then combined by conjunctions and disjunctions 
to built more complex, problem specific error functionals.
While conjunction of quadratic concepts
leads to classical quadratic regularization functionals,
disjunctions, 
representing ambiguous priors,
result in non--convex error functionals.
These go beyond classical quadratic regularization approaches
and correspond, in Bayesian interpretation,
to non--gaussian processes.
Numerical examples show that the resulting stationarity equations, 
despite being in general
nonlinear, inhomogeneous (integro--)differential equations,
are not necessarily difficult to solve.
Appendix \ref{Statistics} 
relates the formalism of statistical mechanics to statistics and 
Appendix \ref{Bayesian}
describes the framework of Bayesian decision theory.
\end{abstract}
 % AMS classif. codes go here
\end{frontmatter}
%%%*************** Text entry area *************

\tableofcontents

\section{Introduction}

In the setting of empirical learning training data are used
to predict the outcome of future test situations.
The principal problem of empirical learning
can already be seen in a noise free toy example:
Assume we have measured $h(x_1)$ = 2 (training data)
and have to predict $h(x_2)$ (test data).
Clearly, this is a hopeless task unless
we know some relations between $h(x_1)$ and $h(x_2)$
\cite{Wolpert-1996}.
If, however, we know a relation, for example
$h(x_2) - h(x_1)$ = 5, the task becomes solvable.
Such relations do not have the form of standard training data
and must be provided by what we will call {\it a priori information}.

The simple example clarifies two points which are also valid 
for more complex scenarios:
1. learning is technically merely a reformulation of
available (training and a priori) knowledge, and
2. the success of learning 
is essentially based on empirical control of 
the implied dependencies between test and training data.

One may now distinguish two principal possibilities
of implementing a priori information in learning systems:
1. The dependencies between test and training data
can be implemented by restricting the space of 
possible functions $h(x)$, for example
by choosing a parameterized form for $h(x)$.
Examples include linear regression models or (finite) neural networks.
hereby it is often difficult to interpret such implemented priors
in terms of dependencies of function values $h(x)$
\cite{Wolpert-1994}.
2. Dependencies between test and training data
may also be directly expressed in terms of the functions values $h(x)$
itself, like we have done in the above toy example.
This is the approach used for regularization terms 
in empirical risk minimization
and for stochastic prior processes in Bayesian statistics. 
Here, for example,
a smoothness term 
like
\be
\int \!dx\, \left(\frac{\partial h(x)}{\partial x}\right)^2 
\ee
within an error functional can provide the necessary dependencies.

To control and adapt the generalization behavior
of learning it seems helpful
to treat a priori information as explicit as possible.
We choose in this paper therefore 
the second possibility of {\it explicit implementation} of
a priori information in terms of the function values $h(x)$.

Furthermore, 
we consider especially low data situations
where the number of available training data is small
compared to the complexity of $h$.
We do this because those are the cases where a priori information becomes 
the essential input.
Contrasting the well known uninformative priors
of Bayesian statistics \cite{Berger-1980}
one may call this a situation with {\it informative priors}.

As a typical problem, consider an image completion task 
where only some of the pixels are given and the complete
image has to be completed.
If we expect for example the image of a face
then a priori information to be implemented 
includes the expectation that a face has two eyes,
a mouth and a nose and typical distances between that constituents.
Similar problems are times series predictions
with a priori informations concerning typically expected
relatively well defined structures.
Informative priors are also useful
in object recognition or pattern classification
when the object is relatively complex 
compared to the number of available training data.
For a face detector, for example, a priori information 
can be implemented in form of
a crude a priori model of what a face is
which is then refined by the available training data.

In practice, a priori information 
is usually given in qualitative
rather than in quantitative form.
The approach used in this paper
assumes a priori information stated
in form of logical statements
like $h$ has (probably) property $A$ AND $B$
(e.g., a face has eyes AND mouth)
or property $A$ OR $B$
(eyes may be open OR closed).
In a first step properties of $h$ are quantified by so called
quadratic concepts introduced in Section \ref{Quadratic concepts}.
Quadratic concepts consist hereby of a prototype or
function template $t$ for $h$,
and a corresponding distance measure defining $||t-h||$. 
In a second step
an error functional is constructed
implementing the logical statements  
which represent the given a priori information.
This is done in Section \ref{Combinations}.
An optimal approximation 
is finally found by minimizing
the error functional (Section \ref{Learning}).

Combining quadratic concepts 
by AND (Section \ref{and})
yields well known quadratic regularization functionals.
Not common is the explicit inclusion
of `continuous data' or template functions $t$
representing approximate reference models for $h$.

OR--like combinations, however,
give rise to non--convex error functionals
going beyond classical regularization approaches
(Section \ref{OR}).
Numerically especially interesting
are OR--like combinations of quadratic concepts
with equal distance and only differing template functions.
Those can be treated without
calculating normalization factors which are difficult to obtain.
This is exemplified in Section \ref{exam-two}.
Section \ref{Landau--Ginzburg}
discusses an alternative implementation
of OR--like combinations.

In case the number of properties combined by OR is too
large to be treated exactly additional approximations are necessary
which are discussed in Section \ref{continuous}.
This is for example the case if an expected structure
can be arbitrarily translated or otherwise
continuously transformed.
An eye, for example, can appear in different shapes/scales/positions
and only those variants most consistent with the training data
will then be included explicitly.

The proposed approach provides a 
{\it interface between symbolic methods 
and statistics}\cite{Aleksander-Morton-1993}:
A priori information is decomposed 
by symbolic/logical operations
into components, 
simple enough to be quantified in form of quadratic concepts.
Quadratic concepts are then combined, 
modeling the symbolic operations, resulting in an error functional 
which is then treated by non--symbolic methods.

Finally, let us add three remarks
concerning the numerical
requirements of the approach, 
the language of statistical physics,
and the interpretation of error functionals.

Explicit implementation of a priori information
is in generally numerically extensive.
Due to increasing computational resources
numerical methods used in statistical physics or field theory, 
e.g., Monte Carlo calculations,
become more and more applicable 
to learning problems.
Up to now this is mainly the case
for  one or two dimensional problems,
like for example in Bayesian image reconstruction
\cite{Poggio-Torre-Koch-1985,Geman-Geman-1984,Winkler-1995,Zhu-Mumford-1997}.
In particular, the stationarity equations to be solved to 
minimize the presented error functionals can in general
be nonlinear inhomogeneous integro--differential equations.
They are however not necessarily difficult to solve
as can be seen in Section \ref{exam-two}.
The reason is that the nonlinearity, in particular the number of minima,
is under explicit control
and there are limiting cases (at `high and low temperatures')
where the equations become linear.
In any case, the presented error functionals
can be utilized as a well defined starting point for further 
approximations.

Due to the background of the author
the paper is written mainly
in the language of statistical mechanics.
Of course, there exists alternative formulations
such as function approximation theory 
or empirical processes, which might be better suited for some purposes.
The advantage of using a statistical mechanics formulation, however,
is that it provides easy contact to
approximations (e.g., saddle point approximation, perturbation theory, 
high temperature expansions
\cite{Itzykson-Zuber-1985,Negele-Orland-1988,Zinn-Justin-1989})
and numerical algorithms 
(e.g., Monte Carlo algorithms 
\cite{Binder--1992,Binder-Heermann-1992,Gelman-Carlin-Stern-Rubin-1995,Neal-1996})
well known from statistical mechanics or statistical field theory
and especially useful for large systems.
In Appendix \ref{Statistics} the language of statistics
is related to that of statistical mechanics.

Like regularization approaches
in empirical risk minimization 
we use an error functional
to find an optimal approximation.
We remark, however, that the 
error functional has a different interpretation
from the view point of empirical risk minimization 
and that of Bayesian statistics. 
In empirical risk minimization the error functional
represents the regularized form of an empirical risk
with data terms being an empirical estimate of an expected risk.
In the Bayesian interpretation
the error function is related to the
negative logarithm of the posterior density of $h$
given the training data.
Minimizing the error function
is then equivalent to a maximum a posteriori approximation
under log--loss.
As we use the Bayesian interpretation
to built up error functionals
the necessary background is provided
in Appendix \ref{Bayesian}.

\section{Quadratic concepts}
\label{Quadratic concepts}
\subsection{Definitions}
\label{Definitions}

In this Section quadratic concepts are defined
as fundamental building blocks to construct regularization functionals.
We concentrate on function approximation or regression problems.
Hereby an unknown function or true state of Nature  ${h}_N$
is approximated by a function $h$ 
chosen from a model space ${\cal H}$
using training data 
$D_T$ = $\{(x_i,y_i)|i\le 1\le n\}$ = $(x_{\!{}_T},y_{\!{}_T})$.
Training data are assumed to consist
of $n$ pairs of independent variables $x$ 
(describing the kind of measurement performed)
and dependent variables $y$ (responses or measured values). 
In classical regularization theory \cite{Tikhonov-1963,Wahba-1990}
the error to be minimized is given by a
regularization functional $E(h)$ (see Appendix \ref{Bayesian}
for a Bayesian interpretation) which 
contains the  mean--square error terms for training data
and an additional prior term, in many cases related to smoothness.
We denote prior information by $D_0$
so $D=D_T\cup D_0$
represents all available data we have.
Application to density estimation problems (having an additional
normalization constraint and logarithmic terms replacing the 
mean--square data terms)
or classification problems (being a special case of function approximation
with integer $y$) poses no principal new problems
and will be reported elsewhere.

\begin{Exa} 
(Image completion)
Consider an image completion or image reconstruction task 
\cite{Poggio-Torre-Koch-1985,Geman-Geman-1984,Winkler-1995,Zhu-Mumford-1997}.
Hereby an image ${h}$ is drawn
from a set of images ${\cal {H}}$.
Then, given noisy observations $y_i(x_i)$ 
for some pixels $x_i$ of ${h}$
a reconstructed and completed
image $y=h(x)\in {\cal H}$ should be returned by the learning system.
For grey level images ${h}$ is a vector of real numbers
containing a two dimensional
$n_1 \times n_2$ array of grey level values $g(k,l)$, i.e., 
with $j$--components 
${h}_j = g(k,l)$ with
$1\le k \le n_1$,
$1\le l \le n_2$,
$1\le j \le n_1 n_2$
and, for example, $j = k + n_1 (l-1)$.
%We choose ${\cal H} = {\cal {H}}$
%both given by the set of all possible 
%grey level $n_1 \times n_2$ images.
Assume now one knows that the image is that of a face.
For the class of faces experts can contribute information
by verbally describing prototypical forms of 
constituents (e.g.\ eyes, nose, mouth),
their variants (e.g.\ open vs.\ closed mouth, 
translated, scaled, or deformed eyes)
and relations (e.g.\ typical spatial distances).
The related concepts 
like `eye' or `typical distance between eyes',
are here linguistic or `fuzzy' variables \cite{Klir-Yuan-1996} which
must be quantified to enter a regularization functional.
As prototype or template $t$ of an `eye'
an image or drawing of an eye can be chosen.
A distance $d(t,h)$ 
between a reconstructed image $h$ and an `eye' template $t$
can be defined pixel-wise, e.g.\ by
\be 
d^2(t,h) = \sum_x (h(x)-t(x))^2
,
\ee
or, more generally, by a real symmetric, positive definite kernel
\be
d^2_K(t,h) = 
\sum_{x,x^\prime} (h(x)-t(x)) K(x,x^\prime) (h(x^\prime)-t(x^\prime)).
\ee
The sum over pixels $x$ may be 
restricted to regions where eyes are expected.
Furthermore, 
eyes may be open OR closed,
appear in different sizes and at varying locations.
Hence an eye is represented not by a single template 
but by a set of templates
describing the variants in which an eye can appear.
For continuous variations we will call such a set 
a deformable or adaptive template.
It is described, for example, by scaling,
translation, or deformation  parameters.
% can be introduced.
%i.e.\ giving typical variations in which the prototype may appear.
Then the image of a face, 
incomplete and disturbed by noise,
should be reconstructed 
%the reconstructed image $h(x)$ should approximate 
by approximating the given noisy data points of the
incomplete image AND typical constituents of a face
and their spatial relations in either one of their variants.
Thus, a missing eye should be reconstructed
at a typically expected position
depending, for example, on the approximated position of mouth and nose,
and in a form depending on the form of another, possibly visible eye.
\end{Exa}

The model treated in Section (\ref{exam-two})
represents a simple example for such a task
(for a one--dimensional image or time series, respectively).
For comparison, we discuss shortly a possible use 
of template functions for classification tasks
(not treated in this paper):

\begin{Exa}
\label{face-detecion}
(Face detection)
Consider a face detection task.
Hereby the independent variable $x$ is drawn
from a set of images ${\cal X}_R$
by an unknown mechanism
which should be approximated by a function $h(x)$.
Then,
given an image $x\in {\cal X}_R$ the output $y\in \{0,1\}$ 
of the learning system
should indicate whether it is the image of a face, 
e.g.\ by $y=h(x)=1$,
or not, e.g.\ by $y=h(x)=0$. 
In contrast to the previous example
now $x$ is a vector of real numbers
containing a two dimensional array of grey level values.
A function template $t$ now represents a prototoype
for the classification function $h$. 
Classification templates $t_j$ entering the error functional
can now, for example,
be constructed by reference to image templates $t^x$
representing prototypical face/non--face input vectors $x$ or parts of it.
Thus,
$E(h)$ = $E(h,\{t_j\})$ where classification templates $t_j$ 
are defined with the help of image templates  $t_j$ = $t_j(x,\{t^x_i\})$.
For example face templates $t^x$ for images $x$
can be constructed as described in the previous example.
Then, in a second step statements like:
`A face has eyes, nose and mouth
in either one of several variations'
are expressed in terms of distances $d^x_i$ from image $x$
to image templates $t^x$. 
A classification templates $t$
could for example be a simple threshold function
responding with $t(\{d^x_i\})$ = $1$ 
if some function of distances $\{d^x_i\}$ to typical faces
are below a certain threshold.
Finally, distances $d(h,t)$ between classification functions $h\in {\cal H}$
have to be chosen
to build an error functional $E(h)$ to be minimized.
(We refer to Appendix \ref{Approximation-problems}
for the distinction between specifying a reconstruction model,
i.e., the probability of face vs.\ non--face given an image,
and specifying a generative model,
% $p(x|y,\tilde {h})
i.e., the probability of an image provided 
it represents a face or non--face.)
\end{Exa}

In general, a {\it concept} is based upon
\begin{itemize}
\item[1.] a {\it template} $t(x)$ representing a prototype for function $h$
and 
\item[2.] a {\it distance} $d(h,t)$ measuring the
similarity of a function $h$ to the template $t$.
For practical reasons, a quadratic distance is especially convenient.
Concepts with quadratic distances will be called quadratic concepts.
\item[3.] A template $t$ can be restricted to a subset of ${\cal X}_R$
with the help of a {\it projection operator}.
\item[4.] Complex concepts are constructed 
as {\it combination of quadratic concepts}.
In this paper variants of AND and OR--like combinations are used.
\end{itemize}

Now we come to the formal definition of a quadratic concept.
Let ${\cal H}$ be a Hilbert space of possible hypothesis functions $h$,
and let angular brackets denote scalar products
and matrix elements of symmetric operators, e.g.\
$\scp{t}{h}$ = $\int \!d^dx\, t(x) h(x)$
and
$\mel{t}{K}{h}$ 
= $\int \!d^dx \,d^dx^\prime t(x)K(x,x^\prime) h(x^\prime)$
for $d$--dimensional $x$.
(We will often write in the following
simply $\ind{x}$ also for $d$--dimensional integrals.)
%the following definitions. 
%An operator $O$ is orthogonal if $O^T O$ = $I$,
%where $O^T$ is the transpose and 
%operator $K$ is symmetric if
%A real operator has real matrix elements.
%An 
Recall also that a real operator $K$ is positive definite 
(semi positive definite)
if it can be diagonalized,
i.e., in terms of eigenfunctions $\Phi_k$ of $K$
(with transpose $\Phi^T_k$) 
\be
K = \sum_k \lambda_k \Phi_k \Phi_k^T , 
\ee
and all eigenvalues $\lambda_k>0$ ($\lambda_k\ge 0$)
are positive (non--negative).
Thus, $K=O^T D_K O$ with $O$ orthogonal
(i.e., $O^T O = I$ with identity $I$ and transpose $O^T$)
and diagonal operator $D_K>0$ $(D_K\ge 0)$.
Positive definite operators define a scalar product
by $\scp{t}{h}_K$ = $\mel{t}{K}{h}$.
Positive (semi) definite operators $K$ can be decomposed 
$K=W^TW = (OW)^T (OW)$, 
with real $W$, invertible if $K$ positive definite,
and arbitrary orthogonal $O$.
%, i.e.\ $O^T O = I$
%with identity $I$.
A quadratic concept is defined as follows:
\begin{Def}
%\begin{defn}
(Quadratic concept)
A quadratic {\it concept} is a pair $(t,K)$
consisting of a {\it template function} $t(x)\in {\cal H}$ 
and a real symmetric, positive semi--definite operator,
the {\it concept operator} $K$ with 
eigenfunctions ({\it features}) $\Phi_k$
and eigenvalues (feature weights) $\lambda_k$.
%A positive semi definite operator can always be written
%$K = W^T W$ with real $W$ and transpose $W^T$
%which are invertible if $K$ is positive definite.
We call $W$ a {\it concept filter} if $K = W^T W$.
The operator $K$ defines a {\it concept distance}:
\begin{equation}
d^2 (h) 
= d^2_K (t,h) 
= \mel{h-t}{K}{h-t} 
= ||h-t||^2_{K}
\ee
\be
= \scp{W(h-t)}{W(h-t)} 
= ||W(h-t)||^2
\ee
\be
= \sum_k \lambda_k \scp{h-t}{\Phi_k}\scp{\Phi_k}{h-t}
= \sum_k \lambda_k \,\, || \scp{h}{\Phi_k}-\scp{t}{\Phi_k}||^2
,
\end{equation}
on subspaces where it is positive definite.
The maximal subspace in which the positive semi--definite
$K$ is positive definite
is the {\it concept space} $H_K$ of $K$.
The corresponding hermitian projector $P_K$ in this subspace $H_K$
%, i.e.\ $P_K(H) = H_K$, $P_t(H\setminus H_K) = K$, 
%and $P_K^2 = P_K$
is the {\it concept projector}.
%\end{defn}
\end{Def}

There exists a correspondence
between quadratic forms and gaussian processes
\cite{Wahba-1990,Lifshits-1995,Neal-1996}.
Let $x$ be elements of (the dual of) ${\cal H}$,
i.e., $h(x)-t(x)$ = $\scp{x}{h-t}$ 
= $\mel{K^{-1} x}{K}{h-t}$
= $\scp{K^{-1} x}{h-t}_K$,
is a bounded functional according to the Riesz representation theorem.
In the context of stochastic processes 
${\cal H}$ is known as reproducing kernel Hilbert space for $x$
with reproducing kernel $K^{-1}$ \cite{Wahba-1990}.
(The term reproducing kernel Hilbert space
does not name a certain subclass of Hilbert spaces.
Indeed, all separable Hilbert spaces are isomorphic for equal dimension.
It characterizes the representation by the coordinates $x$.
For example, the space ${\cal L}_2$ of square integrable functions $h(x)$
is not a reproducing kernel Hilbert space
with respect to the coordinates $x$. This is reflected by the fact that
functions $h(x)\in{\cal L}_2$ are not even defined pointwise.)
For reproducing Hilbert spaces the scalar product with kernel $K$
can be related to a covariance of zero mean gaussian variables.
%can be interpreted
%as a Hilbert space of gaussian variables
%
Interpreting $t$ as data $y_{\!{}_D}$ 
obtained in situation $x_{\!{}_D}=K$
(i.e., measuring features $\Phi_k$ weighted by $\lambda_k$)
we write 
\be
p(y_{\!{}_D}|x_{\!{}_D},h) = p(t|K,h)\propto e^{-d^2_K(t,h)/2}
\label{gauss}
\ee
for a distribution of $t$
with mean $h$ and covariance operator $K^{-1}$.
(For a interpretation as posterior $p(h|D)$ 
see Section \ref{likelihoods-and-posteriori}.)
The kernel function $K^{-1}(x,x^\prime)$
of the covariance operator
is also known as
Green's function of $K$, propagator, 
or two-point correlation function.
Hence, an error or energy functional 
\be
E(h) = \frac{d^2(h)}{2}
\ee 
corresponds up to a constant to a negative log--probability.
For approximation problems minimizing an error functional
$E(h)$ can,
from a Bayesian point of view, 
be interpreted as a maximimum--a--posteriori approximation
(see Appendix \ref{Approximation-problems}).

We remark that we will not
discuss in the following problems of infinite spaces.
This holds especially for the nonlinear models 
in the next sections
for which the question of a continuum limit 
is highly non-trivial \cite{Itzykson-Drouffe-1989,Zinn-Justin-1989}.
Hence, if necessary,
integrals can in the following be considered as convenient notation for
sums. Analogously,
derivative operators can be replaced by their discretized
lattice versions.
This reflects also the fact that finally
numerical calculations have to be done in a finite dimensional space.

The next two examples 
show that the definition of a quadratic concept
includes the standard regularization functionals.
The first example is a discrete concept with `trivial' concept distance,
the second a continuous concept with `trivial' template function.

\begin{Exa}
%\begin{exmp}
(Data template)
A standard mean--square error term used for regression is
\begin{equation}
(y_j-h(x_j))^2 
= \int_{-\infty}^\infty \!d^dx \, \delta (x-x_j) (h(x)-t_j(x))^2
= \mel{h-t_j}{P_j}{h-t_j}.
\label{mse}
\end{equation}
Thus, such a mean--square error term
corresponds to a quadratic concept with 
%concept projector and 
concept operator with kernel 
$P_j (x,x^\prime)$ = $\delta(x-x_j)\delta(x-x^\prime)$
and (data) template $t_j$ is the constant function
$t_j(x) \equiv y_j$.
The measured features are $h(x)$ = $\scp{h}{x}$.
%\end{exmp}
\end{Exa}

%\begin{exmp}
\begin{Exa}
\label{example2}
(Prior template)
A typical smoothness functional in regularization theory, 
corresponding to the Wiener measure for stochastic processes
\cite{Doob-1953,Gardiner-1990,van-Kampen-1992,Lifshits-1995}
and to the kinetic energy 
or a free massless scalar Euclidean field
in physics \cite{Glimm-Jaffe-1987,Itzykson-Drouffe-1989,Zinn-Justin-1989}, is
\begin{equation}
\int_{-\infty}^\infty \!d^dx\, 
\sum_{l=1}^d \left( \frac{\partial h(x)}{\partial x_l}\right)^2
= - \mel{h - t_0}{\Delta}{ h-t_0}.
\label{smooth}
\end{equation}
Here partial integration has been used
under the assumption of vanishing boundary terms.
This quadratic concept has the zero function $t_0(x)\equiv 0$ as template. 
It is a sum of terms with concept filters 
$\partial/\partial x_l$
which are the generators of infinitesimal translations
in $d$ dimensions.
Hence, the functional represents a measure 
of approximate infinitesimal translational symmetry.
The concept operator is
the negative (semi) definite $d$--dimensional laplacian
with kernel 
\be
\Delta (x,x^\prime ) = 
\delta (x-x^\prime)\sum_{l=1}^d \frac{\partial^2}{\partial
  x_l^2}.
\ee
%Thus, (\ref{smooth}) can be interpreted as a specific distance of $h$ from 
%the zero function $t_0$.  % but now over all $x$.
%\end{exmp}
\end{Exa}

In the following we will mainly be interested
in the non--standard case of a combination of discrete training data 
with several prior concepts with non--zero template functions.
Firstly, we discuss in more detail the two main ingredients of a concept:
templates and distances.

\subsection{Templates}
\label{templates}

Template functions can be constructed in various ways.
%and used in various contexts.
The following list gives
some potential applications of template functions
in different contexts:
\begin{itemize}
\item[1.] (Direct construction by experts) A template can be 
directly constructed by experts.
%for example, using fuzzy techniques.
For financial time series, for example,
often an expected trend is included \cite{Hull-1989}.
\item[2.] (Combination and extension of arbitrary learning methods) 
More generally, a template $t(x)$ can be the output of an arbitrary,
parametric or non--parametric
learning method trained for the same problem.
For example, the prototype $t(x)$ can be
a rule--based expert system, a regression tree, a neural network
or a simple linear regression
(for a collection and comparison 
of methods see for example \cite{StatLog-1994}).
Because a prior concept
can depend on training data $D_T$ (see Appendix \ref{postandlike})
such a $t(x)$ can be obtained by training with the same data $D_T$ 
which are also used to determine the optimal approximation $h^*$.
If one wants $t(x)$ to be independent of the training data $D_T$
it can alternatively be obtained by training
with an independent set of training data.
\item[3.] (Transfer) The template $t(x)$ can also be the result 
of a learning algorithm for a similar problem and therefore
be used to transfer knowledge between tasks.
Such a transfer template can be adapted and restricted to
certain subspaces by using concept projectors.
The new solution $h$ adapts the transfer template $t$ to the new situation 
according to the new training data
and other, additional prior templates.
\item[4.] (Learning history)
In all cases a template $t(x)$
can be seen to contain in a compressed form the {\it learning history}
prior to the new training data. 
This allows for example to construct on--line learning procedures.
Hereby an intermediate solution $h$ is obtained by using only 
a part $D_1\subset D_T$ of the available training data $D_T$.
Then the intermediate solution $h$ is chosen as additional template $t$ 
and learning is continued with a new data set $D_2\subset D_T$.
\item[5.] (Sampling)
A template corresponds to the mean of a gaussian process.
For a finite number of $x$
it can therefore be approximated by a sample mean,
if samples for $h(x)$ are available.
This can be, for example, 
a set of complete images (or images of constituents) 
for usage in image reconstruction.
Thus, let $t_\alpha$ denote a sample for (a part of) $h(x)$.
Then, if $p(h)$ can be approximated by a gaussian,
the empirical mean $t = \sum_\alpha t_\alpha$ 
is a natural candidate for template.
Similarly, multimodal distributions $p(h)$ can be approximated
by a mixture of gaussians.
In that case not only one but a set of centers, i.e., templates $t_i$
has to be chosen (see Section \ref{Combinations}).
The centers $t_i$  can be obtained by clustering methods.
\end{itemize}

Clearly,
these fields have long publication histories
and the advantages and disadvantages of
using templates still have to be investigated.
In this paper we concentrate mainly on the first possibility.

\subsection{Covariances and symmetries}
\label{Symmetry}

The concept kernel, respectively its inverse
the covariance operator, are often related to 
approximate symmetries of a problem.
Sometimes, also a finite rank approximation can be obtained by sampling.

Frequently prior information has the form
of symmetries which $h(x)$ has to fulfil, exactly or approximately.
We already have seen in Example \ref{example2}
that approximate symmetry under infinitesimal translations
is related to smoothness.
The implementation of approximate symmetries
requires the definition of a distance to exact symmetry.

We can write for a positive (semi)definite concept operator
$K= W^T W$ = $(I-(W-I))^T ( I -(W-I))$
with identity $I$, 
%invertible 
and concept filter $W$. %, and $F^T$ the transpose of $F$. 
Now consider operators $S$ which
just change the argument $x$ of $h(x)$ into $\sigma(x)$, i.e.,
\begin{equation}
S h(x) = h(\sigma (x) ).
\end{equation}
If the transformation $\sigma (x)$ is one--to--one
then $S$ permutes the function values
and $S$ is a symmetry transformation.
Hence, for $K=(I-S)^T(I-S)$, i.e., $S=W-I$ and a zero template $t\equiv 0$
the corresponding squared $K$--norm  
\be
||h||_K^2 = \mel{h}{K}{h} = \scp{h-Sh}{h-Sh} = ||h-Sh||^2_I = d^2_S(h)
\ee
compares
$h$ with $Sh$ and measures therefore a degree of symmetry.

A bit more generally, we call a concept operator
$K = (I-S)^\dagger K_S (I-S)$
a symmetry concept operator, if $S$ is a symmetry operator.
Here $K_S$ can be some positive (semi--)definite operator
usually taken equal to the identity $I$,
and the hermitian conjugate $S^\dagger$ is equal to the transpose $S^T$
for real matrices.
With zero template we have 
\begin{equation}
d_S^2= \mel{h - S h}{K_S}{h-Sh}
=\mel{h}{(I-S)^\dagger K_S (I-S)}{h}.
\end{equation}
Continuous symmetries are represented by Lie groups
which locally can be written as exponential
of $m$ generators $s$ = $(s_1,\cdots, s_m)$
parameterized by a vector $\theta$
\begin{equation}
S(\theta)= e^{\scp{\theta}{s}} = 
\sum_{k=0}^\infty \frac{1}{k!} \scp{\theta}{s}^k.
%commuting s=
%\sum_{k_1,\cdots,k_m=0}^\infty 
%\prod_{j=1}^m \frac{(z_j s_j)^{k_j}}{k_j!} ,
\end{equation}
using the scalar product notation
$\scp{\theta}{s}$ = $\sum_j \theta_j s_j$
also for vectors of matrices or operators $s$.
The infinitesimal generators $s$ form the corresponding Lie algebra.
In particular, the group of $d$--dimensional translations is generated
by the gradient operator $\nabla$.
%with the components $\partial/\partial x_j, 1 \!\le\! j \!\le\!d$.
This can be verified by recalling the multidimensional Taylor formula 
for expansion of $h$ at $x$ 
\begin{equation}
S(\theta) h (x) = e^{\scp{\theta}{\nabla}} h(x)
%\frac{\partial h}{\partial x_j}} 
= \sum_{k=0}^\infty 
\frac{\scp{\theta}{\nabla}^{k}}{k!} h(x)
= h(x+\theta).
\end{equation}
Up to first order the expansion of the exponential function reads
$S \approx  1+\theta s$.
Thus,  we can define a distance to an  infinitesimal symmetry by
\begin{equation}
d_s^2 (h)
= \mel{\frac{h - (1 + \theta s) h}{\theta}}{K_s}
      {\frac{h - (1 + \theta s) h}{\theta}} 
=
\mel{s h}{K_s}{s h},
\end{equation}
with infinitesimal symmetry operator
$K$ = $s^\dagger K_s s$.
 For translations and $K_s$ equal to the identity $I$ 
this results exactly in
(\ref{smooth}).
% with zero template, i.e.,  $t_0\equiv 0$.

In cases
where samples $t_\alpha$ for $h$ are available, 
like sometimes in image reconstruction or in times series prediction,
a finite rank approximation $\hat K^{-1}$
of 
\be
K^{-1}(x,x^\prime) = 
\ind{h} p(h)h(x) h(x^\prime)
\ee
can be obtained
by the empirical estimator
\cite{Poggio-Girosi-1998}
%$\rightarrow$ cite Poggio
%
\be
\hat K^{-1}(x,x^\prime) = 
\sum_\alpha t_\alpha (x) t_\alpha(x^\prime).
\ee
An optimal $h^*$ 
in the subspace where $\hat K^{-1}$ is positive definite
and therefore invertible
can be found by singular value decomposition.
Alternatively,
more prior concepts with concepts operators $K_i$
can be added so the sum $\sum_i K_i$ becomes strictly positive definite.

%\subsection{Likelihoods and posterior probabilities}
\subsection{From posterior probability to likelihood energy}
\label{likelihoods-and-posteriori}

In this Section we discuss the interpretation of
quadratic concepts and error functionals 
in terms of posterior probability and likelihoods 
(see also Appendix \ref{Bayesian}).
We will need this interpretation 
in the following
to combine quadratic concepts
to more complex error functionals.

Typically, hypotheses $h$ are defined by specifying
the probabilities $p(y_{\!{}_D}|x_{\!{}_D},h)$
of finding data $y_{\!{}_D}$ in situations $x_{\!{}_D}$ under $h$.
These data generating probabilities $p(y_{\!{}_D}|x_{\!{}_D},h)$, 
or ($x_{\!{}_D}$--conditional) 
likelihoods of $h$ under $y_{\!{}_D}$,
are related to posterior probabilities
$p(h|D)$ we are interested in
%looking for the most probable $h$ for given data $D$,
%Analogously, inversion 
according to Bayes' rule
\be
p(h|D)
=\frac{p(D|h)p(h)}{p(D)}
=\frac{p(y_{\!{}_D}|x_{\!{}_D},h)p(h)}{p(y_{\!{}_D}|x_{\!{}_D})},
\label{Bayes-rule}
\ee
with 
\be
p(y_{\!{}_D}|x_{\!{}_D})=\int_{\cal H}\!dh\,p(y_{\!{}_D}|x_{\!{}_D},h)p(h),
\ee
assuming $p(h|x_{\!{}_D})$ = $p(h)$
and shorthand notation
\be
\int_{\cal H} dh \cdots 
= \int_{{\cal H}} \prod_x  dh(x) \cdots 
= \int_{{\cal H}_1} dh(x_1) \int_{{\cal H}_2} dh(x_2) \cdots
= \left( \prod_x  \int_{{\cal H}_x} dh(x) \right) \cdots 
.
\ee
%and mean $t$.
The terms of Eq.(\ref{Bayes-rule}) 
are in the Bayesian context often referred to as
\be
{\rm posterior} = \frac{{\rm likelihood} * {\rm prior}}{{\rm evidence}}
.
\ee

For complete data which means
for uniform (possibly improper, i.e., non--normalizable) $p(h)$,\footnote{
The probability $p(h)$ is known as prior probability of $h$.
In this paper we treat prior information
explicitly and analogous to standard training data. That means
we assume $p(h)$ to be $h$--independent
and collect all prior information in $D_0$.
They enter the formalism through 
$p(h|D_0)$ or $p(D_0|h)$, respectively.}
yields
(see also Appendices \ref{model}-\ref{postandlike} 
identifying ${h}$ with $h$ 
%for an approximation problem 
according to Appendix \ref{Approximation-problems}) 
\be
p(h|D) 
= \frac{p(y_{\!{}_D}|x_{\!{}_D},h)}{\ind{h}p(y_{\!{}_D}|x_{\!{}_D},h)}
= \frac{p(y_{\!{}_D}|x_{\!{}_D},h)}{Z_L}
\propto p(y_{\!{}_D}|x_{\!{}_D},h).
\label{eqb}
\ee
Being interested in the dependency on $h$ for given 
data $D$
we can skip the $h$--independent factor $p(D)$
or ${\ind{h}p(D|h)}$
and calculate instead of $p(h|D)$ the inverse
$p(D|h)$. 
%If hypotheses $h$ are, as usually, defined by specifying
%$p(D|h)$ this is more convenient than calculating
%$p(h|D)$.
For the special case (\ref{gauss})
of one quadratic concept with template $t$,
so $y_{\!{}_D} = t$ and $x_{\!{}_D}$ corresponds to $K$,
we find for the denominator in (\ref{eqb})
because of the translational invariance of the gaussian measure
\be
Z_L = 
\ind{h}p(t|K,h)
=
\frac{\ind{h} e^{-d^2/2}}{\ind{t} e^{-d^2/2}}
=
\frac{\ind{h} e^{-\frac{1}{2}\mel{h-t}{K}{h-t}}}
{\ind{t} e^{-\frac{1}{2}\mel{h-t}{K}{h-t}}}
=1,
\label{Z_L}
\ee
and thus $p(h|D) = p(y_{\!{}_D}|x_{\!{}_D},h)$.

To obtain error functionals
we will now express probabilities in terms
of energies and related quantities (see Appendix \ref{Logprob}).
%refering to the formalism
%described in Appendices \ref{Statistics} and \ref{Bayesian}. 
%The reader less interested in this discussion 
%can directly proceed with Proposition \ref{and-prop}.
%In terms of energies o
For example,
the data generating probabilities or ($x_{\!{}_D}$--conditional) 
likelihoods of $h$
can be written
in terms of ($x_{\!{}_D}$--conditional) likelihood energies 
$E(y_{\!{}_D}|x_{\!{}_D},h)$
and inverse likelihood temperature $\beta_{\!{}_L}$
\be
p(y_{\!{}_D}|x_{\!{}_D},h) 
= \frac{e^{-\beta_{\!{}_L} E(y_{\!{}_D}|x_h)}}{Z({\cal Y}_D|x_{\!{}_D},h)}
= e^{-\beta_{\!{}_L} \big( E(y_{\!{}_D}|x_{\!{}_D},h) - F({\cal Y}_D|x_{\!{}_D},h)\big)}
,
\label{gener}
\ee
and analogously the posterior probability $p(h|D)$ becomes
in terms of posterior energy $E(h|D)$
and inverse posterior temperature $\beta_{\!{}_P}$
%related to the first according to 
%Bayes' rule (\ref{Bayes-rule})
\be
p(h|D) 
= \frac{e^{-\beta_{\!{}_P} E(h|D)}}{Z({\cal H}|D)}
= e^{-\beta_{\!{}_P} \big( E(h|D) - F({\cal H}|D)\big)}
\ee
or in terms of likelihoods
\bea
p(h|D)&=& 
\frac{p(h)}{p(y_{\!{}_D}|x_{\!{}_D})}p(y_{\!{}_D}|x_{\!{}_D},h)
%= \frac{p(y_{\!{}_D}|x_{\!{}_D},h)}{Z_L(h,D)}
%\nonumber\\&=& 
= \frac{1}{Z_L(h,D)} \frac{e^{-\beta_{\!{}_L} E(y_{\!{}_D}|x_{\!{}_D},h)}}{Z({\cal Y}_D|x_{\!{}_D},h)} 
%= p(y_{\!{}_D}|x_{\!{}_D},h) e^{c_L(h,D)} 
\nonumber\\&=& 
e^{-\beta_{\!{}_L} \big( E(y_{\!{}_D}|x_{\!{}_D},h) - F({\cal Y}_D|x_{\!{}_D},h)\big) + c_L(h,D)} 
%=
%e^{-\beta_{\!{}_L} E(y_{\!{}_D}|x_{\!{}_D},h)+ c(h,D)} 
,
\label{post-likeli}
\eea 
with
%\be
%c(h,D) = \beta_{\!{}_L} F({\cal Y}_D|x_{\!{}_D},h) + c_L(h,D)
%,
%\ee
%and
\be
Z_L (h,D) = \frac{p(y_{\!{}_D}|x_{\!{}_D})}{p(h)}
,\quad
c_L (h,D) = \ln p(h) - \ln p(y_{\!{}_D}|x_{\!{}_D})
,
\ee
free energies, % (see Appendix \ref{Logprob}), 
%in particular 
\be
{F({\cal H}|D)} = -\frac{1}{\beta_{\!{}_P}} \ln {Z({\cal H}|D)}
,\quad
{F({\cal Y}_D|x_{\!{}_D},h)} = 
-\frac{1}{\beta_{\!{}_L}} \ln {Z({\cal Y}_D|x_{\!{}_D},h)}
,\ee
and normalization factors or partition sums 
\be
Z({\cal H}|D)
=
\int_{\cal H} \!dh\,
e^{-\beta_{\!{}_P} E(h|D)}
%=\int_{\cal D} \left( \prod_x^l dt_j(x)\right) e^{-\beta E_j(t_j|h)}
,\quad
Z({\cal Y}_D|x_{\!{}_D},h)
=
\int_{{\cal Y}_D} \!dy_{\!{}_D} \,\,
e^{-\beta_{\!{}_L} E(y_{\!{}_D}|x_{\!{}_D},h)}
.
\ee
Thus,
the posterior energy can be expressed
by the likelihood energy
\be
E(h|D)
=F({\cal H}|D)
+\frac{\beta_{\!{}_L}}{\beta_{\!{}_P}} 
\Big( E(y_{\!{}_D}|x_{\!{}_D},h)-F({\cal Y}_D|x_{\!{}_D},h)\Big)
+\frac{1}{\beta_{\!{}_P}} c_L(h,D)
%+\frac{1}{\beta_{\!{}_P}} \ln \frac{p(h)}{p(y_{\!{}_D}|x_{\!{}_D})}
.
\ee

For complete data, i.e.,  $h$--independent $p(h)$
\be
Z_L(h,D) = Z_L(D) 
=\ind{h} p(y_{\!{}_D}|x_{\!{}_D},h)
=\ind{h} \frac{e^{-\beta_{\!{}_L} E(y_{\!{}_D}|x_{\!{}_D},h)}}{Z({\cal Y}_D|x_{\!{}_D},h)}.
\ee
In that case we have already seen
that maximizing the likelihood $p(y_{\!{}_D}|x_{\!{}_D},h)$
with respect to $h$ 
is equivalent to maximizing the posterior $p(h|D)$.
However, minimizing the posterior energy is not necessarily equivalent
to minimizing the (conditional) likelihood energy.
This is due to the possibility of 
a $h$--dependent normalization of the likelihood energy.
If, however, in addition to complete data
also the likelihood normalization 
is  $h$--independent, i.e.,
$Z({\cal Y}_D|x_{\!{}_D},h)$  = $Z({\cal Y}_D|x_{\!{}_D})$, 
then
\be
Z_L 
%=\ind{h} p(y_{\!{}_D}|x_{\!{}_D},h)
= \frac{\ind{h}e^{-\beta_{\!{}_L} (y_{\!{}_D}|x_{\!{}_D},h)}}{Z({\cal Y}_D|x_{\!{}_D})}
= \frac{Z(y_{\!{}_D}|x_{\!{}_D},{\cal H})}{Z({\cal Y}_D|x_{\!{}_D})}
\ee
and the posterior becomes
according to Eq.(\ref{post-likeli})
\be
p(h|D) 
%=\frac{e^{-\beta^{\prime\prime} E(D|h)}}
%     {\ind{h}e^{-\beta^{\prime\prime} E(D|h)}}
=\frac{e^{-\beta_{\!{}_L} E(y_{\!{}_D}|x_{\!{}_D},h)}}{Z(y_{\!{}_D}|x_{\!{}_D},{\cal H})}
\propto e^{-\beta_{\!{}_L} E(y_{\!{}_D}|x_{\!{}_D},h)}
,
\label{post-likeli2}
\ee
where 
$Z(y_{\!{}_D}|x_{\!{}_D},{\cal H})$ 
=
$\ind{h}e^{-\beta_{\!{}_L} E(y_{\!{}_D}|x_{\!{}_D},h)}$
is an integral over the conditional variable $h$.
Thus, {\it for complete data and $h$--independent likelihood normalization
maximizing the posterior is equivalent to minimizing
the ($x_{\!{}_D}$--conditional) likelihood energy.}
%Thus,
%for uniform (possibly improper)
%$p(h)$, i.e., $h$--independent $c_L$,
%and $h$--independent $Z({\cal Y}_D|x_{\!{}_D},h)$, 
%and therefore $h$--independent $F({\cal Y}_D|x_{\!{}_D},h)$,
%minimizing the likelihood energy $E(y_{\!{}_D}|x_{\!{}_D},h)$
%is equivalent to a minimization of the posterior energy $E(h|D)$
%we are interested in.
%with respect to hypothesis $h$.

\begin{Rem}
%\begin{rem}
(Mixed representation by likelihood and posterior energies)
\label{mixed-rem}

Up to here we expressed 
posterior energies completely by likelihood energies.
It is often also useful, however,
to express the posterior energy by a sum 
of likelihood energy and posterior energy terms.
A prior term for example, may describe
a $h$ (and not data) generating process
directly given by $p(h|D_0)$.
Let the data $D$ = $(x_{\!{}_D},y_{\!{}_D})$ 
be therefore divided in measured data 
$D_L$ = $(x_L,y_L)$ which enter as likelihoods
and a generating prior $D_P$ = $(x_{\!{}_P},y_{\!{}_P})$ 
describing the posterior of the $h$--generating process.
Thus,
$D=D_L\cup D_P$
and 
$x_{\!{}_D}=x_L\cup x_{\!{}_P}$,
$y_{\!{}_D}=y_L\cup y_{\!{}_P}$.
Then, according to Bayes' theorem
and $p(h|x_L,D_P)$ = $p(h|D_P)$
\be
p(h|D) = \frac{p(y_L|x_L,h) p(h|D_P)}{p(y_L|x_L,D_P)}
=
\frac{e^{ -\beta_{\!{}_L} E(y_L|x_L,h) -\beta_{\!{}_P} E(h|D_P) + \ln p(y_L|x_L,D_P) } }
{Z(Y_L|x_L,h) \, Z({\cal H}|D_P)}
.
\ee
In this case,
minimizing the mixed energy
$\beta_{\!{}_L} E(y_L|x_L,h) + \beta_{\!{}_P} E(h|D_P)$
is equivalent to maximizing the posterior $p(h|D)$
if the normalization of the likelihood terms $Z(Y_L|x_L,h)$
is $h$--independent.
\end{Rem}

\section{Combinations of quadratic concepts}
\label{Combinations}
\subsection{AND: Classical regularization and gaussian processes}
\label{and}

In the classical situation of regularization theory
the aim is to approximate all available data,
training data $D_T$ as well as prior information $D_0$.
In logical terms, the aim is to approximate
$D_0$ AND $D_{T_1}$ AND $D_{T_2}$ AND $\cdots$ $D_{T_n}$.
A logical AND of events corresponds to a
product for probabilities 
$p(A, B) = P(A)p(B|A)$
or $p(A, B) = P(A)p(B)$ for independent events.
A product for probabilities corresponds to a sum for log--probabilities.
This holds also for concepts, if we interpret
concepts as linear functions of log--probabilities, i.e., 
as energies (see Appendix \ref{Statistics})
or errors (see Appendix \ref{Approximation-problems}).
In the following we will identify the events
$A$, $B$, etc.\ with data in the form of template functions $t_j$.

Consider a set of quadratic concepts $d^2_j$
with templates $T_N=\{t_j|1\le j\le N\}$ 
and concept operators $K_N=\{K_j|1\le j\le N\}$.
Assume we have data
$y_{\!{}_D}=t_1$ AND $t_2$ AND $\cdots$, i.e.,
\be
p(h|D)=p(h|T_N,K_N)\propto 
p(y_{\!{}_D}|x_{\!{}_D},h)=p(T_N|K_N,h) 
\ee
\be
=\prod_j^N p(t_j|K_j,h)
=\frac{e^{-\beta_{\!{}_L} \sum_j^N E(t_j|K_j,h)}}{Z({\cal T}^N|K_N,h)}
= e^{-\beta_{\!{}_L} E(T_N|K_N,h)}
,\ee
with 
$E(T_N|K,h)=E(y_{\!{}_D}|x_{\!{}_D},h)=\sum_j E(t_j|K_j,h)
=\sum_j^N d_{j}^2(h)/2$.
In writing $p(t_j|K_j,h)$ 
%= $p(t_j|h)$
we suppressed the dependency on the temperature $1/\beta_{\!{}_L}$.
The normalization factor
$Z({\cal T}^N|K_N,h)$ 
factorizes
\be
Z({\cal T}^N|K_N,h)
=\prod_j^N Z({\cal T}|K_j,h)
\ee
with
\be
Z({\cal T}|K_j,h)
= \int \left( \prod_x dt_j(x)\right) 
e^{-\beta_{\!{}_L} E(t_j|K_j,h)}
.
\ee
For quadratic concepts
the integrals are gaussian 
and therefore independent of the mean $h$, i.e.,
\be
Z({\cal T}^N|K_N,h)  = Z({\cal T}^N|K_N).
\ee
Thus, according to the results of the last Section
we may minimize 
the likelihood energy $E(y_{\!{}_D}|x_{\!{}_D},h)= E(T_N|K_N,h)$
instead of the posterior energy $E(h|D)$.
Writing now for simplicity $E(T_N|K_N,h) = E(h)$,
the following proposition is obtained by straightforward
calculation:

\begin{Pro}
%\begin{prop}
\label{and-prop}
(Probabilistic AND)
%An AND--like combination of quadratic concepts 
%is obtained by adding
A sum of squared distances 
$E_j$ = $d_j^2 (h)/2 = d(t_j,h)/2$
can be written
\[
E(h)=\sum_j^N E_j(h) =\frac{1}{2}\sum_j^N d^2_j(t_j,h) 
%\]
%\[
= \frac{1}{2} \sum_j^N \mel{h-t_j}{K_j}{h-t_j}
\]
\[
= \frac{1}{2} \left(
\mel{h}{K}{h}
+\sum_j^N \mel{t_j}{K_j}{t_j}
\right)
-\Big\langle h \Big|  \sum_j K_j t_j \Big\rangle
\]
\[
= \frac{1}{2} \left(
\mel{h-t}{K}{h-t}
+\sum_j^N \mel{t_j}{K_j}{t_j}
-\mel{t}{K}{t}
\right)
\]
\[
= \frac{d^2 (0,h)}{2} - \scp{\tilde t}{h} + \frac{N}{2} M_2 
\]
\be
= \frac{N}{2} \left( \bar d^2 (t,h) + V\right) 
= \frac{d^2 (t,h)}{2}+E_{min}
\ee
as sum of one squared distance
$d^2(t,h)$ from template average $t$
and a $h$--independent minimal energy $E_{min}$.
Squared distances are defined as
\be
d^2(0,h) = \mel{h}{ K }{h}
,
\ee
\be
d^2(t,h) 
= \mel{h\!-\!t}{ K }{h\!-\!t}
,
\ee
\be
\bar d^2(t,h) 
= \frac{d^2(t,h)}{N} 
= \mel{h\!-\!t}{ \bar K }{h\!-\!t}
,
\ee
with template average 
\be
t = K^{-1} \tilde t
,\quad
\tilde t = \sum_{j}^N K_j t_j
,
\ee
concept operators
\be
K = \sum_{j} K_j
,\quad
\bar K = \frac{K}{N} = \frac{1}{N} \sum_{j} K_j
,
\ee 
and $h$--independent minimal energy 
and template variance $V$ 
\be
E_{min} (T_N) = \frac{N}{2} V
,\quad
V = M_2-M_1^2,
\ee
with
\be
M_2 = \frac{1}{N}
%\left(
\sum_{j}^N \mel{t_{j}}{K_j}{t_j} 
,\quad
M_1^2 = \mel{t}{\bar K}{t}
%\right)
.
\ee
%which has  up to a factor $2N$
%the structure of a variance.
The linear stationarity equation for a functional 
$E= d^2(t,h)/2 +E_{min}$ reads 
\be
0  = K(t - h)
= \tilde t-Kh
.\ee 
For positive definite, i.e., invertible $K$,
this has solution
%$h=\overline{t}=K^{-1}\tilde t$.
$h=t=K^{-1}\tilde t$.
%\vspace{0.25cm}
%\end{prop}
\end{Pro}

We see that for quadratic concepts
addition does not lead to something really new:
The sum of quadratic functions is a quadratic function,
or in a probabilistic interpretation,
a product of gaussians is a gaussian.
{\it It is also interesting to note that 
for such additive combinations all
template functions $t_i$ corresponding to infinite data
can be eliminated from the formalism
by combining them in one term with template average $t$
and solving for a shifted
$h^\prime = h-t$.
%Then the necessary infinite amount of data
%is represented by a zero function and not explicitly visible
%within the formalism.
This elimination of `infinite data' templates
will not be possible in the nonlinear
combinations presented in the next sections.}

\begin{Rem}
(Normalization)
The normalization factor
$Z({\cal T}^{N}|K_N,h)$  to obtain $p(T_N|K_N,h)$
being a product of $q$--dimensional gaussian integrals
can be calculated explicitly
(see Eq.(\ref{exactgauss}))
\be
Z({\cal T}^{N}|K_N,h)
= 
\prod_{j}^{N} 
\int 
\left( \prod_{x}^{q} dt_{j}(x) \right)
e^{-\frac{\beta_{\!{}_L}}{2}d^2_j(t_{j},h)}
%\right) 
\ee
\be
=
\prod_{j}^{N} \left(
\pi^{\frac{q}{2}}
{\beta_{\!{}_L}}^{-\frac{q}{2}}
\left(\det K_{j}\right)^{-\frac{1}{2}}
\right)
=
\pi^{\frac{qN}{2}}
{\beta_{\!{}_L}}^{-\frac{qN}{2}}
\left(\det \prod_j^{N}K_{j}\right)^{-\frac{1}{2}}
.
\label{gauss1}
\ee
%which shows explicitly the independence of $h$.
For gaussian integrals 
the exact solution (\ref{gauss1})
coincides with a saddle point approximation
(see \ref{spa}).
Eq.(\ref{gauss1}) differs from a $T$--dependent
normalization over $h\in {\cal H}$
which would be necessary to obtain $p(h|D)$
if we would interpret a sum of quadratic concepts as posterior energy
$E(h|D)$ = $\sum_j E(h|t_{j},K_j)$
= $\sum_j d_{j}^2/2$,
\[
Z({\cal H}|D)
=\ind{h} e^{-\beta_{\!{}_P} E(h)}
=
\int 
\left( \prod_x^{q} dh(x) \right)
e^{-\frac{\beta_{\!{}_P}}{2}\sum_j^{N}d^2_j(t_{j},h)}
%\right) 
%\ee
\]
\be
=e^{-\beta_{\!{}_P} E_{min}} \ind{h} e^{-\beta_{\!{}_P} \frac{d^2(0,h)}{2} }
=e^{-\beta_{\!{}_P} E_{min}}
\pi^{\frac{q}{2}}
{\beta_{\!{}_P}}^{-\frac{q}{2}}
\left(\det \sum_j^{N} K_{j} \right)^{-\frac{1}{2}}
\label{gauss2}
.\ee
%While the difference in normalization does not matter for minimizing
%$E(h)$, it does if sampling according to $p(h|T)$
%is required.
Denoting by
$\avi{\cdots}{K}$
a gaussian average over $h$ with covariance 
$\propto K^{-1}$ = $(\sum_j K_j)^{-1}$
we recognize the moment generating function for $h(x)$
\[
M(\beta_{\!{}_P} \tilde t) = e^{\beta_{\!{}_P} E_{min}} Z({\cal H}|D)
=
\avi{e^{\beta_{\!{}_P} \scp{\,\tilde t\,}{\,h\,}}}{K}
\]
\be
=
\sum_{n=0}^\infty
\frac{\beta_{\!{}_P}^n}{n!} \int dx_1\cdots dx_n \,
\tilde t(x_1)\cdots \tilde t(x_n)
\avi{h(x_1)\cdots h(x_n)}{K}
.
\ee
Hence (functional) derivatives of $M(\beta_{\!{}_P} \tilde t)$
with respect to $\beta_{\!{}_P} \tilde t(x)$
generate  moments of $h(x)$
(see Section \ref{generating-functions},
in particular Wick's theorem \ref{Wick}).
\end{Rem}

\begin{Rem}
(Kernel methods)
Practically important is the case where the stationarity
equation can be solved in a space
with dimension $\tilde n\le n$
being the number of different $x$ values in the training data.
This can be much less
then the space necessary for a reasonable
discretization of the whole function $h$.
To see this we consider the classical situation of
mean--square data terms and  one additional prior concept
\be
E(h)=%\sum_k
\mel{h-t_T}{K_T}{h-t_T} + \mel{h-t_0}{K_0}{h-t_0}
\label{classical}
\ee
with
\be
\mel{h-t_T}{K_T}{h-t_T} + V
= \sum_j^n\mel{h-t_j}{K_j}{h-t_j}
= \sum_j^n \left(y_j-h(x_j)\right)^2 
.
\label{data-concept}
\ee
Thus, 
the operator $K_T = \sum_j K_j$ is diagonal,
commutes with the projector $P_T$
into the space ${\cal X}_T$ of training data
and has  matrix elements
\be
K_T (x,x^\prime ) 
= \delta (x-x^\prime ) \sum_j^n \delta (x-x_j)
=\delta (x-x^\prime ) \, \delta (0) n_x
, \quad
n_x = \sum_j^n \frac{\delta (x-x_j)}{\delta (0)},
\ee
containing, for discrete $x$ 
where $\delta (0) = \delta_{0,0} =1$ represents a Kronecker--$\delta$,
the number $n_x$ of data for every $x$.
For continuous $x$ the factor $\delta (0)$ becomes infinite
but will cancel in the following calculations.
The data template 
\be
t_T
= K_T^{-1} \sum_j K_j t_j
,\quad
t_T (x) 
= \frac{\sum_j^n \delta(x-x_j) y_j(x_j)}{\sum_j^n \delta (x-x_j)} 
= \frac{1}{n_x}\sum_j^{n_x} y_j(x_j)
\label{data-template}
\ee
is the average of $y$ values for given $x$.
Here $K_T^{-1}$ is defined on ${\cal X}_T$
and $t_T(x)$ = $0$ for $x$ not in the training data, i.e, with $n_x$ = $0$.
The stationarity equation 
\[ 0 = K_T (h-t_T) + K_0 (h-t_0) \]
yields
%\[ h = O_0^{-1} O_T (t_T - h)+ t_0) \]
%\raisebox{-0.2mm}{\shadowbox{
\be
h = K_0^{-1} a + t_0,
\ee 
with $a=K_T(t_T-h)$.
For known 
$K_0^{-1}$
it is sufficient to solve
\be
a = [I+K_T K_0^{-1}]^{-1} [K_T (t_T-t_0)]
\ee
for a function defined on the space ${\cal X}_T$
with dimension Tr$P_T$ = $\tilde n\le n$.  
The formulas can be adapted to the case
where $K_0$ has zero eigenvalues 
so it is not invertible over the whole space \cite{Wahba-1990}.
Classical examples of such {\it kernel methods} 
are gaussian Radial Basis Functions which use
as concept operator the pseudo--differential operator
\be
K_0 = 
\sum_{m=0}^\infty (-1)^{m} \frac{\sigma^{2m}}{m!\,2^m} \Delta^m
,
\ee
where $\Delta^m$ is the $m$--iterated laplacian
and which has a gaussian inverse.
They also include
piecewise linear interpolation ($K_0 = -\Delta$) 
and various spline methods (e.g., $K_0 = \Delta^2$)
\cite{Wahba-1990,Poggio-Girosi-1990,Girosi-Jones-Poggio-1995}.
\end{Rem}

\begin{Rem}(Robust error functions and support vector machine)
\label{svm}
As a generalization of quadratic concepts
one can allow nonquadratic functions $U$ of 
filtered differences $W(h-t)$ \cite{Zhu-Mumford-1997}.
In the simplest case 
$U$ is only applied to the data concept (\ref{data-concept})
of a functional of the classical form (\ref{classical}).
Then the nonlinear stationarity equation can also 
be restricted to the 
$\tilde n$--dimensional  space ${\cal X}_T$.
The data term can for example be
replaced by a gaussian mixture model (see Section \ref{OR}).
Robust error functions have flat regions and are therefore
insensitive to, i.e., robust against, 
changes in that flat region.
Also numerically flat regions in the error surface can be useful
because there the gradient vanishes and those regions do not
contribute to the stationarity equations.
Robust error functions
can for example down--weight large errors.
Typical cases are
filters used in image processing where edges represent
large discontinuities (regions with low smoothness) but are relatively likely.
Fig.\ref{robust} shows examples obtained by gaussian mixtures.
\begin{figure}
\begin{center}
\raisebox{-80mm}[0mm][20mm]{
%\hspace{-1cm}\includegraphics[scale = .60 ]{robust.ps}
\hspace{-1cm}\includegraphics[scale = .60 ]{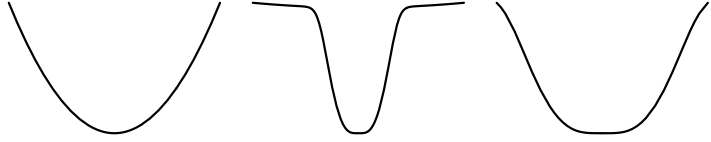}
}
\end{center}
\caption{Gaussian vs.\ two `robust' gaussian mixtures:
Left: Logarithm of one gaussian (parabola),
Middle: Logarithm of mixture with two gaussians 
with different variance and equal mean
(insensitive for large deviations),
Right: Logarithm of mixture with three gaussians 
with different variances and different means.
}
\label{robust}
\end{figure}
\begin{figure}
\begin{center}
\raisebox{-80mm}[0mm][20mm]{
%\hspace{-1cm}\includegraphics[scale = .60 ]{insens.ps}
\hspace{-1cm}\includegraphics[scale = .60 ]{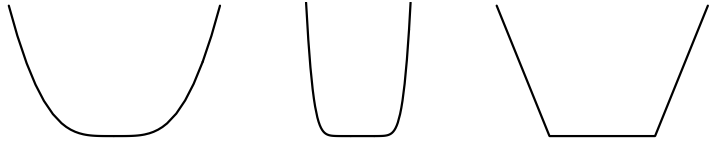}
}
\end{center}
\caption{Three 
robust error functions which are 
insensitive to small errors.
Left: Logarithm of mixture with two gaussians with equal variance 
and different means,
Middle: Logarithm of mixture with 11 gaussians with equal variance 
and different means,
Right: $\epsilon$--insensitive error.
}
\label{insens}
\end{figure}
Another variant of robust error functions is insensitive to small errors.
An example is an $\epsilon$--insensitive error 
(zero between $\pm\epsilon$, linear outside, see Fig.\ref{insens}) 
for the data term.
For example, expanding $h$ in a basis of
eigenfunctions $\Phi_k$ of the prior concept operator
\be
 K_0 = \sum_k \lambda_k \Phi_k \Phi_k^T 
,\quad
h(x) = \sum_k n_k \Phi_k (x)
\ee
yields for functional (\ref{classical})
\be
E(h) = \sum_i \left(\sum_k n_k \Phi_k (x_i)-y_i\right)^2
+\sum_k \lambda_k |n_k |^2 
.
\ee
Replacing the mean--square error data term
by an $\epsilon$--insensitive error
results in a standard quadratic programming problem
and is equivalent to Vapnik's support vector machine
\cite{Vapnik-1995,Girosi-1997,Smola-Schoelkopf-1998,Smola-Schoelkopf-Mueller-1998}.
\end{Rem}

\subsection{OR: Mixture models}
\label{OR}

Assume we believe that a function 
can be similar to either one of two templates.
Often a list of possible alternatives can be given.
For example, 
one may simply state that a face is that of a women or that of a man,
eyes may usually be open or closed,
or we may be able to give a list of possible pattern prototypes
for a time series.
For example, 
electrocardiograms, or similarly earthquakes,
have typical patterns which can 
appear in distinguishable variants.
Thus, we discuss here the case where
we want to approximate
$t_1$ OR $t_2$ OR $\cdots$ $t_n$.
For probabilities we
have 
$p(A \,{\rm OR}\, B) = P(A) + p(B) - P(A,B)$
where the last term vanishes for exclusive events.
For log--probabilities $L$ of exclusive events 
this means
\be
L(A \,{\rm OR}\, B) = \ln \left( e^{L(A)}+e^{L(B)}\right).
\ee
Relating now concepts to log--probabilities
by interpreting them as energies or errors, respectively, 
shows that alternative quadratic concepts 
lead to gaussian mixture models.

To be specific, let us discuss the cases of 
discrete input, output and generation noise
(See Section \ref{Complex-models}).
{\it Input noise} means that the measurement device 
producing outcome $y$ is not known exactly, i.e.,
\be
%p(y|x,{h}) = \int\!d\theta_x\,p(y|\theta_x,{h}) p(\theta_x|x)
p(y|x,h) = \sum_{i} p(y|x_i,h) \, p(x_i|x)
.
\label{input-noise}
\ee
For gaussian $p(y|x_i,{h})$ hypothesis $h$ defines the mean and
the independent variable $x_i$
defines the covariance $K^{-1}_i/\beta_{\!{}_L}$
of the measurement producing $y$.
Thus, model (\ref{input-noise})
%describes a situation where a measured value may have produced
%by measurement devices with different covariances 
%and leads thus
leads to a mixture of gaussians with different covariances.
Similarly, 
under {\it output noise} a measured value cannot be read off exactly,
\be
p(y|x,h) = \sum_{i} p(y|y_i)\, p(y_i|x,h) 
.
\ee
Thus having found $y$ the true measurement result
was one of the $y_i$.
For gaussian $p(y_i|x,{h})$ 
where $y_i$ is represented by a template function $t_i$
this leads to a likelihood for $h$ being a mixture of gaussians
with different means.
In particular, consider a situation of ambiguous data
where a set of $y_i$ cannot be distinguished by a 
measurement procedure so all $y_i$
lead to outcome $y$.
In that case
\be
p(y|y_i) = \frac{1}{N_i}\sum_i^{N_i} \delta(y-y_i)
,
\ee
%with $N_i$ = $\int \!dy\, \sum_i \delta(y-y_i)$, 
so the likelihood
becomes a simple sum over alternatives
\be
p(y|x,h) = \frac{1}{N_i}\sum_{i}^{N_i} p(y_i|x,h) 
.
\ee
{\it Generation noise} on the other hand means that
the $h$--producing probability is modeled as a mixture
\be
p(h) = \sum_i p(i) p(h|i)
, 
\ee
where $i$ could determine
mean and covariance of a gaussian $p(h|i)$.

For a maximum posterior approximation
we have to maximize the posterior density 
\be
p(h|D) =
p(h|D_L, D_P) =
\frac{p(y_{\!{}_L}|x_{\!{}_L},h) p(h|D_P)}
     {p(y_{\!{}_L}|x_{\!{}_L},D_P)}
.
\label{posterior-l-p-gen}
\ee
Including input, output, and generation noise
the posterior becomes
\be
p(h|D) \propto 
\sum_{ijk} 
p(k)p(h|k)
p(y|i)p(j|x)
p(i|j,h)
,
\ee
skipping the $h$--independent factor 
$p(y_{\!{}_L}|x_{\!{}_L},D_P)$.
We will especially consider the case where
$p(i|j,h)$ = $\prod_l^{N_L} p(i_l|j_l,h)$
factorizes into $N_L$ independent components
\be
p(h|D) \propto 
\sum_{ijk} 
p(k)p(h|k)
p(y|i)p(j|x)
\prod_l^{N_L} 
p(i_l|j_l,h)
,
\label{factorpost}
\ee
with $i$ = $\{ i_l|0\le l\le N_L\}$
and
$j$ = $\{ j_l|0\le l\le N_L\}$.
In particular, if also input and output noise factorize, i.e.,
$p(y|i)p(j|x)$ = $\prod_l p(y_l|i_l)p(j_l|x_l)$,
this would read
\bea
p(h|D) &\propto& 
\sum_{l^\prime}^{N_L} \sum_{i_{l^\prime}j_{l^\prime}k} 
p(k)p(h|k)
\prod_{l}^{N_L}  
p(y_l|i_l)
p(i_l|j_l,h)
p(j_l|x_l)
\nonumber\\&&
%p(h|D) \propto 
=
\sum_k p(k)p(h|k)
\prod_{l}^{N_L}  
\sum_{i_{l}j_{l}} 
p(y_l|i_l)
p(i_l|j_l,h)
p(j_l|x_l)
.
\eea
Choosing now quadratic concepts
for $h$--dependent (likelihood and generation) energies,
Eq.(\ref{factorpost}) becomes
\be
p(h|D) \propto 
\sum_{ijkm} 
p(k,m)p(h|K_k,t_m) 
p(T|i)p(j|K)
\prod_l 
p(t_{i_l}|K_{j_l},h) 
,
\ee
where
$T$ = $\{t_l|0\le l\le N_L\}$,
$K$ = $\{K_l|0\le l\le N_L\}$,
and 
separate summation variables 
for mean and covariance 
of the $h$--generating process have been introduced.

To find an error functional to be minimized
we express the posterior in terms of energies and free energies,
the latter determined by normalization factors.
Two kinds of free energies have to be included 
in an error functional.
\begin{itemize}
\item[1.]
Free energies depending on variables
for which we want to maximize the posterior.
may not be skipped.
%have to be included.
% in the error functional. 
%In our case this is $h$, so that
Thus, in general free energies of all likelihood terms
have to be included.
For the special case of gaussians, however,
normalization of likelihood terms is $h$--independent.
\item[2.]
Furthermore, 
it is often easier and more meaningful
to specify instead of a joint probability, e.g., $p(h,k)$, 
conditional and marginal probability, e.g., $p(k)$ and $p(h|k)$.
In that case also free energies depending on summation
variables $i$, $j$, $k$, $m$ have to be included.
Otherwise, such free energies would contribute
to marginal energies like $p(k)$,
and terms like $E(k)$ could not be interpreted
as energy for marginal $p(k)$.
Systems  specified by conditional energies
are also known as  
{\it disordered systems} (see Section \ref{disorder}).
\end{itemize}

Thus, in terms of energies,
\bea
p(h|D) &\propto& 
\sum_{ijk} 
e^{
 -\beta_h E(k) 
}
e^{
 -\beta_{\!{}_P} 
  \left(E(h|k) - F({\cal H}|k)\right) 
}
e^{
 -\beta_y 
  \left(E(y|i) - F({\cal Y}|i)\right) 
}
\nonumber
\\&&\times
e^{
 -\beta_{\!{}_L} 
  \sum_l\left(E(i_l|j_l,h)-F({\cal I}_{l}|j_l,h)\right)
}
e^{
 -\beta_x
  \left( E(j|x) - F({\cal J}|x)  \right)
}
,
\eea
for $i_l\in {\cal I}_l$,
$j\in {\cal J}$,
$y\in {\cal Y}$,
$h\in {\cal H}$.
In particular for quadratic concepts
this becomes,  
%\bea
%p(h|D) &\propto& 
%\sum_{ijkm} 
%e^{
% -\beta_h E(k,m) 
%}
%e^{
% -\beta_{\!{}_P} 
%  \left(E(h|K_k,t_m) - F({\cal H}|K_k,t_m)\right) 
%}
%e^{
% -\beta_y 
%  \sum_l\left(E(t_l|t_{il}) - F({\cal T}_l|t_{il})\right) 
%}
%\nonumber
%\\&&\times
%e^{
% -\beta_x
%  \sum_l\left( E(K_{jl}|K_l) - F({\cal K}_{jl}|K_l)  \right)
%}
%e^{
% -\beta_{\!{}_L} 
%  \sum_l\left(E(t_{il}|K_{jl},h)-F({\cal T}_{il}|K_{jl},h)\right)
%}
%.
%\eea
%
%assuming $p(k)$, $p(T|i)$, $p(j|K)$ 
%can be expressed easily as probabilities 
%and
choosing
$\beta_{\!{}_P}$ = $\beta_{\!{}_L}$ = $\beta$,
\be
p(h|D) \propto 
\sum_{ijkm} 
p(k,m)p_{ij}
e^{
 -\beta\left(
  \left(E(h|K_k,t_m) - F({\cal H}|K_k,t_m)\right) 
  +\sum_l\left(E(t_{i_l}|K_{j_l},h)-F({\cal T}_{i_l}|K_{j_l},h)\right)\right)
}
,
\label{postdensmix}
\ee
%\\&&\times
with
$p_{ij} \propto p(T|i)p(j|K)$.
Notice that despite $p(T|i)$ 
is not normalized over
$i$ we could nevertheless choose $p_{ij}$ to be normalized
over $i$ and $j$ by taking
$p_{ij}$ = $p(T|i)$$p(j|K)/\sum_i p(T|i)$.
For 
$k$-- and $j$--independent covariances
this becomes 
\be
p(h|D) \propto 
\sum_{im} 
p(t_m) \,p_i\,
e^{
 -\beta \left( E(h|K_P,t_m)
  +\sum_l E(t_{i_l}|K_{l},h)
  \right)
}
,
\ee
with 
$p_{i} \propto \prod_l p(T|i)$.
The posterior can be written completely in 
likelihood form by 
using that for quadratic concepts
the free energy is 
$h$--independent
$F({\cal H}|K_k,t_m)$ = $F({\cal T}_{m}|K_k,h)$
and
$p(h|K_P,t)$ = $p(t|K_P,h)$ 
under uniform prior.
The $h$--generating energy $E(h|K_P,t_m)$
can therefore be included in the sum over $l$.
Hence, combining $i$, $j$, $k$, $m$ 
into one multi--index $i$
and writing 
$t_{i_l}$ = $t_{il}$,
$K_{i_l}$ = $K_{il}$
Eq.(\ref{postdensmix}) reads
\be
p(h|D) \propto 
\sum_{i}  p_{i}\,
e^{ -\beta
     \sum_l^{N_L+1}
      \left(E(t_{il}|K_{jl},h)-F({\cal T}_{il}|K_{jl},h)\right)}
.
\label{mixture-posterior}
\ee
Hence, we obtain:

\begin{The} 
%\begin{thm} 
(Mixture model)
\label{mix-th}
Alternative quadratic concepts can be implemented
by the mixture model
\begin{equation}
E_M(h)
=-\ln p(h|D)
%\frac{p_i\, p(y_{\!{}_D}|x_{\!{}_D},h,i)}{p(D|x_{\!{}_D})} \right)
=-\ln \left(\sum_i^{N} p_i\, e^{-\beta E_i(h) + c_i}\right)
,
\label{mixture}
\end{equation}
%related to the mixture coefficients and normalization conditions
%$c_i$ = $c_i(\beta)$ = $-\ln\sum_i e^{-\beta E_i}$ %= $\beta F (h,{\cal I})$ 
where the component energies
\be
E_i(h) = \sum_{j}^{N_i} E(t_{ij}|K_{ij},h) = \sum_{j}^{N_i} E_{i{j}} (h)
\ee
are additive combinations
of quadratic concepts
\be
E_{i{j}} (h) = \frac{d^2_{i{j}}(h)}{2}
,\quad
d^2_{i{j}}(h) = \mel{h-t_{i{j}}}{K_{i{j}}}{h-t_{i{j}}}.
\ee
If $i$--dependent 
the normalization integrals
\be
c_i =    - \sum_{j}^{N_i} \ln \left( \int dt_{ij}
              e^{-\beta\, E_{ij}}
\right)+c
,
\label{c-i}
\ee
have to be calculated
up to an arbitrary constant $c$ 
so they do not interfere with mixture probabilities $p_i$.
The model has the 
stationarity condition\footnote{If $h$--dependent the $c_i$ = $c_i(h)$
additional terms arising from $\delta c_i/\delta h$
would contribute to the stationary equations.
}
%for $E_M$
\begin{equation}
0   %= <t_i(h) - K_i(h) h>_M
= t_M(h) - K_M (h) h, 
%\label{statEq2}
\end{equation}
with 
\be
%K_M = \sum_i p(D,i|h)  K_i
K_M = \sum_i a_i(h)  K_i
,\quad
K_i = \sum_{j} K_{i{j}}
,\quad
a_i(h) = p_i\, e^{-\beta E_i(h)+ c_i}
\ee
and
\be
t_M = \sum_i a_i(h)  K_i t_i 
= \sum_i a_i(h)  \tilde t_i
,\quad
t_i = K_i^{-1} \tilde t_i
,\quad
\tilde t_i = \sum_{j} K_{i{j}} t_{i{j}}.
\ee
%\end{thm}
\end{The}

{\it Proof:}
The form of $E_M$ follows directly from
Eq.(\ref{mixture-posterior}).
The stationary equation is obtained by straightforward calculation.
\hfill {\it q.e.d.}

It is in general nonlinear and can have multiple solutions.
Indeed, the model (\ref{mixture}) is in contrast to 
classical regularization functionals in general
non--convex.
The mixture model energy $E_M$ 
has with respect to the summation variable $i$
the form of a free energy for
a system at finite temperature.
Notice, also, the difference to mixture models
like they are used frequently in density estimation.
In such approaches $h$ is assumed as gaussian mixture.
In contrast, model (\ref{mixture})
represents a mixture model for the posterior density
and does not restrict $h$ to be a gaussian mixture.

\begin{Rem}
(Separate saddle point approximation for each component)
Alternatively, a maximum posterior approximation 
(which would be exact for gaussian integrals)
can also be applied to the $i$ components separately.
%The result leave the space ${\cal {H}}$ of 
%distributions $p{y|x,{h})$
Then, however, in general
a second minimization step has to be performed
also for approximation problems (see Appendix \ref{Approximation-problems}).
Notice that also for separate saddle point approximations
weighting factors $(\det K_i)^{-1/2}$ have to be calculated
in case of unequal component covariances
(see \cite{De-Bruijn-1981,Bleistein-Handelsman-1986,Negele-Orland-1988}
and Appendix \ref{map}).
\end{Rem}

\begin{Rem} (Low and high temperature limits)
In the interpretation of error (or energy) minimization
as saddle point approximation of a Bayesian risk
(see Appendix \ref{Bayesian})
the result becomes exact for zero temperatures,
provided that only one dominating stationary point survives
at zero temperature
(see Appendix \ref{spa}).
Thus we expect error minimization to be 
good at low enough temperature, i.e., large $\beta$.
At low temperatures, however, the stationarity equations generally 
have many solutions, 
making it difficult to find the dominating one.
(For a more detailed discussion of stable low temperature
solutions for the mixture model see \cite{Lemm-1996}).
In practice one often uses annealing methods
which start solving the stationarity equations at high temperature
and iteratively adapt the found solution to lower and lower temperatures
(see Appendix \ref{annealing}). 
%as soon as the included stationary points dominate.
Interestingly,
a saddle point approximation is for 
gaussian functions
also exact
at arbitrary temperatures $1/\beta$.
Figures \ref{map0.1} and \ref{map0.5}
in Appendix \ref{Bayesian}
show that for
high temperatures a sum of two gaussians with different centers
becomes approximately a gaussian again.
Thus,  
in that case a saddle point approximation is also
a good approximation at high temperatures.
More generally, one may perform a
(moment or high temperature) expansion of the exponential in powers of $\beta$,
\be
\sum_i p_i\, e^{-\beta E_i}
= \avi{e^{-\beta E_i}}{\cal I}
= \avi{1-\beta E_i+\frac{\beta^2}{2}E_i^2+\cdots}{\cal I}
.\ee
Thus, at high enough temperature
minimizing $E_M = -c_i-\ln\avi{e^{-\beta E_i}}{\cal I}$
with $i$--independent $c_i$
becomes minimizing 
$\avi{E_i}{\cal I}=\sum_i p_i\, E_i$.
This is just the AND--case of Proposition \ref{and-prop}.
At medium temperatures
larger differences to a full Bayesian approach have to be expected.
(Especially at `phase transitions' 
where solutions of the stationarity equations
vary strongly with temperature.) 
\end{Rem}

\subsection{An example with ambiguous prior}
%\begin{exmp}
\label{exam-two}
%(Model with two alternative concepts)
Consider the following situation with ambiguous prior:
Assume we want to implement that a function can be similar
to prototype $t_a$ OR another prototype $t_b$.
Thus, we take the two
prototypes as template functions $t_a(x)$ and $t_b(x)$.
Assume further, we choose the same
concept distances $d_a(h,t_a)$, $d_b(h,t_b)$ by taking 
for both templates the same concept operator
$K_{a,0}$ = $K_{b,0}$ = $K_0$.
In the following we consider a smoothness related sum of iterated laplacians
\be
K_0 = 
  \sum_l 
\lambda_l (-{\Delta})^l
,
\label{model-K}
\ee
where we understand
$(-{\Delta})^0 = I$.
For equal mixture probabilities
$p(1)=p(2)$ we obtain for (\ref{mixture})
\be
p^M = p^M_1 +p^M_2 
\propto 
e^{-E_M} 
= e^{-\beta E_1} + e^{-\beta E_2} 
= e^{-\frac{\beta}{2}(d_D^2+d_a^2)} 
            + e^{-\frac{\beta}{2}(d_D^2+d_b^2)}
,
\label{mix-model}
\ee
with mean--square data concept
$d^2_D = \mel{h-t_T}{K_T}{h-t_T}$
as in Eq.(\ref{data-concept}),
$d^2_a = \mel{h-t_a}{K_0}{h-t_a}$ and analogously
for $b$.
The stationarity condition for Eq.(\ref{mix-model}) is
\be
h
%= p^M_1 \overline{t}_1 + p^M_2 \overline{t}_2 .
= p^M_1 t_1 + p^M_2 t_2, 
\label{stat2}
\ee 
with component template averages
\be
t_1=(K_T+K_0)^{-1}\left( K_T t_T + K_0 t_a\right)
,\quad
t_2=(K_T+K_0)^{-1}\left( K_T t_T + K_0 t_b\right)
.
\ee
Because $p^M_1$ and $p^M_2$
are not functions but only two temperature dependent
convex mixing coefficients
the solutions for arbitrary temperatures
have to be on the one dimensional line
spanned by the two single components solutions
%$\overline{t}_1$ and $\overline{t}_2$.
$t_1$ and $t_2$.
The stationarity condition (\ref{stat2}) can be rewritten  
\be
h = 
%\overline{t} +{\rm tanh}\left( \frac{\beta}{2}(E_2- E_1)\right)
%\frac{\overline{t}_1 \!\! - \overline{t}_2}{2} .
t +{\rm tanh}\left( \frac{\beta}{2}(E_2- E_1)\right)
\frac{t_1 \!\! - t_2}{2} 
,
\ee
with total template average
\be
t=(K_T+2 K_0)^{-1}\left( K_T t_T + K_0(t_a+t_b)\right).
\ee
In the one--dimensional space of solutions spanned by the two
component averages, the equation is equivalent to that of 
the celebrated ferromagnet.
Thus, the two template mixture model shows the typical ferromagnetic
bifurcation 
(related to a `phase transition' of the underlying physical system)
switching with decreasing temperature
from one to two stable solutions.

\begin{figure}
\begin{center}
\raisebox{-20mm}[65mm][0mm]{
\includegraphics[scale = .45 ]{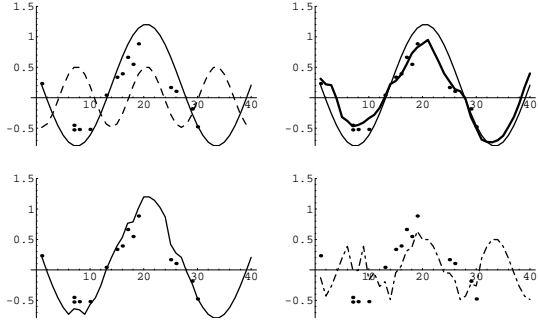}
}
\vspace{-1cm}
\caption{
Example with ambiguous priors.
The upper left diagram shows the two templates $t_a$ and $t_b$
and data $D_T$ drawn from the interval $[0,30]$.
The upper right diagram shows 
the true state of Nature $h_N$ (thick line)
in comparison with $t_a$.
The second row shows two
solutions $t_1$ and $t_2$ for vanishing smoothness coefficients
$\lambda_1 = 1$, $\lambda_i = 0$, $i>1$.
} 
\label{header}
\end{center}
\end{figure}

\begin{figure}
\begin{center}
%\hspace{-2cm}
%\begin{minipage}[b]{7.2cm}
%\begin{center}
%\raisebox{-20mm}[65mm][0mm]{
\raisebox{-30mm}[65mm][10mm]{
%\hspace{-0.5cm}
%\includegraphics[scale = .45 ]{mixpic.ps}
%\includegraphics[scale = .45 ]{figure3.ps}
\includegraphics[scale = .5 ]{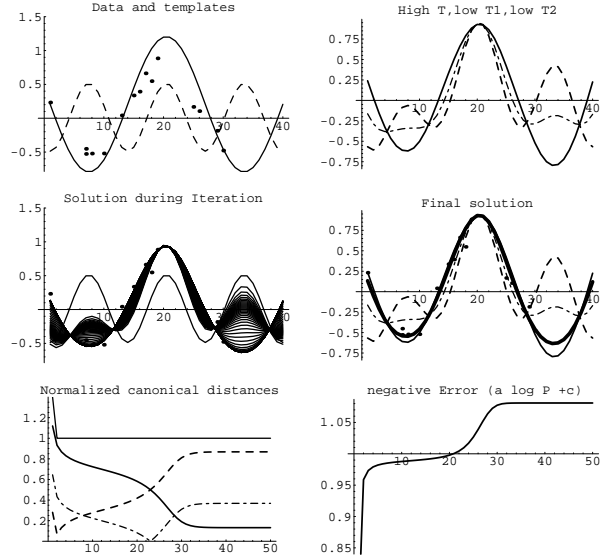}
}
%\end{center}
\caption{Mixture model (\protect\ref{mix-model}) 
near a bifurcation. 
%(`phase transition').
The chosen templates are
$t_a (x)=-\sin\left( \frac{3\pi(x-1)}{m-1}\right)+a_1$
and
$t_b (x)=\sin^2\left( \frac{3\pi(x-1)}{m-1}\right)+a_2$
with $m=40$ and the constants $a_i$ adjusted
to obtain zero mean over the interval $[0,40]$.
15 data points have been sampled on the interval $[0,30]$
from a function similar to $t_a$.
The value $\beta=0.105$ is near the critical value 
shortly before the second solution occurs.
The parameters for $K_0$ of Eq.(\protect\ref{model-K}) are
$\lambda_0=0.1$,
$\lambda_1=\alpha^2\lambda_0$,
$\lambda_2=\alpha^4\lambda_0$,
$\lambda_i=0$, $i>2$,
with
$\alpha=(m-1)/(3\pi)$.
Row 1. Left: Shown are the templates $t_a$, $t_b$, (dashed)
and the training data (dots).
Right: The solution in the high temperature limit $t$ 
(template average of data with $t_a$ and $t_b$)
$\beta\rightarrow 0$ (dashes with dots)
and the two low temperature solutions $t_1$, $t_2$
which average the data with one of the 
templates $t_a$ or $t_b$, respectively.
Row 2. Left: The evolving solution during iteration
with $A=K_M$ and relaxation factor $\eta=1$ 
(see Section \protect\ref{Learning})
starting from `wrong' initial guess $h(x) = t_b(x)$.
Right: The final solution compared with the high temperature 
and the two low temperature solutions.
%
%commented out wegen buf-size fehlermeldung
%
%(For the generalization behavior
%see the interval $[31,40]$ without data).
Row 3. Left:
Shown are the
%`normalized canonical distances' are
squared distances of the solution $h$ 
to the low and high temperature solutions 
$t_1, t_2, t$
during iteration:
$n_{t_i}^2(h)$ =
$\left\langle {h-t_i}\left|{K_T+K_0}\right|{h-t_i}\right\rangle/d^2(t_1,t_2)$,
for $t_i=t_1$(dashes), $t_i=t_2$(thick) and $t_i=t$(dashes with dots))
.
%normalized with respect to the distance of the low temperature solutions
%$\hat d_{12}(h)$ =
%$\left\langle {t_1-t_2}\left|{K_T+K_0}\right|{t_1-t_2}\right\rangle$.
After one iteration step the sum $n^2_{t_1}(h)+n^2_{t_2}(h) = 1$ (thin)
meaning that the solution moves along the one dimensional line connecting 
the two low temperature limits $t_1$ and $t_2$.
Right: The negative error $-E_M (h)$
% = %a \ln p_M(h) +c$ 
during iteration.
}
\label{mix-fig}
%\end{minipage}
%\hspace{-1cm}
%\begin{minipage}[b]{6cm}
%\begin{center}
%\raisebox{-20mm}[65mm][0mm]{
%\hspace{-1cm}
%\includegraphics[scale = .45 ]{priorpic2.ps}
%}
%\end{center}
%\caption{}
%\label{mix2}
%\end{minipage}
%
%{\footnotesize
%}
\end{center}
\end{figure}

Fig.\ref{mix-fig} shows a numerical example of the
two--template situation with 
\be
t_a (x)=-\sin\left( \frac{3\pi(x-1)}{m-1}\right)+a_1
,\quad
t_b (x)=\sin^2\left( \frac{3\pi(x-1)}{m-1}\right)+a_2
,
\ee
with $m=40$ and the constants $a_i$ adjusted
so the mean over the interval $[0,40]$ becomes zero
(see Fig.\ref{header}).
The shown results have been obtained by the EM (expectation--maximization)
algorithm (see Section \ref{Learning}).

\begin{figure}
\begin{center}
%\raisebox{-120mm}[0mm][75mm]{
\raisebox{-160mm}[0mm][75mm]{
\hspace{-1cm}
\includegraphics[scale = .70]{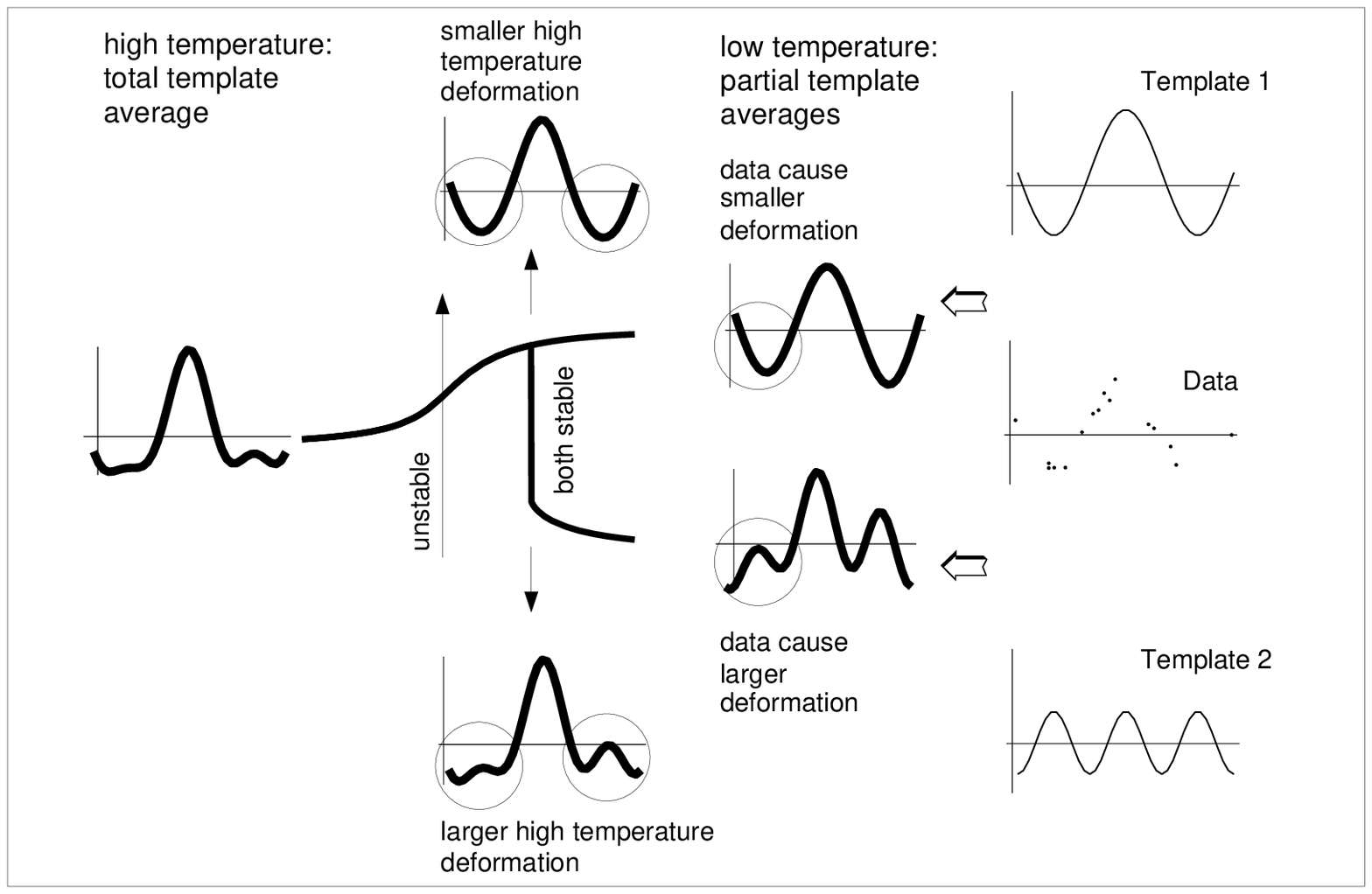}}
 
\end{center}
\caption{Scheme of temperature ($\beta^{-1}$) 
dependence of solutions for the error functional $E^M$ of model 
(\protect\ref{mix-model}).
The two prior templates $t_i$ 
(template 1 = $t_a$, template 2 = $t_b$)
and the training data $D_T$ (dots)
are shown on the right hand side.
At low temperature two solutions exist,
being the template average of the data $D_T$
with either one of the templates $t_1$ or $t_2$.
Going to higher temperatures the less well fitting solution
disappears at a critical temperature $1/\beta^*$.
The better fitting solution survives and transforms
for higher temperatures slowly into the high temperature solution
which is the template average of
all three templates: data $D_T$, $t_1$ and $t_2$.
%dependence of minima $h$ of 
%error functional $E^M[h]$ in the model of Example 8 
%  with data template (training data) 
%  and two continuous prior templates 
%  combined by OR and laplace operator as smoothness distance. 
%  For low temperatures two local minima appear 
%  corresponding to one of the two continuous prior templates 
%  deformed by the training data.
%  At high temperature all templates are effectively AND--ed 
%  and the solution is the total template average of all three templates.
%($\lambda_D=100$, $\lambda_0=1$, $\lambda_2=\alpha^2$,
%$\lambda_4=\alpha^4$, $\alpha=39/ (3\pi)$).)
}
\label{scheme}
\end{figure}

Fig.\ref{scheme} summarizes the temperature dependence of the model.
While the two {\it low temperature limits} $\beta\rightarrow \infty$
are given by the single component solutions
%$\overline{t}_1$ and $\overline{t}_2$,
$t_1$ and $t_2$,
the {\it high temperature limit} $\beta\rightarrow 0$
is the total template average $t$.
All three solutions
$t_1$, $t_2$ and $t$
correspond to a quadratic minimization problem (or a gaussian process)
with therefore linear stationarity equation.
%\end{exmp}

The fact
that the solutions for arbitrary temperatures
are in the convex hull spanned by the low temperature solutions $t_i$
is easily generalized to more then two alternative templates
with equal covariances.
This is a numerical important observation,
because the low (and high) temperature limits
are for quadratic concepts determined by linear equations.
As explained in Section \ref{and}
those linear equations can be solved in a 
$\tilde n\le n$ dimensional space 
given by the training data $D_T$.
If the number of components is small the optimal mixture
of low temperature solutions $t_i$
may then for example be obtained by cross--validation or similar techniques.
Then it is not necessary to discretize the whole function $h$
which is costly or even practically impossible for higher dimensional $x$.
This makes such calculations feasible 
also for higher dimensional ${\cal X}_R$ spaces.

\subsection{OR: Landau--Ginzburg regularization and interacting systems}
\label{Landau--Ginzburg}

The interpretation of quadratic concepts as energies
is not the only possibility.
Their quantitative relation to probabilities may be different.
For example, combination of concepts
may also be modeled by fuzzy logical operations.
%or similar techniques.
Those exist in many variations
but coincide on their boundaries with Boolean operations
\cite{Klir-Yuan-1995,Klir-Yuan-1996}.
We choose for concept distances $d(h,t)$
the logical interpretation
of $d^2(h,t)=0$, i.e., $h=t$, as `true'  
and of $d^2(h,t)=\infty$  as `false'.
For such variables a typical implementation of a logical OR 
is a product
\be
d^2_{(A\, {\rm OR}\, B)} = d^2_A d^2_B.
\ee
A product implementation for alternative concepts
is especially convenient
because then arbitrary combinations
of quadratic concepts by (an additive) AND and (a multiplicative) OR
are polynomial expressions in the concept distances.
The stationarity equations of such polynomial models are easily obtained
by setting the functional derivatives with respect to $h$ to zero
and given in the following theorem:

\begin{The}
%\begin{thm}
(Landau--Ginzburg regularization)
The polynomial model
\begin{equation}
E_{LG} 
=  \frac{1}{2\beta_{LG}} 
\sum_{i=1}^{N} 
\prod_{j=1}^{N_i}
\left(\beta_{LG} \, d_{ij}^2\right)
=  \frac{1}{2} \sum_{i=1}^{N} 
\beta_{LG}^{N_{i}-1} \prod_{j=1}^{N_i}
d_{ij}^2
,
\label{landau}
\end{equation}
%with a strictly monotonic increasing function $g$.  
has stationarity equation
\begin{equation}
K_{LG} (h) h = t_{LG}(h).
\end{equation}
Here
\be
K_{LG}(h) = \sum_i^N\sum_{j}^{N_i} M_{ij}(h) K_{ij}
\ee
 and 
\be
{t}_{LG}(h) =  \sum_i^N\sum_{j}^{N_i} M_{ij}(h) K_{ij} t_{ij} 
,
\ee
with 
\be
M_{ij} (h) = \beta_{LG}^{N_{i}-1} \prod_{k\ne j}^{N_i} d_{ik}^2(h)
,
\ee
and
\be
d^2_{ij} (h)= \mel{h-t_{ij}}{K_{ij}}{h-t_{ij}}
.
\ee
\end{The}
%\end{thm}
{\it Proof:}
The stationary equation follows
using the product rule for derivatives.
\hfill {\it q.e.d.}

In the high temperature limit $\beta_{LG}\rightarrow 0$
only quadratic terms survive and the stationarity
equation becomes linear.
The polynomial model $E_{LG}$ resembles the phenomenological Landau--Ginzburg 
treatment of phase transitions in physics
\cite{Landau-Lifshitz-1980,Binney-Dowrick-Fisher-Newman-1992}.

\begin{Rem}
(Mixed likelihood and posterior interpretation of error/energy)
A product $\prod_j d^2_{j}$ of quadratic concepts or log--probabilities 
does not implement a probabilistic OR for exclusive events $i$
with gaussian probability according to $d^2_i$.
% $i$ with $p(y_{\!{}_D}|x_{\!{}_D},h,i) = d^2_{i}$.
But like for $E_M$ of the mixture model, one may also try to interpret
additive parts $E_i$ of $E_{LG}$ = $\sum_i E_i/ (2\beta_{LG})$  
as probabilistic AND for independent events.
Hereby only terms $E_i$ depending on only one template 
$\{t_{ij}\}$ = $\{t_i\}$
can be interpreted as likelihood energies
$E_i(y_i|x_i,h)$.
Indeed, for an error $E_{LG}$ of form(\ref{landau}), being composed
out of squared distances $d^2_{ij}(h,t_{ij})$, the normalization 
$Z({\cal Y}_i|x_i,h)$ 
= $\ind{y_i} e^{-\beta_{\!{}_L} E_i(y_i|x_i,h)}$ 
for additive parts $E_i$ depending on only one $t_{i}$
is $h$-- and $t_{i}$--independent.
This can be seen using
$
d^2(h,t) = d^2(h-t)$  = $d^2(t-h) = d^2(t,h)
,
$
so that also arbitrary functions $g(d^2)$ of squared distances
fulfil $g(h,t)=g(d^2_{ij}(h,t)) = g(h-t) = g(t-h)$.
For such functions 
$\ind{h} g(h-t)$
%= $\ind{x} g(x)$
= $\ind{t} g(t-h)$
= $\ind{t} g(h-t)$
for unrestricted or periodic domain of integration.
Thus, an integral $Z({\cal Y}_D|x_{\!{}_D},h)$
depending on only one $t$ is both, $h$-- and $t$--independent.
For terms $E_i$, however, depending on different $t_{ij}$,
like for example a product $d^1_1d^2_2$,
this is not true in general.
Within a probabilistic interpretation such terms must 
either be interpreted
as posterior energy, % $E(h|D)$, 
or 
the $h$ (and possibly $i$)--dependent
logarithm of the normalization constant
%$F({\cal Y}_D|x_{\!{}_D},h)$ = $-\ln Z({\cal Y}_D|x_{\!{}_D},h)/\beta$
$\ln Z({\cal Y}_D|x_{\!{}_D},h)$
has to be subtracted.

%We divide therefore the terms $E_i$ in
%$N_L$ terms interpreted as likelihood energies $E(y_i|x_i,h)$
%and $N_P $ terms interpreted as posterior energies
%$E(h|D_i)$
%and corresponding likelihood 
%$D_L$ = $(x_L,y_L)$ 
%and posterior data 
%$D_P$ = $(x_{\!{}_P},y_{\!{}_P})$ as in Remark \ref{mixed-rem}.
%For
%($x_{\!{}_D}$--, $h$--)conditionally independent data $y_i$, 
%i.e.,
%$p(y_{\!{}_D}|x_{\!{}_D},h)$ = $\prod_i p(y_i|x_i,h)$
%one obtains
%\be
%p(h|D) =
%\frac{Z_P}{Z_L}
%\frac{e^{-\beta \left( \sum_i^{N_L} E(y_{L,i}|x_{L,i},h) 
%                       + \sum_k^{N_P} E(h|D_{P,k})     \right)}}
%       {\left(\prod_i^{N_L} Z({\cal Y}_{L,i}|x_{L,i},h)\right)
%        \left(\prod_k^{N_P} Z({\cal H}|D_{P,k})\right)}
%\propto e^{-\beta E_{LG}(D,h)}
%,
%\ee
%with $\beta_{LG}$--dependent $E_{LG}$,
%and
%\be
%Z_P = \prod_k^{N_P} \ind{h} p(y_{P,k}|x_{P,k},h),\quad
%Z_L = \ind{h}p(y_{\!{}_D}|x_{\!{}_D},h)
%\ee
%for uniform $p(h)$ = $p(h|x_{\!{}_D})$ = $p(h|x_k)$.
%As already discussed, choosing a likelihood interpretation only for
%terms depending on single templates the normalization
%$\prod_i^{N_L} Z({\cal Y}_{L,i}|x_{L,i},h)$ becomes $h$--independent
%and
%\be
%E_{LG} (D,h) = 
%\left( \sum_i^{N_L} E(y_{L,i}|x_{L,i},h) + \sum_k^{N_P} E(h|D_{P,k})  \right)
%.
%\ee
\end{Rem}

\begin{Exa}
%\begin{exmp}
(The two concept model again)
The two--template example of Section \ref{exam-two}
may alternatively be parameterized as follows:
\be
E_{LG} = 
d^2_D + \beta_{LG} d^2_a d^2_b.
\ee
The stationarity equation is cubic with either one or two stable
solutions
depending on $\beta_{LG}$.
Variations are
\be
E_{LG} = 
(d^2_D + d^2_a) (d^2_D+d^2_b),
\ee
or
\be
E_{LG} = 
d^2_D + d^2_a + d^2_b
+\beta_{LG}
\left( d^2_D + d^2_a d^2_b\right)
.
\ee
In the latter formulation $\beta_{LG}$
interpolates between OR at low temperatures and AND at high temperatures,
similar to the situation for mixture models.
For numerical results of the Landau--Ginzburg model 
for the example of Section \ref{exam-two}
see \cite{Lemm-1996}.
\end{Exa}
%\end{exmp}

\subsection{OR: Combination of methods and continuous transformations}
\label{continuous}

The number of components $i$ of a mixture model 
$\sum_i p_i e^{-\beta E_i+c_i}$ can be too large to be treated exactly.
Consider a set of template functions $t(\theta)$, 
with function values denoted $t(x,\theta)$, parameterized
by a continuous parameter (vector) $\theta\in \Theta$.
This results in a continuous mixture model
$\ind{\theta} p(\theta)e^{-\beta E (\theta) + c(\theta)}$ 
where the sum $\sum_i$ is replaced
by an integral $\ind{\theta}$.
Thus, analogous to the Bayesian integral over 
$h$ (or ${h}$, see Appendix \ref{Bayesian}),
the $\theta$--integral has to be solved
by an approximation.
This approximation can be of 
low temperature type (restricting to the most important contributions,
e.g.\ saddle point approximation)
%relying on an expansion of the
%exponent around a stationary point) 
or high temperature type
(starting from the mean, e.g.,
cumulant or moment expansion relying on an expansion of the exponential)
or a Monte Carlo integration (random sampling).
From the fact that the norm of a gaussian does not depend on its mean
it follows that changing templates
does not change the normalization constant 
as long as the covariance $K^{-1}$ is unaltered.
Hence, in cases with $\theta$--independent $K$ 
also the partition sum $Z(\theta)$ and thus $c(\theta)$
is $\theta$--independent.
For $\theta$--dependent $K$ 
the replica method or supersymmetric approaches
(see Appendix \ref{disorder})
can be useful in some cases.

Two important situations have to be mentioned where
continuously parameterized or {\it adaptive templates} appear naturally.
\begin{itemize}
\item[1.]
(Approximate structural models
or combining and extending arbitrary learning methods)
We have already discussed in \ref{templates}
that a template can be 
given by an arbitrary parameterized function,
for example regression models like
decision trees or neural networks 
\cite{Hertz-Krogh-Palmer-1991,Bishop-1995b,Ripley-1996,Golden-1996}.
Then $\theta$ denotes the parameter vector of such a model
and the integral $\ind{\theta}$
is a Bayesian integral over 
the possible parameter values
with prior probability $p(\theta)$.
The usual way to proceed
is to use a learning algorithm
to find an optimal approximation $\theta^*$.
This corresponds
to a saddle point approximation of the $\theta$--integral.
Including such templates $t(\theta)$
in the regularization functional means
that several different approximation methods can be combined
and restrictions given by their parameterization
can be overcome.
Using a combined saddle point approximation for $h$ and $\theta$
leads to stationarity equations which couple
the optimal $h^*$ and the optimal $\theta^*$.
Such a simultaneous saddle point approximation for $h$ and $\theta$
uses the same training data $D_i$ 
to determine both, $h^*$ and $\theta^*$.
%For equal covariances the combination of several
%methods becomes a temperature dependent
%linear combination.
(This can be compared with boosting methods which have received much attention
recently 
\cite{Schapire-Freund-Bartlett-Lee-1997}.)
Using 
%parameterized function templates in this forms
%means assuming structural
%models for the data generating process.
parameterized function spaces $t(x,\theta)$ as adaptive templates
corresponds to the prior assumption that the structure of 
the data generating process is approximately captured by 
their parameterization.
For example, $t(x,\theta)$ can model
a probable hierarchical organization of the generating process.
\item[2.]
(Approximate continuous symmetries)
Typically, templates $t(x)$ can appear in several variations $t(x,\theta)$.
These variations may be described by a continuous parameter vector $\theta$.
For example, one may wish to include 
translated, rotated, scaled or otherwise deformed `eye'--templates
in the regularization functional.
Again, a combined saddle point approximation can be used
to find the best fitting variant of the template $t^*$ = $t(\theta^*)$
and an optimal approximation $h^*$ simultaneously.
\end{itemize}

Consider now 
a general case of input, output, and generation noise.
Furthermore let 
$i$ = $(i_x,i_y,i_h)$
be a discrete mixture variable for which we want to 
treat the summation exactly,
and $\theta$ = $(\theta_x,\theta_y,\theta_h)$ 
a continuous mixture variable 
to be treated in maximum a posteriori approximation.
It follows from Eq.(\ref{posterior-l-p-gen}) 
that 
\be
p(h|D) \propto 
\sum_{i}  \int\!d\theta \,
p(i_h)p(h|i_h)
p(y|i_y,\theta_y)p(i_x,\theta_x|x)
\prod_l^{N_L} 
p(i_{y_l},\theta_{y_l}|i_{x_l},\theta_{x_l},h)
.
\label{factorpost2}
\ee
Especially for quadratic concepts,
\bea
p(h|D) &\propto& 
\sum_{i} \int\!d\theta \,
p(i_h,\theta_h)p(h|K_{i_h,\theta_h},t_{i_h,\theta_h}) 
p(T|i_x,\theta_x)p(i_y,\theta_y|K)
\nonumber\\&&\times
\prod_l p(t_{i_l,\theta_{x_l}}|K_{j_l,\theta_{y_l}},h) 
.
\eea
To obtain an error functional for minimization we write this 
in terms of energies and free energies
\bea
p(h|D) &\propto& 
\sum_{i} \int\!d\theta \,
e^{
 -\beta_h E(i_h,\theta_h)  
}
e^{
 -\beta_{\!{}_P} 
  \left(E(h|i_h,\theta_h) - F({\cal H}|i_h,\theta_h)\right) 
}
\nonumber
\\&&\times
e^{
 -\beta_y 
  \left(E(y|i_y,\theta_y) - F({\cal Y}|i_y,\theta_y)\right) 
}
e^{
 -\beta_x
  \left( E(i_x,\theta_x|x) - F({\cal I}_x,\Theta_x|x)  \right)
}
\nonumber
\\&&\times
e^{
 -\beta_{\!{}_L} 
  \sum_l\left(E(i_{y_l},\theta_{y_l}|i_{x_l},\theta_{x_l},h)
             -F({\cal I}_{l},\Theta_{y_l}|i_{x_l},\theta_{x_l},h)\right)
}
,
\eea
where
$\theta_x\in \Theta_x$,
$\theta_y\in \Theta_y$.
In particular, for quadratic concepts,
choosing
$\beta_{\!{}_P}$ = $\beta_{\!{}_L}$ = $\beta$
and writing the generation probability $p(h|K_k,t_m)$
in likelihood form under uniform prior
\be
p(h|D) \propto 
\sum_{i} \int\!d\theta \,
p_{i,\theta}\,
e^{
 -\beta
%\left(
%  \left(E(h|K_{i_h,\theta_h},t_{i_h,\theta_h}) 
%  - F({\cal H}|K_{i_h,\theta_h},t_{i_h,\theta_h}) \right) 
%  +
 \sum_l^{N_L+1}
    \left(E(t_{il}(\theta_{y_l})|K_{jl}(\theta_{x_L}),h)
    -F({\cal T}_{{il},\theta_{y_l}}|K_{il}(\theta_{x_l}),h)\right)
%\right)
}
,
\label{postdensmix2}
\ee
with
$t_{i_{y_l},\theta_{y_l}}$ = $t_{il}(\theta_{x_l})$,
$K_{i_{x_l},\theta_{x_l}}$ = $K_{il}(\theta_{y_l})$,
and 
\be
p_{i,\theta} = \frac{p(i_h,\theta_h)
                     p(T|i_y,\theta_y)p(i_x,\theta_x|x)
                     }{\sum_{i_y}\int\!d\theta_y\,p(T|i_y,\theta_y)}
.
\ee
Hence, instead of maximizing the posterior probability $p(h|D)$
we can minimize an error functional
%being a weighted sum of conditional likelihood energies
\begin{equation}
E 
=  
-\ln\sum_i 
%p(i)
\int \! d\theta \, 
%p(\theta|i)\,
p_{i,\theta}\,
e^{-\beta E_i(\theta ) + c_i(\theta,h)}
.
\end{equation}
%where we have written $p_{i\theta}$ = $p(i)p(\theta|i)$.
For $h$--, $\theta$--, $i$--independent 
normalization factor 
$Z({\cal T}_{{il},\theta_{y_l}}|K_{il}(\theta_{x_l}),h)$
also 
\be
c_i (\theta,h) = -\ln Z({\cal T}_{{il},\theta_{y_l}}|K_{il}(\theta_{x_l}),h)
+ c
,
\ee 
with arbitrary constant $c$,
is $\theta$--, $i$-- or $h$--independent 
and can thus be skipped.
This is the case for gaussians with
$\theta$--, $i$--independent covariances $K^{-1}/\beta$.

Consider for example 
component energies $E_i$ 
\begin{equation}
E_i (\theta) = E_{i,1} + E_{i,2}(\theta)
,\quad
E_{i,1} = \sum_{j}^{N_i} \frac{d^2_{ij}}{2}
, \quad
E_{i,2} (\theta ) = \frac{d^2_i(\theta )}{2}
,
\ee
where we separated an $\theta$--independent part,
with squared distances 
\be
d^2_{ij} = \mel{h-t_{ij}}{K_{ij}}{h-t_{ij}}
, \quad
d^2_i(\theta ) = \mel{h-t_i(\theta)}{K_i}{h-t_i(\theta )}
.
\label{twocon}
\ee

Performing for differentiable $E_i(\theta)$
the integral in saddle point approximation 
and assuming $p_{i,\theta}$ to be, compared to $E_i(\theta)$,
slowly varying at the stationary point, e.g.\ being uniform, 
(otherwise the derivative of $p_{i,\theta}$ has to be included)
results in the stationarity equation
\begin{equation}
0 = \frac{dE}{d\theta}
\Rightarrow
0 = 
\sum_i^{n_i} p_{i,\theta}\, e^{-\beta E_i(\theta )+c_i} 
\left(\frac{\partial E_{i,2}(\theta)}{\partial \theta}\right)
%\mel{\frac{\partial t_i(\theta)}{\partial \theta} }{K_i}{h-t_i(\theta )}
,
\label{transstat}
\end{equation}
with
\be
\frac{\partial E_{i,2}(\theta)}{\partial \theta}
= \mel{ \frac{\partial t_i(\theta)}{\partial \theta}}{K_i}{t_i(\theta )-h}
%= \mel{ s_it_i(\theta)}{K_i}{t_i(\theta )-h}
.
\label{theta-eq}
\ee

This equation has to be solved simultaneously
with the stationarity equation for $h$.
Thus, to find the optimal $h^*$
two (sets of) coupled stationarity equations have to be solved.
As those are usually nonlinear, a self--consistent solution
has to be found
by iteration, starting from some initial guesses $h^0$ and $\theta^0$.
The iteration can be performed, for example,
according to the following steps:
\begin{itemize}
\item[1.] Obtaining $h^0$:
Restrict the problem first to the training space ${\cal X}_T$,
i.e., the space of $x_k$
which are present in the training data $D_T$.
For equally weighted, standard  mean--square data terms
a natural initial guess for $h(x_k)$ in that subspace  
is the observed average of $y_k$ for the given $x_k$,  i.e.,
$h^0 (x_k)$ 
%= $\sum_{j|x_j = x_k}^{n_k} \frac{y(x_k)}{n_k}$ 
= $\frac{1}{n_x}\sum_i^{n_x} y_i(x_i)$
= $t_T (x_k)$   
with $n_x$ the number of training data for  $x$.
% = \sum_i \delta (x-x_i)$.
       %$\avi{y(x_k)}{\mbox{emp}}$.
\item[2.] Obtaining $\theta^1$:
In the second step  $h^0$ is inserted into
Eq.(\ref{theta-eq}), 
projected  to the training subspace ${\cal X}_T$,
and iterated with initial guess $\theta^0$
to obtain a new $\theta^1$
optimal in ${\cal X}_T$ for that $h^0$.
\item[3.] Obtain full $h$:
The intermediate solution $\theta^1$
is then used to solve for $h(x)$ for all $x\in {\cal X}_R$.
\item[4.] Continue iteration:
The stationarity equations are solved by
iteration until self--consistency is reached.
\end{itemize}

Consider the special case
%of only one component, i.e., $N_i=1$.
%Then 
of choosing 
%$K=P_T$ to be the projector into the space ${\cal X}_T$ 
$K$ proportional to the projector $P_T$ into the space ${\cal X}_T$ 
of training data and using as initial guess
for $h$ the data template (see Eq.(\ref{data-template}) )
$h^0 (x_k)$ = $t_T(x_k)$.
%$\avi{y(x_k)}{\mbox{emp}}$.
This
is equivalent to a standard mean--square error minimization
\be
0
= \sum_k \frac{\partial t_i(x_k,\theta)}{\partial \theta} 
(t_i(x_k,\theta) - y(x_k)).
\label{nn}
\ee
For example,
the template $t_i(x,\theta)$ can represent a neural network with
weights and biases included in $\theta$.
In that case Eq.(\ref{nn}) could be solved by backpropagation.
It is important to note
that in our context the resulting network $t_i(\theta^*)$, optimal
on the training data, is only the initial guess 
to be used for further iteration to obtain an optimal $h^*$.
In later stages of the iteration 
$t_i(\theta)$ can for example be retrained
by virtual examples drawn from $h(x)$.

In the case of continuous symmetry
transformations $S_i(\theta )$ 
the transformed template
is given by 
\be
t_i(\theta )= S_i(\theta )t_i = e^{\theta s_i} t_i
,
\end{equation}
where $\theta$ is the parameter vector of a continuous Lie group 
with vector of generators $s$ (see \ref{Symmetry}).
This yields the derivative
$\partial t_i(\theta)/\partial \theta$ = $s_i t_i (\theta )$
= $s_i S_i (\theta) t_i$
hence
\be
\frac{\partial E_{i,2}(\theta)}{\partial \theta}
=\mel{s_it_i(\theta )}{K_i}{t_i(\theta )-h}
%\ee
%and
%\begin{equation}
%\mel{s_it_i(\theta )}{K_i}{h-t_i(\theta )}
= \mel{S_i(\theta )s_it_i}{K_i}{h-S_i(\theta ) t_i}
\ee
for $s_i S_i = S_i s_i$.
Using $S^{-1}(\theta )$ = $S(-\theta )$,
this can also be written as
\be
\scp{s_i t_i}{S^T_i(\theta ) K_i S_i(\theta ) 
   \left(t_i-S_i(-\theta )h\right)}.
\end{equation}
For translations, for example, 
one finds with $S(\theta ) t(x)$ = $t(x+\theta )$
and $S^T (\theta )$ = $S^{-1}(\theta )$ = $S(-\theta )$
for vanishing commutator $[K_i,s]=K_is-sK_i=0$
\begin{equation}
\mel{st(\theta )\,}{K_i}{t(\theta )-h}
= \int \! dx \, dx^\prime \, 
\left( t(x^\prime )-h(x^\prime-\theta)\right)
K(x,x^\prime)
\frac{dt}{dx^\prime}.
\end{equation}

\begin{figure}
\begin{center}
\fbox{
\begin{minipage}{30mm}
\setlength{\unitlength}{4mm}
\linethickness{1pt}
%\begin{picture}(10,9)(-0.5,-2.5)
\begin{picture}(10,9)(-0.5,-2.5)
\put(0,0){\drawline(0,0)(1,0)(2,2)(3,0)(4,0)(5,-2)(6,0)(7,0)}
\thicklines
%\put(0,0){\drawline(0,2)(2,2)}
%\put(0,0){\drawline(2,2)(5,2)}
\put(0,2){\vector(1,0){2}}
\put(2,2){\vector(-1,0){2}}
\put(2,2){\vector(1,0){3}}
\put(5,2){\vector(-1,0){3}}
\put(2.5,5.5){\makebox(1,1){Adaptive}} %{\tiny Elastic}}
\put(2.5,4.0){\makebox(1,1){template}}
\put(0.7,2.3){\makebox(1,1){$\theta_1$}}
\put(3.0,2.3){\makebox(1,1){$\theta_2$}}
\put(0,0){\drawline(0,1.6)(0,2.4)}
\put(0,0){\drawline(2,1.6)(2,2.4)}
\put(0,0){\drawline(5,1.6)(5,2.4)}
\end{picture}
\end{minipage}
}
\end{center}
\caption{
The figure shows one of the templates
depending on two parameters
used for model (\protect\ref{adap-example}).
Parameter $\theta_1$ describes the location of the first maximum,
$\theta_2$ the distance between maximum and minimum.
}
\label{fig-temp}
\end{figure}
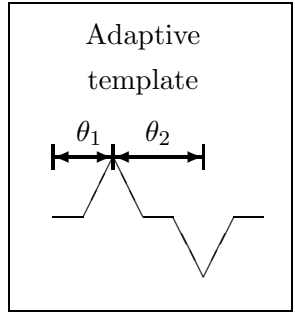

\begin{figure}
\begin{center}
%\raisebox{-110mm}[0mm][0mm]{
\raisebox{-120mm}[0mm][75mm]{
\includegraphics[scale = .70 ]{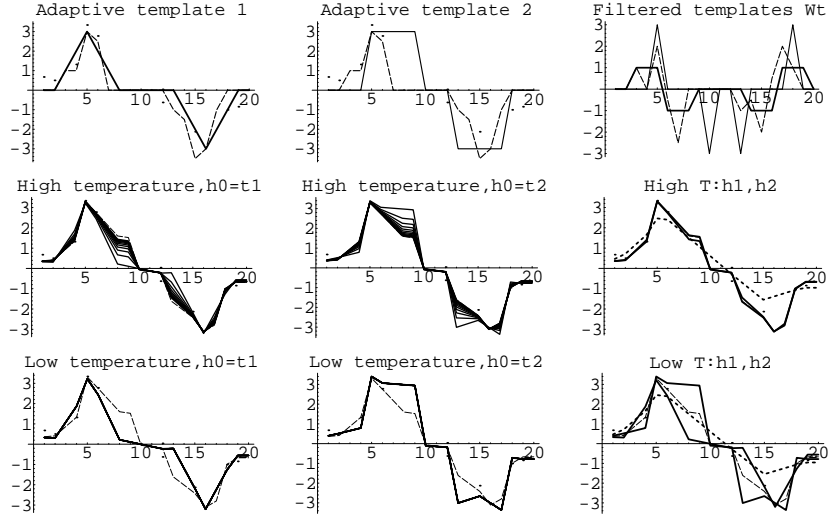}}
\end{center}
\caption{The figure presents numerical results 
for the model (\protect\ref{adap-example}) with two adaptive templates
$t_1(\theta_1,\theta_2)$ (shown in Fig.\protect\ref{fig-temp}
and another
$t_2(\theta_1^\prime,\theta_2^\prime)$ on a mesh with 20 points.
The figures in the first row show 
1. ten data points (dots), the true
function ${h}$ (dashes) from which the data have been sampled
with gaussian distribution (with $\sigma = 0.5$)
and $t_1$ (thick) with $\theta_1$, $\theta_2$ 
optimized for the given data, 
2. the same with $t_2$, 
and 3. the linearly transformed templates $Wt_i$ and $W{h}$.
Hereby $W$ is the derivative operator
which is a concept filter for the negative laplacian.
The second row shows for a high temperature case (here $\beta = 0.1$)
1. the solution $h$ during iteration for initial guess $h^0=t_1$ (thick),
and the high temperature limit (thin dashes).
The high temperature limit is the template average
of data and both adaptive templates, and is the 
limiting case where the OR--like mixture becomes a gaussian AND.
2. The same for initial guess $h^0=t_2$.
3. The final solutions for both cases and 
the classical solution for only one laplacian prior concept with zero
template $t_0\equiv 0$ (thick dashes).
Notice, that in the high temperature case the two solutions coincide.
The third row shows the same for a low temperature case (here $\beta = 0.5$)
Notice that here the two solutions evolving from the two 
initial guesses $h^0$ = $t_1$ and $h^0$ = $t_2$
do not coincide.
 }
\label{fig-adap}
\end{figure}

\begin{Exa}
%\begin{exmp}
\label{exam-cont}
(Adaptive templates)
Figure \ref{fig-temp} shows a simple example
of an adaptive template depending on two parameters.
In Figure \ref{fig-adap} numerical results are presented 
for a corresponding one--dimensional two template model 
\be
E (\theta,\theta^\prime)
= -\ln \left(
e^{-\beta E_1 (\theta)} + e^{-\beta E_2(\theta^\prime)}
\right),
\label{adap-example}
\ee
with mean--square data terms and 
the negative laplacian $-\Delta$ as concept operator $K$, i.e.,
\bea
E_1(\theta ) &=& \frac{1}{2}\sum_k (y_k-h(x_k))^2 
    + \frac{1}{2}\ind{x} 
      \left(\frac{\partial (h(x)-t_1(x,\theta_1,\theta_2))}
       {\partial x}\right)^2
,
\\
E_2(\theta^\prime ) &=& \frac{1}{2}\sum_k (y_k-h(x_k))^2 
    + \frac{1}{2}\ind{x} 
      \left(\frac{\partial 
       (h(x)-t_2(x,\theta_1^\prime,\theta_2^\prime))}{\partial x}\right)^2
.
\eea
The initial guesses $h^0$, $\theta^0$, ${\theta^\prime}^0$ 
have been obtained 
by projection into the space ${\cal X}_T$
according to the method discussed above.
Then $h$ and $\theta$ have been updated alternately
using complete search for $\theta$  and $\theta^\prime$ 
and an expectation--maximization (EM) algorithm for $h$.
%\end{exmp}
\end{Exa}

\section{Learning}
\label{Learning}

\begin{figure}
\begin{center}
%\raisebox{-110mm}[0mm][0mm]{
\raisebox{-140mm}[0mm][75mm]{
%\hspace{-1cm}\includegraphics[scale = .70 ]{RvsG.ps}}
\hspace{-1cm}\includegraphics[scale = .70 ]{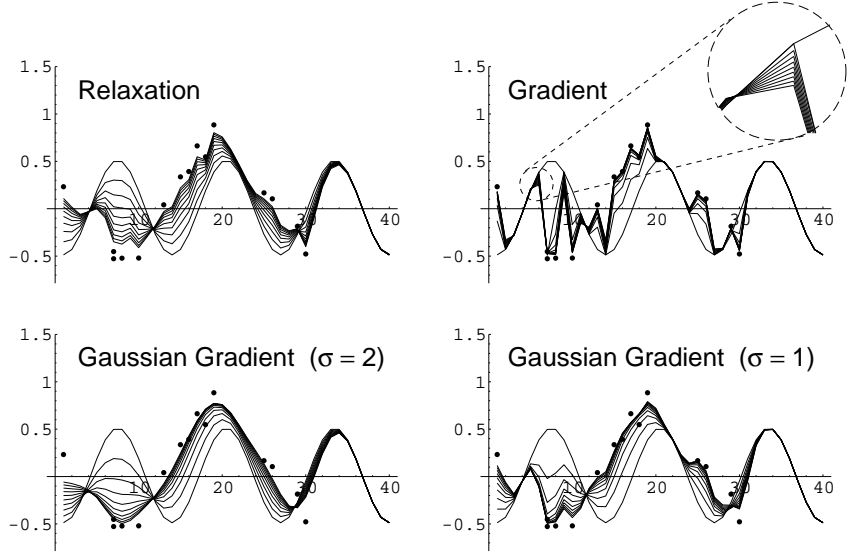}}
\end{center}
\caption{Comparison of learning algorithms
for mixture model (\protect\ref{mix-model}). 
Shown are the first iterations 
(starting with template $t_b$, see Fig.\protect\ref{scheme})
for the four iteration matrices 
$A=K_M$ (EM for mixture model (\protect\ref{mix-model}), labelled relaxation) 
$A=I$ (gradient)
and two gaussian $A^{-1}$ with standard deviations
$\sigma=2$, $\sigma=1$.
The gradient has obviously difficulties
capturing the long distance correlations
and requires an extremely small relaxation factor (step width) $\eta$.
The gaussian algorithm, which like the gradient requires no inversion,
performs relatively well on a global scale
without reaching the fine structure as fast.
}
\label{RvsG}
\end{figure}

The stationarity equations of the presented models 
are in general nonlinear, inhomogeneous
integro--differential equations.  
One may remark, that
similar equations appear for example in quantum
mechanical scattering theory,
where, similarly to templates or data, the inhomogeneities represent
measurable asymptotic states
(``channels'') of the system \cite{Lemm-1995}.  
Nonlinear equations have to be solved by iteration 
\cite{Fletcher-1987,Bazaraa-Sherali-Shetty-1993,Bertsekas-1995}.
Consider the equation to be solved written in a form
\be
K(h) h = t(h).
\ee 
Then an iteration procedure or {\it learning algorithm}
is obtained by selecting an operator $A$,
usually positive definite, 
and a relaxation factor $\eta$,
to be used with the updating rule 
\be
h^{k+1}
= h^{k} + \eta A^{-1} (t - K h^k).
\ee
For $\eta$ small enough
and a positive definite $A$ the function to be minimized decreases 
till reaching a local minimum.
$A$ may depend on the iteration step $k$ and $h^k$.
The gradient algorithm, for example, is obtained
when taking $A=I$ equal to the identity.
It does require matrix multiplication but no inversion.
A gaussian $A^{-1}$ on the other hand can approximate
local correlations induced by differential operators.
Choosing $A=K_M$ corresponds for error functional $E_M$ to the
expectation--maximization (EM) algorithm 
and Newton's method takes the negative Hessian.
Figure \ref{RvsG} compares 
for the mixture model (\ref{mix-model})
gradient algorithm,
an EM algorithm (labelled relaxation), 
and iteration with 
gaussian $A^{-1}$  with the two different 
standard deviations $\sigma=2$, $\sigma=1$.
It shows that the gradient has extreme
difficulties 
with long range correlations.
The gaussian $A^{-1}$ performs well, at least
at the beginning of the iteration.
That means it captures well the covariance structure of 
that particular problem.
This is useful, because in this case $A^{-1}$
is given and so no inversion is needed.
It can be seen that the gaussian algorithm 
especially at  larger variance takes longer  
to adapt the fine structure of the function.
This suggests to change the variance during iteration or
to combine it with the gradient algorithm.
The EM algorithm,
which works here quite well,
requires at every step inversion of the $h$--dependent
$K_M$.
The performance of specific algorithms is clearly problem dependent.
Recently, multiscale or multigrid methods have been become
particularly popular.

\section{Conclusion}

A new and relatively general method has been proposed to construct
problem specific error functionals (or posterior densities)
utilizing complex, `informative' a priori knowledge.
For that purpose a priori information is decomposed
into simple components representing constraints, 
like measured training data or approximate symmetries,
which the function to be approximated is expected to fulfil.
The constraints are represented by quadratic error terms
which are combined using logical operations
(conjunctions and disjunctions).
Conjunctions of quadratic concepts
lead to classical quadratic regularization functionals 
or, in Bayesian interpretation, to gaussian processes.
Disjunctions, 
representing situations with ambiguous data or ambiguous priors, 
result in nonconvex models
going beyond classical regularization 
and gaussian process approaches.
Two variants to treat ambiguous a priori informations
have been discussed in more detail:
mixture models for posterior densities
and polynomial models 
related to the Landau--Ginzburg treatment of phase transitions.
For simple numerical examples
the feasibility of the approach has been demonstrated.

The presented approach might especially be useful 
for complex learning tasks with relatively few available training data.
It seems also worth to study its relations
to knowledge transfer and combination of learning systems
in more detail.

\begin{appendix}
\section{Statistics and Statistical Mechanics}
\label{Statistics} 

Both disciplines, statistics and statistical mechanics,
deal with probabilistic models. 
Their differences in language and methods 
can be traced back to
differences in their typical applications.

Statistical mechanics has been developed for 
extremely large systems, like they appear in condensed matter physics.
Typical systems of statistical mechanics
are of high regularity, defined on a two or three dimensional grid
with local variables having the same range of possible values 
(e.g.\ $\pm 1$ for spins or real numbers for scalar fields), 
and only local interactions.
There are relatively few prototypical systems,
which have been studied extensively,
for example, the celebrated Ising model.
A complex machinery has been developed
to obtain results with very high accuracy
but requiring long and costly calculations.

Statistics, on the other hand, 
is mainly interested in the solution of 
a larger variety and  more application oriented problems.
Compared to prototypical systems studied in statistical mechanics,
the corresponding models are therefore
often smaller but less regular.
The main practical problem consists
in constructing adequate models.
The models are usually not expected to  
allow very precise predictions.
Hence,  their is no need to achieve, 
by costly calculations, a numerical accuracy
which is beyond the validity of the model.
Needed are fast, flexible, and robust algorithms.
%for daily use.

However, due to the increasing computing power becoming widely available,
the gap between the two disciplines gets smaller.
Their is now a growing and seminal interaction between 
the two disciplines, like 
in the development of
Monte Carlo methods or graphical models.
Typical areas where methods of statistical mechanics
can be applied easiest have a large amount of 
quantitative data of the same kind,
organized on a one, two, or three dimensional grid
with dependencies dominated by local interactions.
This, for example, is the case in
Bayesian image reconstruction 
or when predicting financial time series.
As it becomes therefore important to understand 
the languages of both approaches
we will discuss in the following the relations between 
concepts of statistics and statistical mechanics. 

We begin with some remarks concerning
the construction 
%of  probabilistic models requires definition
of probability distributions:
\begin{itemize}
\item[1.] (Normalization constants or partition sums)
The specification of a probability $p(x)$
starts with unnormalized numbers $Z(x)$.
To ensure the normalization condition $\ind{x} p(x) =1$
a normalization constant $Z=\ind{x} Z(x)$ 
has to be calculated.
For simple systems like a dice this is easy.
For large systems, like in typical systems studied in statistical mechanics,
this can be a highly non--trivial task.
Thus, 
unnormalized functions $Z(x)$ instead of probabilities $p(x)$
are the natural starting point for large systems.
Despite the fact that the $Z$ are not probabilities
because not normalized we will call them in the following  
{\it `unnormalized probabilities'}
or {\it partition sums}.
\item[2.] (Log--probabilities, information and energy)
A large system is usually constructed out of simpler subsystems, 
with the probabilities of the subsystems 
combined by multiplication,  
according to $p(x_1,x_2)=p(x_1)p(x_2|x_1)$.
Sums, however, are easier to handle than products 
and represent a somewhat more intuitive concept.
%and often also easier to intuitively for humans better understandable.
So it is easier to deal with infinite sums,
or in the continuous case, with integrals,
than with infinite or continuous products.
This means, that large systems are easier
constructed in terms of {\it logarithms of probability} $\ln p(x)$.
%as for those products of probabilities become sums.
We will show in the following
how {\it information} is related to 
expectations of the logarithmus of probabilities
and {\it energy} to expectations of the logarithm of
unnormalized probabilities $Z$.
\item[3.] (Conditional probabilities and disordered systems)
Instead of directly
constructing a complicated probability distribution $p(x)$
it is often easier to
break $p(x)$ down in simpler parts. % which are easier to specify.
This is done by selecting conditions $y$
under which $p(x|y)$ and $p(y)$ are relatively easy to specify.
The total probability is then obtained by
combining the alternatives $y$ according to $p(x)=\ind{y} p(y) p(x|y)$.
For example, it can be more easily to 
specify probabilities of symptoms for given specific diseases 
and probabilities of diseases than 
to write down directly a probability for the symptoms
averaged over all diseases in one step.
This, however, also means that due to the normalization requirements
$\ind{x} P(x|y)=1$, 
%$\forall y$
a normalization constant $Z(y)$ over $x$ must be calculated
for every $y$.
%value of condition $y$.
In statistical mechanics such models are called {\it disordered systems}. 
Such averages over
energy functions occur for example for spin glasses
\cite{Mezard-Parisi-Virasoro-1987,Fisher-Hertz-1991,Parisi-1992}.
\end{itemize}

Thus, while statistics
can often be formulated directly in terms of probabilities,
statistical mechanics uses a formulation in terms of 
logarithms of unnormalized probabilities. 
In the following their relations to the %These logarithms are related to 
concepts of energy and free energy will be discussed.
%We also remark that 
%from a Bayesian point of view
%the energy corresponds to error function
%for approximation problems

\subsection{Probability}

Especially for large and complex systems,
it is convenient to work with
the logarithm of `unnormalized probabilities' 
instead directly with probabilities.

\subsubsection{Normalization factors or partition sums}

Let $X$ be a random variable with possible values $x\in{\cal X}$
and assume a probability measure $p(A)$ defined
on a $\sigma$--algebra of possible events ${\cal A}$
being subsets of a set ${\cal X}$.
The event $x$ may be represented by a vector of real numbers.

Denoting unnormalized probabilities 
by $Z(x)\propto p(x)$
we write
\be
p(x)=\frac{Z(x)}{Z({\cal X})}
\ee
and for general events $A$ with $A\subseteq {\cal X}$,
i.e., including $A={\cal X}$
\be
Z(A) = \int_{x\in A} dx\, Z(x)  
\propto p(A)= \int_{x\in A} dx\, p(x)
, \quad p(A) = \frac{Z(A)}{Z({\cal X})}.
\ee
Introducing a second random variable $Y$
and using the compatible normalization
\be
Z(y) = Z({\cal X},y) 
= \ind{x} Z(x,y)  
\propto p(y)= \ind{x} p(x,y)
\label{cond-norm})
\ee
gives 
\be
Z({\cal X},{\cal Y}) 
= \ind{x} \ind{y} z(x,y)
= \ind{y} z(y)
= Z({\cal Y})   
\ee
and therefore
\be
p(y) = \frac{Z(y)}{Z({\cal Y})} = \frac{Z(y)}{Z({\cal X},{\cal Y})}
\quad {\rm and}\,\,
p(x|y) = \frac{p(x,y)}{p(y)} = \frac{Z(x,y)}{Z(y)}.
\ee
%Note that a compatible
%$Z(y)$ requires the calculation of
%the integral in Eq.(\ref{cond-norm})
Choosing $Z(y)$ to be not equal but proportional to $Z({\cal X},y)$
one obtaines
\be
p(x|y) = \frac{p(x,y)}{p(y)} 
= \frac{Z(x,y)\,Z({\cal Y})}{Z(y)\,Z({\cal X},{\cal Y})}.
\ee
In slight generalization of the language of statistical physics
we call unnormalized probabilities $Z$ also {\it partition sums}.

\subsubsection{Log--probabilities, bit numbers, and free energies}
\label{Logprob}

Log--probabilities $L$ are defined by
\be
L(x) = \ln p(x)
,\quad 
p(x) = e^{L(x)} 
\ee
and for $A\subseteq {\cal X}$
\be
L(A) = \ln p(A)
,\quad 
p(A) = e^{L(A)}.
\ee
In terms of log--probabilities $L$ a product
like 
\be
p(x,y) = p(x) p(y|x)
\ee
becomes a sum
\be
e^{L(x,y)} = e^{L(x)+L(x|y)}.
\ee
Log--probabilities are widely used in practice due to the fact that
it is often more convenient to deal with sums than with products.
Also the `quenching' effect of the logarithm can be important
in numerical calculations
when $p(x)$ varies over several orders of magnitudes.

Common are especially negative log--probabilities $b$
also called bit numbers 
\be
b(x) = -  \ln p(x)
,\quad 
p(x) = e^{- b(x)} 
\label{bit-num-def}
\ee
and for $A\subseteq {\cal X}$
\be
b(A) = -  \ln p(A)
,\quad 
p(A) = e^{- b(A)}.
\ee

Analogously the {\it free energy} is defined
for unnormalized probabilities $Z$ by
\be
F(x) = -  \frac{1}{\beta}\ln Z(x)
,\quad 
Z(x) = Z(x | \beta) = e^{- \beta F(x)}
,\quad 
p(x) = e^{-\beta\left( F(x) - F({\cal X})\right)}.
\label{free-e-def}
\ee
and for $A\subseteq {\cal X}$
\be
F(A) = -  \frac{1}{\beta}\ln Z(A)
,\quad 
Z(A) = Z(A | \beta) = e^{- \beta F(A)}
,\quad 
p(A) = e^{-\beta\left( F(A) - F({\cal X})\right)}.
\label{vec-ref}
\ee
The dependency on $\beta$ will in the following 
not always be denoted explicitly.
For $x\in {\cal X}$
the factor $e^{-\beta F(x)}$ is also known
as Boltzmann weight of $x$. 
The reasons for the introduction of the parameter
$\beta$ will be discussed later in detail.

\subsection{Random variables}

Information and energy are averages of special random variables.
We discuss now their connection to
bit numbers and free energy.

\subsubsection{Averages}

Recall the definition of an expectation or average of 
a random function $g(x)$ over ${\cal X}$
\be
g_X = \avi{g(x)}{\cal X} = \ind {x} p(x) g(x).
\ee
Analogously, we define an expectation over 
a subset $A\subseteq {\cal X}$ %as follows 
\be
g_X(A) = \avi{g(x,A)}{A} = \ind {x} p(x|A) g(x,A) 
= \int_{x\in A} \! dx\, \frac{p(x)}{p(A)} g(x,A) 
\label{A-average}
\ee
using $p(x,A)=p(x)$ and $p(x|A) = p(x)/p(A)$ for $x\in A$.
Using a second random variable $Y$
this can be generalized to 
a conditional expectation of a random function $g(x,y)$
over event $Y=y$ 
\be
g_X(y)  = \avi{g(x,y)}{{\cal X}|y} = \int dx\, p(x|y) g(x,y).
\label{y-average}
\ee
An average of form (\ref{A-average})
is obtained by choosing in (\ref{y-average})
a random variable $Y$ taking the value $y$
everywhere on $A$ but not on its complement.

\subsubsection{Information}

A random variable $C$ on ${\cal X}$
corresponds to a function $C(x)$ defined for every 
$x\in {\cal X}$.
A special example is a transformation $T$ of $p(x)$ 
which defines a corresponding random variable
by $C(x) = T(p(x))$ for ${x\in \cal X}$.
We will call in the following $C$
the canonical random variable of the transformation $T$ on ${\cal X}$.
Specifically, for every distribution $p(x)$
the corresponding bit number
$b(x) = -\ln p(x)$ can be considered
as random function
with the property of being defined not only
$x$-- but $p(x)$--dependent.  
%Bit numbers $b(x)$ interpreted  
The canonical random variable on ${\cal X}$ for bit numbers,
i.e., for the transformation $-\ln p(x)$, 
will be called {\it information} 
$I(x) = b(x) = -L(x)$.
(For an axiomatic approach, properties of information
 and the definition of related quantities
see for example 
\cite{Balian-1991,Beck-Schloegl-1993,Deco-Obradovic-1996}.)
Accordingly, Eqs.(\ref{bit-num-def})
can be written
\be 
I(x) = -  \ln p(x) ,\quad p(x) = e^{- I(x)}.
\ee
%Additional random variables besides $X$
%are considered as conditions
%\be 
%I(x|A) = -  \ln p(x|A) = I(x) - I(A) = b(x,A),\quad p(x|A) = e^{- I(x|A)}
%\ee
%\be 
%I(x|y) = -  \ln p(x|y) = b(x,y) ,\quad p(x|y) = e^{- I(x|y)}.
%\ee
Like any random variable the bit number or information $b(x)$
can be averaged
over the whole set ${\cal X}$, 
over a subset $A\subseteq {\cal X}$,
or conditioned on $Y=y$. One finds for the average or first moment
of the bit number or information $b$ the {\it average information} $I_X$ 
\be
I_X({\cal X}) =
\avi{b(x)}{\cal X} 
= -\avi{\ln p(x)}{\cal X} 
%=   \ind{x} p(x) b(x) = - \ind{x} p(x) \ln p(x)
\ee
\be
I_X(A)= 
\avi{b(x)}{A} 
= -\avi{\ln p(x)}{A} 
%=   \ind{x} p(x|A) b(x) = - \ind{x} p(x|A) \ln p(x)
\label{info-A}
\ee
\be
I_X(y) =
\avi{b(x,y)}{{\cal X}|y} = -\avi{\ln p(x,y)}{{\cal X}|y} 
\label{info-y}
\ee
%\bea
%I_X(y) =
%\avi{b(x,y)}{x|y} = -\avi{\ln p(x,y)}{x|y} 
%&=&   \ind{x} p(x|y) b(x,y)
%\label{info-y}\\
%&=& - \ind{x} p(x|y) \ln p(x,y).
%\nonumber\eea
In Eq.(\ref{info-A}) we used $b(x,A) = b(x)$ for $x\in A$
and $p(x|A) \ln p(x,A) = 0\ln 0=0$ for $x\notin A$
and in Eq.(\ref{info-y}) 
we allowed $y$--dependent $p(x,y)$.
It follows in accordance with the definition
from Eq.(\ref{info-A}) for $A=\{x\}$
that $I_X(A=\{x\}) = I_X(x) = I(x) = b(x)$ 
%for the events $X=x$.
or analogously
$I_X(Y=X=x) = b(x,x) = b(x)$ from Eq.(\ref{info-y}).
While $I$ and $b$ coincide on events $x\in {\cal X}$,
in general 
%, e.g.\ on events $A\ne \{x\}$,
the difference between  bit number and average information, 
i.e., the difference between transformed probability and 
expectations of the corresponding canonical random variable, 
is given by
\[
I_X(y) -b(y) = \avi{b(x,y)}{{\cal X}|y} - b(y)
= -\avi{\ln p(x,y)}{{\cal X}|y} + \ln p(y)
\]\be
= -\avi{\ln \frac{p(x,y)}{p(y)}}{{\cal X}|y}
= -\avi{\ln p(x|y)}{{\cal X}|y}
= \avi{b(x|y)}{{\cal X}|y}.
\label{i-b}
\ee
Defining the {\it entropy} (conditional information)
\be
H_X(y)= \avi{b(x|y)}{{\cal X}|y}
\ee 
this can be written
\be
I_X(y) -b(y) = H_X(y).
\ee
including
\be
I_X(A) -b(A) = H_X(A).
\ee
The relations
$b({\cal X})=0$ and $I_X(x) = b(x)$
yield the special cases
\be
H_X({\cal X}) = I_X({\cal X}) 
,\quad 
H_X(x) = 0.
\ee

%We will skip the subscript $X$ in the following.

\subsubsection{Energy}

In the same way as $b(x)$ also 
$F(x)$ may be interpreted as random variable 
on ${\cal X}$  called {\it energy} $E(x) = F(x)$.
($E$ is also called (euclidian) {\it action} in field theory).
Hence, energy is 
the canonical random variable of the transformation
$-\frac{1}{\beta} \ln (Z({\cal X}) p(x))$,
and one can write for Eqs.(\ref{free-e-def})
\be
E(x) = -  \frac{1}{\beta}\ln Z(x)
,\quad 
Z(x) = Z(x | \beta) = e^{- \beta E(x)}
,\quad 
p(x) = e^{-\beta\left( E(x) - F({\cal X})\right)}.
\label{canon-e}
\ee
In general $\beta F$ can be decomposed into 
$\beta F = \sum_i \alpha_i E_i$.
(This is the case, for example, in the grand canonical ensemble
of statistical physics where, compared to the canonical ensemble, 
a component corresponding to the particle number
is added.)

The analogue of average informations $I$
are then averages of the 
energy $E(x)=F(x)$ called {\it average energies} $E_X$
\bea
E_X({\cal X}) &= &\avi{F(x)}{\cal X} 
   = -\frac{1}{\beta}\avi{\ln Z(x)}{\cal X} \\
%    &=&   \ind{x} p(x) F(x) = - \frac{1}{\beta} \ind{x} p(x) \ln Z(x)
%\nonumber\\
E_X(A) &=& \avi{F(x)}{A} = -\frac{1}{\beta}\avi{\ln Z(x)}{A} 
\label{energy-A} 
,
\\
%&=&   \ind{x} p(x|A) F(x) = - \frac{1}{\beta}\ind{x} p(x|A) \ln Z(x)
%\nonumber\\
E_X(y) &=& \avi{F(x,y)}{{\cal X}|y} 
        = -\frac{1}{\beta}\avi{\ln Z(x,y)}{{\cal X}|y} 
\label{energy-y} 
%\\
%\hspace{6.5cm}\ee\[\hspace{2cm}
%&=&   \ind{x} p(x|y) F(x,y) = - \frac{1}{\beta}\ind{x} p(x|y) \ln Z(x,y)
%\nonumber
.
\eea
It follows in accordance with the definition
from Eq.(\ref{energy-A}) for $A=\{x\}$
that $E_X(A=\{x\}) = E_X(x) = E(x) = F(x)$ 
%for the events $X=x$.
or analogously
$E_X(Y=X=x) = F(x,x) = F(x)$ from Eq.(\ref{energy-y}).
%Hence, $E$ and $F$ coincide on events $x\in {\cal X}$.
We will skip the subscripts $X$ in the following.
While free energy and energy coincide on events
$x\in {\cal X}$ in general their difference is given by
\[
\beta E(y)-\beta F(y) = 
-\avi{\ln Z(x,y)}{{\cal X}|y} + \ln Z(y)
\]
\be
= -\avi{ \ln \frac{Z(x,y)}{Z(y)} }{{\cal X}|y}
= -\avi{ \ln \left( p(x|y) 
               \frac{ Z({\cal Y}) }{ Z({\cal X},{\cal Y})} \right)}{{\cal X}|y}
.
\ee
Choosing
\be
F(y) 
= -\frac{1}{\beta} \ln Z(y)
= -\frac{1}{\beta} \ln \ind{x} e^{-\beta F(x,y)}
\ee
i.e.,
\be
Z(y) 
= Z({\cal X},y) 
=\ind{x} Z(x,y)\ee
it follows
\be
Z({\cal Y}) 
= \ind{y} Z(y)
= \ind{x}\ind{y} Z(x,y)
= Z({\cal X},{\cal Y}).
\ee
Therefore
\be
-\avi{\ln Z(x,y)}{{\cal X}|y} + \ln Z(y) = -\avi{\ln p(x|y)}{{\cal X}|y}
\ee
and
\be
\avi{F(x,y)}{{\cal X}|y} - F(y)
%= \avi{F(x|y)}{x|y}
= \frac{1}{\beta} \avi{b(x|y)}{{\cal X}|y}.
\label{e-f}
\ee
Assuming $F(x|y)$ here to be defined as
\be
F(x|y)  
=-\frac{1}{\beta} \ln
   \frac{Z(x,y)}{Z(y)}
=-\frac{1}{\beta} \ln
   \frac{e^{-\beta F(x,y)}}{e^{-\beta F(y)}}
= -\frac{1}{\beta} \ln
   \frac{e^{-\beta F(x,y)}}{\ind{x}e^{-\beta F(x,y)}}
,
\ee
i.e.,
\be
Z(x|y) = \frac{Z(x,y)}{Z(y)} = p(x|y) 
,
\ee
one can also write for Eq.(\ref{e-f}) 
\be
\avi{F(x,y)}{{\cal X}|y} - F(y)
= \avi{F(x|y)}{{\cal X}|y}
,
\ee
parallelizing the corresponding
equation(\ref{i-b}) for bit numbers or informations $b$.
%can also be expressed in terms of informations
%\be
%\avi{I(x,y)}{x|y} - I(y)
%= \avi{I(x|y)}{x|y}.
%\ee
Using the definitions of entropy
%H(y) = \frac{1}{\beta} \avi{I(x|y)}{x|y}
$H(y) = \avi{b(x|y)}{{\cal X}|y}$
and energy  $E(y) = \avi{F(x,y)}{{\cal X}|y}$
Eq.(\ref{e-f}) states
that the $\beta$--scaled difference between 
energy and free energy is the entropy
\be
\beta E(y) - \beta F(y) =  H(y).
\ee
This is a generalization of the well known relation
of statistical physics
\be
\beta E({\cal X}) - \beta F({\cal X }) =  H({\cal X})
,
\ee
where the argument ${\cal X}$ is usually skipped.

The following table summarizes some of the relations
(recall that the variable $y$ in the table
can be replaced by $x\in {\cal X}$ or $A\subseteq {\cal X}$
and note that for $Z=\beta=1$
free energy and energy become identical to 
bit number and information):

\begin{center}
\begin{tabular}{|c|c|c|}
\hline
$\!\!\!\!\!$
Transformed &{Averages of canonical} &{}\\
probability& random variable& Difference\\
\hline
$\!\!\!\!\!$
bit number & information & entropy 
\\
%\rule[lift]{breite}{hoehe}
$\!\!\!\!\!$
\rule[-3mm]{0mm}{8mm}
$\displaystyle 
b(y)=-\ln p(y)$ &
$\displaystyle 
I(y)=-\avi{\ln p(x)}{{\cal X}|y}$ &
\parbox{4cm}{
 $H(y)=-\avi{\ln p(x|y)}{{\cal X}|y}$\\
 $= I(y)-b(y)$ 
 \rule[-2mm]{0mm}{2mm}
}
\\
\hline
$\!\!\!\!\!$
free energy & energy & entropy 
\\
$\!\!\!\!\!$
\rule[-4mm]{0mm}{10mm}
$\displaystyle 
 F(y)=-\frac{1}{\beta}\ln Z(y)$ &
$\displaystyle 
 E(y)=-\frac{1}{\beta}\avi{\ln Z(x)}{{\cal X}|y}$ &
\parbox{4cm}{
 $H(y)=-\avi{\ln p(x|y)}{{\cal X}|y}$
 \\
 $= \beta( E(y)-F(y))$ 
 \rule[-2mm]{0mm}{2mm}
}
\\
\hline
\end{tabular}
\end{center}

\subsection{Temperature and external fields}

Now we look at the role of the parameter $\beta$.
Its inverse $T=1/\beta$
is called {\it temperature} 
in statistical mechanics. 
The parameter $\beta$ can also be interpreted
as an {\it external source or field}
coupling to the conjugated random variable energy.
The energy can thereby be subdivided
into several components $E_i$
with corresponding conjugated $\beta_i$.
One $\beta_i$, for example, can be proportional to a magnetic field 
(or chemical potential, pressure, $\cdots$)
coupling to a magnetic moment $E_i = M$ (or particle number, volume, $\cdots$).
Also the calculation of moments or cumulants of 
other random variables
is often facilitated by introducing  
an external source coupling to that variable.

%\begin{rem}
We will discuss the following roles of $\beta$: 
\begin{itemize}
\item[1.] $\beta$ is Lagrange parameter determining the expectation 
of the energy $\ave{E}$ = $\ind{x} p(x) E(x)$.
Its variation defines an exponential family.
\item[2.]
$\beta$ is a homotopy parameter
used by annealing methods,
interpolating between easy and difficult to solve problems.
\item[3.]  
$\beta$ represents an external source or field coupling to the energy.
The cumulants of $E$ can be obtained as responses to a changing 
external field,
i.e., as derivatives of $\ln Z$ with respect to $\beta$.
For example,
$
\avi{E}{\cal X} = -({\partial}/{\partial \beta}) \ln Z({\cal X}).
%\label{diss-fluct}
$

\end{itemize}
%\end{rem}

\subsubsection{Maximum entropy and Boltzmann--Gibbs distributions}

It is well known that
minimizing the entropy and fixing  normalization condition
$\ind{x} p(x) =1$ 
and expectations
$\ind{x} p(x) E_i(x) = E_i({\cal X})$ 
by the Lagrange multiplier method
yield Boltzmann--Gibbs (or generalized canonical) distributions.
Indeed, adding the constraints with
Lagrange multipliers $\alpha_i$
and setting to zero the functional derivative of
\be
H(y) - \sum_{i=1}\alpha_i E_i (y) - \alpha_{{}_0} \avi{1}{{\cal X}|y}
\ee
\be
=
- \ind{x} \left( p(x|y) \ln p(x|y) 
- \sum_{i=1} \alpha_i E_i (x,y) p(x|y) - \alpha_{{}_0} p(x|y) 
\right)
,
\ee
with respect to $p(x|y)$
\bea
0\quad &=&\,\,\frac{\delta}{\delta p(x|y)}
\left(
H(y) - \sum_{i=1}\alpha_i E_i (y) - \alpha_{{}_0} \avi{1}{{\cal X}|y}
\right)\\ 
&=&\,\,
-\ln p(x|y)-1-\sum_{i=1}\alpha_i E_i (x,y) - \alpha_{{}_0} 
,
\eea
one finds
\be
p(x|y) 
= e^{-\sum_{i=1}\alpha_i E_i(x,y) -\alpha_{{}_0} - 1 }
= \frac{e^{-\beta F(x|y)}}{Z({\cal X}|y)}
,
\ee
with
\be
\sum_{i=1} \alpha_i E_i(x,y) = \beta E(x,y) = \beta F(x|y)
\quad {\rm and} \quad
Z({\cal X}|y) 
= e^{\alpha_0 +1}.
\ee

For $p(x|y)$ = $p(x)$ 
%$F(x|y)$  = $F(x|{\cal X})$  = $F(x,{\cal X})$  =$F(x)$
this gives Eq.(\ref{canon-e}).
%.
%Hence, for $X$ independent of $y$
%Then Eq.(\ref{canon-e}) is 
Thus any probability distribution $p(x)$
can be seen as result of a maximum entropy 
procedure with normalization constraint
and fixed expectation of the energy $E(x)$.

\subsubsection{Annealing methods}
\label{annealing}

The Lagrange multiplier $\beta$,
respectively the temperature $T=1/\beta$,
determines the average energy.
%Its inverse $T=1/\beta$ is also called {\it temperature}.
Introduction of
several Lagrange multipliers allows the fixation 
of several components $E_i$ of $E(x)$.
Varying $\beta$ defines an exponential family 
with canonical parameter $\beta$ and
and canonical statistic $E$.
In the high temperature limit 
\ $T\rightarrow \infty$, i.e., $\beta\rightarrow 0$, 
all $p(x)$ become equal.
In the low temperature limit $T\rightarrow 0$, i.e.,
$\beta\rightarrow \infty$, only events $x^*$ with maximal
$p(x^*)\ge p(x)$, $\forall x\in {\cal X}$ survive,
while all other events $x$ with
$p(x)< p(x^*)$, $\exists x^*\in {\cal X}$ 
are damped exponentially with decreasing temperature.
The temperature dependency is used by annealing methods
which are specific realizations of general 
homotopy or parameter continuation methods
and very important in practice
\cite{Kirkpatrick-Gelatt-Vecchi-1983,Mezard-Parisi-Virasoro-1987,Rose-Gurewitz-Fox-1990,Yuille-Kosowski-1994}.
They solve a difficult (e.g.\ multimodal) problem at finite or zero temperature
by beginning with an  easier (e.g.\ convex) high temperature
problem and then slowly decrease the temperature.

\subsubsection{Generating functions}
\label{generating-functions}

Moments or cumulants of random variables
can often be conveniently 
calculated by the use of generating functions.
%(or similar characteristic functions).
The $n$th {\it moment} $M_n$ of a random function
$g(x)$, with $E(x)$ being a special case,
is the expectation of its
$n$th power
\be
M_n(g) = \avi{g^n(x)}{\cal X}
,\quad {\rm e.g.\ } \quad
M_n(E) = \avi{E^n(x)}{\cal X}
.
\ee
For a vector valued function $g$ 
with components $g_i$ (e.g.\ $E_i$), $i\in {\cal I}$ the moments 
become the functions 
({\it unconnected correlation functions})
\be
M_{i_1,i_2,\cdots,i_n} 
= \avi{g_{i_1}(x)g_{i_2}(x) \cdots g_{i_n}(x)}{\cal X}
.
\ee
%For continuous index set ${\cal I}$
%the moments are functions
%from $n$ variables.
The {\it cumulants} 
(or {\it connected correlation functions})
are given by \cite{Gardiner-1990,Montvay-Muenster-1994}
\be
C_{i_1,i_2,\cdots,i_n} 
 = \sum_{{\cal P}} (-1)^{m-1} (m-1)! \,\,
M_{j_1,j_2,\cdots,j_{p_1}} 
M_{k_1,k_2,\cdots,k_{p_2}} 
\cdots 
M_{l_1,l_2,\cdots,l_{p_m}} 
\ee
with inverse
\be
M_{i_1,i_2,\cdots,i_n} 
= \sum_{{\cal P}} 
C_{j_1,j_2,\cdots,j_{p_1}} 
C_{k_1,k_2,\cdots,k_{p_2}} 
\cdots 
C_{l_1,l_2,\cdots,l_{p_m}} 
\label{moments}
\ee
where ${\cal P}$ denotes a partition of the
$n$ indizes into non--empty subsets
and $m$ is the number of factors in the summand
and one takes $C_0 = 0$.
For a small number $m$ of components $i$
moments and cumulants may be more conveniently indexed by
``occupation numbers'' $n_i$
\be
M_{(n_1,n_2,\cdots,n_m)} 
= \avi{g_{1}^{n_1}(x)g_{2}^{n_2}(x) \cdots g_{n}^{n_m}(x)}{\cal X}
.
\ee
For scalar function $g$, i.e., in the one component case ${\cal I} = \{ 1 \}$,
we will 
write the $n$th moment 
$M_{1,1,\cdots ,1} = M_{(n)}$ = $M_n$ and
$n$th cumulant
$C_{1,1,\cdots ,1} = C_{(n)}$ = $C_n$,
skipping the bracket for the sake of simplicity.
Hence, for a scalar $g$   %random variable
\be
M_0 = 1,
\quad M_1 = C_1
,\quad M_2 = C_2 + (C_1)^2
,\quad M_3 = C_3 + 3 C_2 C_1 +(C_1)^3,
\ee
\be
C_0 = 0,
\quad C_1 = M_1,\quad C_2 = M_2 - (M_1)^2
,\quad C_3 = M_3 - 3 M_2 M_1 +2(M_1)^3,
\ee
where the second cumulant is the well known variance.
Unlike moments, the cumulants are additive for independent subsystems,
i.e., $p(x_1,x_2)=p(x_1) p(x_2) 
\Rightarrow C_n (p(x_1,x_2))= C_n(p(x_1)) +C_n(p(x_2))$.
Another significant property of cumulants
is the possibility to set consistently $C_n=0$ for all $n>2$
(for gaussian distributions),
which is not possible for moments.
If, however, one $C_n\ne 0$ for $n>2$ then automatically
an infinite number of other $C_m$ do also not vanish
(See for example \cite{Gardiner-1990}).
For a multidimensional gaussian distribution with vanishing 
means $M_i=0$
Eq.(\ref{moments})
reduces to a sum over 
two--point connected correlation functions (or propagators)
$C_{ij}$ 
\be
M_{i_1,i_2,\cdots,i_{2n}} 
= \sum_{\rm Pairings} 
C_{j_1,k_1} 
C_{j_2,k_2} 
\cdots 
C_{j_n,k_n}. 
\label{Wick}
\ee
This relation is known as Wick's theorem.

%For continuous index set ${\cal I}$ 
%the cumulants are called (connected) correlation functions.
For a scalar function $g(x)$ the 
{\it moment generating function} is given by
\bea
M(\gamma)&=&
\frac{Z(\gamma )}{Z}
=\avi{e^{\gamma g(x)}}{\cal X}
=\ind{x} p(x) e^{\gamma g(x)}
\nonumber\\
&=&
\frac{1}{Z}\ind{x} e^{-\beta E(x) +\gamma g(x)}
=\sum_{n=0}^\infty \frac{\gamma^n}{n!} M_n
,
\eea
with
\be
Z(\gamma )= \ind{x} e^{-\beta E(x) +\gamma g(x)}
.
\ee
For vector $g$ (with discrete or continuous index set ${\cal I}$
one has $\gamma g = \sum_i \gamma_i g_i$
(or $\ind{i}$ for continuous $i$)
and $\gamma^n M_n$
has to be understood as
$\avi{\left(\sum_i \gamma_i g_i(x)\right)^n}{\cal X}$, i.e.,
%\be
%\avi{e^{\gamma g(x)}}{\cal X} =
\[
M(\gamma) =
\avi{e^{\sum_i^m \gamma_i g_i(x)}}{\cal X}
=\sum_{n=0}^\infty \frac{1}{n!} 
 \avi{\left(\sum_i^m \gamma_i g_i(x)\right)^n}{\cal X}
\]
\[
=\sum_{n=0}^\infty 
 \sum_{n_1,\ldots,n_m}^n
\delta (\sum_i^m n_i -n)
\frac{
      \gamma_{1}^{n_1}\cdots\gamma_n^{n_m}
     }{n_1!n_2!\cdots n_m!} 
 M_{(n_1,n_2,\cdots,n_m)} 
,
\]
\be
=\sum_{n=0}^\infty \frac{1}{n!} 
 \sum_{i_1,\ldots,i_n}^m\gamma_{i_1}\cdots\gamma_{i_n}
 M_{i_1,i_2,\cdots,i_n} 
.
\label{multidim}
\ee
It is easy to verify that the moments can be obtained by
\[
    \frac{\partial^n }{\partial \gamma^n} 
    M(\gamma ) \Big|_{\gamma = 0} 
   =\frac{\partial^n }{\partial \gamma^n} 
    \avi{e^{\gamma g(x)}}{\cal X} \Bigg|_{\gamma = 0} 
\]
\be
   =\avi{\frac{\partial^n }{\partial \gamma^n}e^{\gamma g(x)}}
         {\cal X} \Bigg|_{\gamma = 0} 
   =\avi{g^n(x) e^{\gamma g(x)}}
         {\cal X} \Bigg|_{\gamma = 0} 
   =\avi{g^n(x)}{\cal X} 
     = M_n (g).
\ee
or in the multidimensional case
\be
    \frac{\partial^n }{\partial \gamma_{i_1}\cdots\partial\gamma_{i_n}} 
    M(\gamma) \Big|_{\gamma = 0} 
    =\frac{\partial^n }{\partial \gamma_{i_1}\cdots\partial\gamma_{i_n}} 
    \avi{e^{\sum_i \gamma_i g_i(x)}}{\cal X} \Bigg|_{\gamma = 0} 
     = M_{i_1,i_2,\cdots,i_n} (g).
\label{multi2}
\ee
The cumulants are generated by differentiating the logarithm 
$\ln M(\gamma)$  = $C(\gamma)$ 
\be
    \frac{\partial^n }{\partial \gamma^n} 
    C(\gamma )\Big|_{\gamma = 0} 
    =\frac{\partial^n }{\partial \gamma^n} 
    \ln \avi{e^{\gamma g(x)}}{\cal X} \Bigg|_{\gamma = 0} 
     = C_n (g)
\ee
\be
    \frac{\partial^n }{\partial \gamma_{i_1}\cdots\partial\gamma_{i_n}} 
    C(\gamma ) \Big|_{\gamma = 0} 
    =\frac{\partial^n }{\partial \gamma_{i_1}\cdots\partial\gamma_{i_n}} 
    \ln\avi{e^{\sum_i \gamma_i g_i(x)}}{\cal X} \Bigg|_{\gamma = 0} 
     = C_{i_1,i_2,\cdots,i_n} (g).
\ee
Hence, $C(\gamma )$
is the {\it cumulant generating function} 
with Taylor expansion around $\gamma=0$
\be
C(\gamma )= \ln M(\gamma )=
    \ln \avi{e^{\gamma g(x)}}{\cal X} 
     = \sum_{n=0}^\infty \frac{\gamma^n}{n!}\, C_n (g)
,
\ee
or in the multidimensional case
\be
C(\gamma )= %\ln M(\gamma )=
    \ln \avi{e^{\sum_i\gamma_i g_i(x)}}{\cal X} 
     = \sum_{n=0}^\infty \frac{1}{n!} 
 \sum_{i_1,\ldots,i_n}\gamma_{i_1}\cdots\gamma_{i_n}
 C_{i_1,i_2,\cdots,i_n}.
\ee

Analogous to $\beta$ the parameter
$\gamma$ can be thought as an external source or field
(e.g., a magnetic field) coupling to $g$.
Even if a field $\gamma$ is not present in `reality'
the cumulants can still be 
calculated as derivatives at $\gamma=0$.
If a nonzero field $\gamma_0$ is present
we can incorporate $\gamma_0 g$  in a new energy
$-\beta E$
=
$-\beta^\prime E^\prime +\gamma_0 g$
replacing a given $\beta^\prime E^\prime$
and proceed as before.
Because sometimes useful, especially to obtain the 
Legendre transform of $C$, 
%to see the dependence on $\gamma_0$,
we will give in the following some of the formulae 
explicitly for nonzero field $\gamma_0$.
Assuming that derivatives and integration
can be interchanged one has
\be
    \frac{\partial^n }{\partial \gamma^n} 
    M(\gamma ) \Big|_{\gamma=\gamma_0} 
%=    \frac{\partial^n }{\partial \gamma^n} 
%    \frac{Z(\gamma )}{Z} \Bigg|_{\gamma=\gamma_0} 
=   \frac{\partial^n }{\partial \gamma^n} 
    \avi{e^{\gamma g(x)}}{\cal X} \Bigg|_{\gamma=\gamma_0} 
=   \avi{g^n(x)e^{\gamma_0 g(x)}}{\cal X} %\Bigg|_{\gamma} 
=   \frac{Z(\gamma_0 )}{Z} M_{n,\gamma_0} (g) 
\label{non-zero}
.
\ee
Including the  $\gamma_0$--term in the expectation
$\avi{\cdots}{{\cal X}|\gamma_0}$
we find for 
\be
M_{\gamma_0}(\gamma )
= \frac{Z(\gamma )}{Z(\gamma_0 )}
= \frac{M(\gamma )}{M(\gamma_0 )}
=\avi{e^{(\gamma-\gamma_0)g}}{{\cal X}|\gamma_0}
\ee
as derivatives
\[
    \frac{\partial^n }{\partial \gamma^n} 
    M_{\gamma_0}(\gamma ) \Big|_{\gamma=\gamma_0} 
=   \frac{\partial^n }{\partial \gamma^n} 
    \frac{Z(\gamma )}{Z(\gamma_0 )} \Bigg|_{\gamma=\gamma_0} 
=   \frac{\partial^n }{\partial \gamma^n} 
    \left( 
    \frac{\ind{x}e^{-\beta E(x) +\gamma g(x)}}
         {\ind{x}e^{-\beta E(x) +\gamma_0 g(x)}}
    \right)
    \Bigg|_{\gamma=\gamma_0} 
\]
\be
%=   \frac{\ind{x}g^n(x) e^{-\beta E(x)+\gamma_0 g(x)}}
%         {\ind{x}e^{-\beta E(x) +\gamma_0 g(x)}}
=\avi{g^n e^{(\gamma-\gamma_0)g}}{{\cal X}|\gamma_0 }\Big|_{\gamma=\gamma_0}
=   \avi{g^n(x)}{{\cal X}|\gamma_0} 
=   M_{n,\gamma_0} (g) 
.
\ee
Cumulants for nonzero fields are generated
by $C_{\gamma_0} = \ln M_{\gamma_0}$
\be
    \frac{\partial^n }{\partial \gamma^n} 
    C_{\gamma_0}(\gamma ) \Big|_{\gamma=\gamma_1}
=  \frac{\partial^n }{\partial \gamma^n} 
    \ln \avi{e^{(\gamma-\gamma_0)g(x)}}
            {{\cal X}|\gamma_0 }\Bigg|_{\gamma=\gamma_1} 
= C_{n,\gamma_1} (g)
,
\ee
where $\gamma_1$ = $\gamma_0$ is possible.
Because an additive constant does not change
the derivatives of a cumulant generating function 
also 
$C(\gamma )$ = $\ln M (\gamma )$ 
= $\ln M_{\gamma_0}(\gamma )+\ln Z(\gamma_0)-\ln Z(\gamma )$
= $\ln M_{\gamma_0}(\gamma ) + \ln M(\gamma_0)$
can be used as cumulant generating function for $\gamma_0\ne0$.
The expansion of the generating functions
$M_{\gamma_0}$ and $C_{\gamma_0}$ around $\gamma_0$ 
in powers of $(\gamma-\gamma_0)$ becomes
\be
M_{\gamma_0}(\gamma )=
    \avi{e^{(\gamma-\gamma_0) g(x)}}{{\cal X}|\gamma_0} 
    = \sum_{n=0}^\infty \frac{(\gamma-\gamma_0)^n}{n!}\, M_{n,\gamma_0} (g)
,
\ee
\[
C_{\gamma_0}(\gamma )= \ln M(\gamma )=
    \ln\avi{e^{(\gamma-\gamma_0) g(x)}}{{\cal X}|\gamma_0} 
\]
\be
    = \sum_{n=0}^\infty \frac{(\gamma-\gamma_0)^n}{n!}\, C_{n,\gamma_0} (g)
    = \sum_{n=0}^\infty \frac{(\gamma-\gamma_1)^n}{n!}\, C_{n,\gamma_1} (g)
,
\ee
and analogously for the multidimensional case.
Moments and cumulants for different
fields $\gamma_0$ and $\gamma_1$ 
are related according to
\be
M_{n,\gamma_0}
=
\frac{Z(\gamma_1)}{Z(\gamma_0)} 
\frac{\partial^n}{\partial \gamma^n} M_{\gamma_1}(\gamma)
\Big|_{\gamma=\gamma_0}
\!\!
=
\frac{ 
      \avi{g^n e^{(\gamma_0-\gamma_1)g}}{{\cal X}|\gamma_1}
 }
 { 
      \avi{e^{(\gamma_0-\gamma_1)g}}{{\cal X}|\gamma_1}
 }
=
\frac{ 
      \sum_{m=0}^\infty \frac{(\gamma_0-\gamma_1)^m}{m!} M_{m+n,\gamma_1}  
 }
 { 
      \sum_{k=0}^\infty \frac{(\gamma_0-\gamma_1)^k}{k!} M_{k,\gamma_1} 
 }
,
\label{diffmoment}
\ee
and
\be
C_{n,\gamma_0}
=
\frac{\partial^n}{\partial \gamma^n} C_{\gamma_1}(\gamma)
\Big|_{\gamma=\gamma_0}
\!\!
=
      \sum_{m=0}^\infty \frac{(\gamma_0-\gamma_1)^m}{m!} C_{m+n,\gamma_1}  
,
\ee
i.e., for the difference
\be
\Delta C_n (\gamma_0,\gamma_1) =
C_{n,\gamma_0} - C_{n,\gamma_1}
=  \sum_{m=1}^\infty \frac{(\gamma_0-\gamma_1)^m}{m!} C_{m+n,\gamma_1}  
.
\label{difflabel}
\ee

Because inverse temperature $\beta$ can be regarded as
a special nonzero field,
the moments and cumulants of the energy
can be obtained by 
\be
 \frac{\partial^n}{\partial (-\beta)^n} Z= Z M_n(E)
\ee
\be
 \frac{\partial^n}{\partial (-\beta)^n} \ln Z= C_n(E)
\ee
like
\be
  \frac{\partial^2}{\partial (-\beta)^2} \ln Z
  = C_2(E)
  = \avi{E^2(x)}{\cal X} - \avi{E(x)}{\cal X}^2.
\ee
The generalization to the multidimensional case 
is analogous to Eqs.(\ref{multidim}), (\ref{multi2}).
%Tables \ref{table1} and \ref{table2} summarize
%some of the results.
Equations for the second cumulants
connecting the derivative with respect to an external field 
(response, dissipation) with a variance (fluctuation)
are also known under the name dissipation--fluctuation theorems.

Moments of a function $g$ of $x$ can be expressed by moments of 
$\Delta x$ = $x-x_0$
by expanding $g(x)$ around $x_0$
\be
g(x) = \sum_{n=0}^\infty \frac{(x-x_0)^n}{n!}
       \frac{\partial^n}{\partial x^n}
       g(x)\Big|_{x=x_0}
     = e^{\scp{\Delta x}{\nabla^\prime}} 
       g(x^\prime)\Big|_{x^\prime=x_0}
     = e^{\scp{\Delta x}{\nabla}} 
       g(x_0)%\Big|_{x^\prime=x_0}
,
\ee
with gradient $\nabla^\prime g(x^\prime)$ 
       = $(\partial/\partial x^\prime) g(x^\prime)$
= $g^{(1)}(x^\prime)$
and analogously
$(\nabla^\prime)^n g(x^\prime)$ 
       = $(\partial^n/\partial (x^\prime)^n) g(x^\prime)$
       = $g^{(n)}(x^\prime)$.
We understand here and in the following the expression
$e^{\scp{\Delta x}{\nabla}}$
or $\scp{\Delta x}{\nabla}^n$
to be ``normal ordered'',
meaning that the derivatives act only to the right 
and not on $\Delta x$.
Analogously in the multidimensional case for example 
 $\nabla^2=\Delta$ 
creates the Hessian matrix.
This yields, 
\be
M_n = \avi{g(x)}{\cal X}
= \avi{g(x_0) + \Delta x \, g^{(1)} (x_0) 
          + \frac{(\Delta x)^2}{2} \, g^{(2)}(x_0) + \cdots}{\cal X}
\ee
\be
= \avi{e^{\scp{\Delta x}{\nabla}} g(x_0)}{\cal X}
= g(x_0) + \avi{\Delta x}{\cal X} g^{(1)} (x_0) 
         + \frac{\avi{(\Delta x)^2}{\cal X}}{2} g^{(2)}(x_0) + \cdots
%=\scp{e^{\avi{\Delta x}{\cal X}}{\nabla}} g(x_0)
.
\ee

We have seen that an expansion in moments or cumulants 
%of a given exponential %$e^{\gamma g}$
depends on the the choice of $\gamma_0$
or, equivalently, on the splitting of the $x$--dependent terms 
in one term $-\beta E(x)$ which defines an expectation 
$\avi{\cdots}{{\cal X}}$ 
and a field term $\gamma g(x)$.
Thus, there is a practically important
freedom in choosing different moment or cumulant expansions
as approximations for the same exponential.
Assume for example that moments for
$p^\prime(x)\propto e^{-\beta^\prime E^\prime(x)}$
can be easily calculated.
For $-\beta E = -\beta^\prime E^\prime + \gamma g(x)$
we can then write 
Eq.(\ref{diffmoment})
for an expectation $\avi{\cdots}{\cal X}$
of a function $q(x)$ under $p(x)\propto e^{-\beta E(x)}$
\be
\avi{q(x)}{{\cal X}}
=\avi{q e^{\gamma g}}{{\cal X}}^\prime
 \frac{Z^\prime}{Z}
=\frac{\avi{q e^{\gamma g}}{{\cal X}}^\prime}
      {\avi{e^{\gamma g}}{{\cal X}}^\prime}
=\frac{\avi{q \left(1+\gamma g+\cdots \right)}{{\cal X}}^\prime}
      {\avi{1+\gamma g+\cdots}{{\cal X}}^\prime},
\ee
with $Z^\prime$ = $\ind{x} e^{-\beta^\prime E^\prime(x)}$
and 
$\avi{\cdots}{\cal X}^\prime=\ind{x}p^\prime(x)\cdots $.
By expansion around $x_0$
this can also be expressed in terms of moments of $\Delta x$
\be
\avi{q(x)}{{\cal X}} =
\frac{e^{\gamma g(x_0)} \left( 
        q(x_0) 
        + \avi{\Delta x}{{\cal X},(\gamma)}^\prime
          \left( q^{(1)}(x_0) + q(x_0)\gamma g^{(1)} (x_0) \right)
          +\cdots \right)}
     { e^{\gamma g(x_0)} \left( 1+
         \avi{\Delta x}{{\cal X},(\gamma)}^\prime
          \gamma g^{(1)} (x_0)+\cdots\right)}
,
\ee
where the prefactor cancels.
This is the basis of saddle point approximation
(or loop expansion)
which will be discussed in Section \ref{spa}
and also the basis of importance sampling
in Monte Carlo calculations
\cite{Binder--1992,Binder-Heermann-1992,Gelman-Carlin-Stern-Rubin-1995,Neal-1996}.  
% -- \ref{loopex}.

To obtain equations which go beyond 
a cumulant expansion it is useful to consider the Legendre
transform of $C(\gamma)$.
The {\it Legendre transform} (or {\it effective action})
$\Gamma(\phi)$ of $C(\gamma)$
is defined by requiring
\cite{Rockafellar-1970,Glimm-Jaffe-1987,Zinn-Justin-1989}
\be
\Gamma(\phi) + C(\gamma)-\sum_i \gamma_i \phi_i
,
\label{legendre-functional}
\ee
to be stationary with respect to variations of fields
$\gamma_i$ (coupling to $g_i$)
at $\phi$ fixed.
(We may remark here that $C(\gamma)$ and thus its Legendre transform
$\Gamma$ depends on the choice of $g_i$.
A typical case is $g_i = x_i$.)
This means
\be
\frac{\partial C(\gamma)}{\partial \gamma_i} = \phi_i
= \avi{g}{{\cal X}|\gamma} = C_{i,\gamma}
,
\label{legendre1}
\ee
using also that $C$ is the cumulant generating function.
Using the chain rule 
one finds by differentiating
functional (\ref{legendre-functional}) with respect to $\phi_i$
\be
\frac{\partial \Gamma(\phi)}{\partial \phi_i} 
= 
-\sum_i 
       \frac{\partial C(\gamma)}{\partial \gamma_i}  
       \frac{\partial \gamma_i}{\partial \phi_i}  
+ \gamma_i
+ \sum_i
       \phi_i \frac{\partial \gamma_i}{\partial \phi_i}  
,
\ee
and thus
with Eq.(\ref{legendre1})
\be
\frac{\partial \Gamma(\phi)}{\partial \phi_i} = \gamma_i
.
\label{legendre2}
\ee
Now set
\be
\Gamma (\phi)
=
\sum_{n=1}^\infty \frac{1}{n!} \sum_{i_1,\cdots,i_n }
                               \Gamma_{i_1,\cdots,i_n }
            \Delta C_{i_1}\cdots\Delta C_{i_n}
,
\ee
with 
$\Delta C_i$ = $C_{i,\gamma}-C_i$ = $\phi_i-C_i$. 
This defines the {\it (proper) vertex functions} $\Gamma_{i_1,\cdots,i_n }$
for which   
\be
\Gamma_{i_1,\cdots ,i_n}
=
\frac{\partial^n \Gamma}{\partial \phi_{i_1}\cdots \partial \phi_{i_n}}
\Bigg|_{\gamma=0}
=
\frac{\partial^{n-1} \gamma_{i_n}(\phi )}
{\partial \phi_{i_1}\cdots \partial \phi_{i_{n-1}}}
\Bigg|_{\gamma=0}
,
\ee
using Eq.(\ref{legendre2})
and noting that $\Delta C = 0$ if $\gamma = 0$. 
Inverting the multidimensional version of Eq(\ref{difflabel})
(setting $\gamma_1 = 0$)
to obtain
%$\Delta \gamma = \gamma_0-\gamma_1$
$\gamma_i (\phi )$
in terms of $\Delta C_{i}$,
and 
defining {\it amputated correlation functions} 
\be
A_{i_1,\cdots ,i_n}
= \sum_{j_1,\cdots j_n} C_{i_1,j_1}^{-1}\cdots C_{n_1,j_n}^{-1} 
                        C_{j_1,\cdots,j_n}
,
\ee
one finds
\bea
\Gamma_{i_1} 
&=& 0,\\
\Gamma_{i_1,i_2} 
&=& C_{i_1,i_2}^{-1},
\label{gam1}
\\
\Gamma_{i_1,i_2,i_3} 
&=& - A_{i_1,i_2,i_3},\\
\Gamma_{i_1,i_2,i_3,i_4} 
&=& - A_{i_1,i_2,i_3,i_4}
 + \sum_{i,j} \left( A_{i_1,i_2,i} C_{i,j}  A_{j,i_3,i_4} \right.
\nonumber
\\&&
\left.
 + A_{i_1,i_3,i} C_{i,j}  A_{j,i_2,i_4} 
 + A_{i_1,i_4,i} C_{i,j}  A_{j,i_2,i_3} 
\right)
,\\
\cdots&&\nonumber
\eea
These equations can easily be inverted 
%to find amputated correlation functions from vertex functions
\bea
C_{i_1,i_2} 
&=& \Gamma_{i_1,i_2}^{-1},\\
A_{i_1,i_2,i_3} 
&=& - \Gamma_{i_1,i_2,i_3},\\
A_{i_1,i_2,i_3,i_4} 
&=& - \Gamma_{i_1,i_2,i_3,i_4}
 + \sum_{i,j} \left( \Gamma_{i_1,i_2,i} C_{i,j}  \Gamma_{j,i_3,i_4} \right.
\nonumber
\\&&
\left.
 + \Gamma_{i_1,i_3,i} C_{i,j}  \Gamma_{j,i_2,i_4} 
 + \Gamma_{i_1,i_4,i} C_{i,j}  \Gamma_{j,i_2,i_3} 
\right)
,\\
\cdots&&\nonumber
\eea
%It can be useful to formulate equations
%in terms of differences
%\be
%\left(\Sigma\right)_{i_1,\cdots i_n}
%=
%\Gamma_{i_1,\cdots,i_n} - \Gamma^0_{i_1,\cdots,i_n}
%\ee
%(the full or perturbed) 
Often it is only possible
to find (full or perturbed) vertex functions
$\Gamma_{i_1,\cdots,i_n}$  
by expanding around known vertex functions
$\Gamma^0_{i_1,\cdots,i_n}$
for a solvable (reference or unperturbed) system.
For example, Eq.(\ref{gam1})
can be reformulated
in terms of the  ``self energy'' 
%$\left(\Sigma\right)_{i_1,i_2}$ 
being the difference 
between perturbed and unperturbed two point vertex functions
\be
\left(\Sigma\right)_{i_1,i_2} = 
\Gamma_{i_1,i_2} - \Gamma^0_{i_1,i_2}
.
\ee
This results in
\be
C_{i_1,i_2} = 
C^0_{i_1,i_2}
- \sum_{j,k} C^0_{i_1,j} \left(\Sigma\right)_{j,k} C_{k,i_2} 
,
\ee
which expresses the full connected two point correlation function 
$C_{i_1,i_2}$
in terms of an unperturbed $C^0_{i_1,i_2}$.
An approximation $(\hat{\Sigma})$ for the self energy
can be obtained for example by perturbation theory
with respect to the unperturbed reference system.
Then a corresponding 
self-consistent solution $\hat C_{i_1,i_2}$
can be found by iteration according to
\be
\hat C_{i_1,i_2}
= C^0_{i_1,i_2}
- \sum_{j,k} C^0_{i_1,j} (\hat{\Sigma})_{j,k} C^0_{k,i_2} 
+ \sum_{j,k,l,m} C^0_{i_1,j} (\hat{\Sigma})_{j,k} C^0_{k,l} 
                 (\hat{\Sigma})_{l,m} C^0_{m,i_2} 
\label{selfenergy})
.
\ee

\vspace{1.0cm}
The following two Tables show generating functions 
%and explicitly 
for zero field
%
%here
%\vspace{0.5cm}
%and for zero field
%, respectively
\nopagebreak
%\begin{table}
%\renewcommand\arraystretch1.0
\begin{center}
\begin{tabular}{|c|c|}
\hline
generating function & derivatives \\
\hline
 \rule[-6mm]{0mm}{14mm}
 $Z(\gamma )$ % $= Z \avi{e^{\gamma g(x)}}{\cal X}$   
 = $\ind{x} e^{-\beta E(x)+\gamma g(x)}$      
 & $\displaystyle 
    \frac{\partial^n }{\partial \gamma^n}Z(\gamma )\Big|_{\gamma = 0} 
    = Z M_n(g)$\\
\hline
 \rule[-6mm]{0mm}{14mm}
 $\ln Z(\gamma )$
 = $\ln \ind{x} e^{-\beta E(x)+\gamma g(x)}$       
 & $\displaystyle 
    \frac{\partial^n}{\partial \gamma^n} \ln Z(\gamma )\Big|_{\gamma = 0} 
     = C_n (g)$\\
\hline
 \rule[-6mm]{0mm}{14mm}
 $M(\gamma )$ 
 = $\frac{Z(\gamma )}{Z}$  
 = $\avi{e^{\gamma g(x)}}{\cal X}$ 
 & $\displaystyle 
    \frac{\partial^n}{\partial \gamma^n}
    M(\gamma)\Big|_{\gamma = 0} 
%\avi{e^{\gamma g(x)}}{\cal X}\Bigg|_{\gamma = 0} 
    = M_n(g)$\\
\hline
 \rule[-6mm]{0mm}{14mm}
 $ C(\gamma )$
 = $\ln M(\gamma )$
 = $\ln \avi{e^{\gamma g(x)}}{\cal X}$ 
 & $\displaystyle 
    \frac{\partial^n }{\partial \gamma^n} 
    C(\gamma)\Big|_{\gamma = 0} 
%\ln \avi{e^{\gamma g(x)}}{\cal X} \Bigg|_{\gamma = 0} 
     = C_n (g)$\\
\hline
\end{tabular}
\end{center}
%\caption{Generating functions
%for zero field.% $\gamma$ = $0$.
%}
%\label{table1}
%\end{table}
\vspace{1cm}
and for nonzero field $\beta$ or $\gamma_0$, respectively
%\begin{table}
\begin{center}
\begin{tabular}{|c|c|}
\hline
generating function & derivatives \\
\hline
 \rule[-5mm]{0mm}{12mm}
 $Z$ 
 = $\ind{x} e^{-\beta E(x)}$   
 & $\displaystyle \frac{\partial^n}{\partial (-\beta^\prime)^n} 
  Z(\beta^\prime) \Big|_{\beta^\prime = \beta}
= Z M_{n}(E)$\\
\hline
 \rule[-5mm]{0mm}{12mm}
 $\ln Z$
 = $\ln \ind{x} e^{-\beta E(x)}$   
 & $\displaystyle \frac{\partial^n}{\partial (-\beta^\prime)^n} 
    \ln Z (\beta^\prime ) \Big|_{\beta^\prime = \beta}
= C_{n} (E)$\\
%\hline
%\end{tabular}
%\end{center}
%
%
%
%
%\begin{center}
%\begin{tabular}{|c|c|}
%\hline
%generating function & derivatives \\
%
%
\hline
 \rule[-6mm]{0mm}{14mm}
 $Z(\gamma )$ % $= Z \avi{e^{\gamma g(x)}}{\cal X}$
 = $\ind{x} e^{-\beta E(x)+\gamma g(x)}$      
 & $\displaystyle 
    \frac{\partial^n }{\partial \gamma^n}Z(\gamma )\Big|_{\gamma = \gamma_0} 
    = Z(\gamma_0) M_{n,\gamma_0}(g)$\\
\hline
 \rule[-6mm]{0mm}{14mm}
 $\ln Z(\gamma )$ 
 = $\ln \ind{x} e^{-\beta E(x)+\gamma g(x)}$      
 & $\displaystyle 
    \frac{\partial^n}{\partial \gamma^n} 
       \ln Z(\gamma )\Big|_{\gamma = \gamma_0} 
     = C_{n,\gamma_0} (g)$\\
\hline
 \rule[-6mm]{0mm}{14mm}
 $M(\gamma )$ 
 = $\frac{Z(\gamma )}{Z}$  
 = $\avi{e^{\gamma g(x)}}{{\cal X}}$ 
 & $\displaystyle 
    \frac{\partial^n}{\partial \gamma^n}
    M(\gamma)\Big|_{\gamma = \gamma_0} 
%\avi{e^{\gamma g(x)}}{\cal X}\Bigg|_{\gamma = 0} 
    = M(\gamma_0 ) M_{n,\gamma_0}(g)$\\
\hline
 \rule[-6mm]{0mm}{14mm}
 $ C(\gamma )$
 = $\ln M(\gamma )$
 = $\ln \avi{e^{\gamma g(x)}}{{\cal X}}$ 
 & $\displaystyle 
    \frac{\partial^n }{\partial \gamma^n} 
    C(\gamma)\Big|_{\gamma = \gamma_0} 
%\ln \avi{e^{\gamma g(x)}}{\cal X} \Bigg|_{\gamma = 0} 
     = C_{n,\gamma_0} (g)$\\
\hline
 \rule[-6mm]{0mm}{14mm}
 $M_{\gamma_0}(\gamma )$ 
 = $\frac{Z(\gamma )}{Z(\gamma_0 )}$  
 = $\avi{e^{(\gamma -\gamma_0) g(x)}}{{\cal X}|\gamma_0}$ 
 & $\displaystyle 
    \frac{\partial^n}{\partial \gamma^n}
    M_{\gamma_0}(\gamma)\Big|_{\gamma = \gamma_0} 
%\avi{e^{\gamma g(x)}}{\cal X}\Bigg|_{\gamma = 0} 
    = M_{n,\gamma_0}(g)$\\
\hline
 \rule[-6mm]{0mm}{14mm}
 $ C_{\gamma_0}(\gamma )$
 = $\ln M_{\gamma_0}(\gamma )$
 = $\ln \avi{e^{(\gamma -\gamma_0) g(x)}}{{\cal X}|\gamma_0}$ 
 & $\displaystyle 
    \frac{\partial^n }{\partial \gamma^n} 
    C_{\gamma_0}(\gamma)\Big|_{\gamma = \gamma_1} 
%\ln \avi{e^{\gamma g(x)}}{\cal X} \Bigg|_{\gamma = 0} 
     = C_{n,\gamma_1} (g)$\\
\hline
\end{tabular}
\end{center}
%\caption{Generating functions for nonzero fields.}
%\label{table2}
%\end{table}

\subsection{Conditional probabilities and disordered systems}
\label{disorder}

We already discussed that it is 
often useful to look for conditions 
under which the energy is easy to specify
and to combine the different possible conditions in a second step
to obtain the complete probability.
Thus, a joint probability $p(x,y)$
can be specified either by a
`joint' (or annealed) energy function $E(x,y)$
with conjugated joint (or annealed) temperature $1/\beta$
or 
a conditional (or quenched) energy $E(x|y)$
with conjugated conditional (or quenched) temperatures $B(y)$
and a mixture energy $E(y)$ with mixture temperature $b$.
Hence,
\be
p(x,y)
= \frac{e^{-\beta E(x,y)}}{Z({\cal X},{\cal Y})}
= p(y)p(x|y)
= \frac{e^{-b E(y)} e^{-B(y) E(x|y)}}
     {Z({\cal Y}) Z({\cal X}|y)}
,
\label{cond-prod}
\ee
with
\be
p(y) = e^{-b E(y)}/Z({\cal Y}),
\ee
\be
p(x|y) = e^{-B(y) E(x|y)}/Z({\cal X}|y)
,
\ee
 and
\bea
Z({\cal X},{\cal Y})&=&\ind{x} \ind{y} e^{- \beta E(x,y)}, \\
Z({\cal Y})         &=&\ind{y}         e^{- b    E(y)  }, \\
Z({\cal X}|y)       &=&      \ind{x}   e^{-B(y) E(x|y)}
.
\eea
One may remark, that choosing 
$-\beta E(x,y)=-b E(y)-B(y)E(x|y)$
produces a joint probability
\be
p^\prime(x,y)=\frac{e^{- \beta E(x,y)}}{Z({\cal X},{\cal Y})}
%=\ind{x} \ind{y} e^{- \beta E(x,y)}, 
=\frac{e^{-b E(y)-B(y)E(x|y)}}
{\ind{y} e^{- b E(y)} Z({\cal X}|y)}
,
\ee
which is different from $p(x,y)$ of Eq.(\ref{cond-prod}).
%\be
%p(x,y)=p(y)p(x|y)=\frac{e^{- \beta E(x,y)}}{Z({\cal X},{\cal Y})}
%%=\ind{x} \ind{y} e^{- \beta E(x,y)}, 
%=\frac{e^{-b E(y)-B(y)E(x|y)}}
%{Z({\cal X}) Z({\cal X}|y)}
%. 
%\ee

If interested only in variable $x$
one integrates (marginalizes) over $y$. 
Working with joint energies this gives
\be
p(x) = \ind{y} p(x,y)
= \frac{\ind{y} e^{-\beta E(x,y)}}{Z({\cal X},{\cal Y})}
= \frac{e^{-\beta E(x)}}{Z({\cal X})}
%= \frac{\ind{x} e^{-\ b E(x)} e^{-\beta(x) E(y|x)}}{Z({\cal X}) Z({\cal Y}|x)}
%= \ind{x} p(x)\frac{e^{-\beta(x) E(y|x)}}{Z({\cal Y}|x)}
,
\ee
whereas working with conditional energies yields
\be
p(x) = \ind{y} p(x,y)
%= \frac{\ind{x} e^{-\beta E(x,y)}}{Z({\cal X},{\cal Y})}
= \ind{y}\frac{ e^{-\ b E(y)} e^{-B(y) E(x|y)}}{Z({\cal Y}) Z({\cal X}|y)}
= \ind{y} p(y)\frac{e^{-B(y) E(x|y)}}{Z({\cal X}|y)}
.
\ee
In the formulation with joint probabilities $E(x,y)$
is the canonical variable.
Its averages can be obtained
by differentiation with respect to $\beta$
\be
E_{ann} 
= E({\cal X})
= \avi{E(x)}{\cal X}
= \avi{E(x,y)}{\cal X, Y}
= -\frac{\partial}{\partial \beta} \ln Z ({\cal X})
.
\ee
In the formulation with conditional probabilities 
the expectation of $E(x|y)$ can be obtained
by differentiation with respect to $B(y)$
\be
E_{quen}
= E({\cal X}|{\cal Y})
= \avi{E({\cal X}|y)}{\cal Y}
= \avi{E(x|y)}{\cal X, Y}
= -\avi{\frac{\partial}{\partial B(y)} \ln Z ({\cal X}|y)}{\cal Y}
.
\ee
Even for $y$--independent $B(y)= B$
the averaging of $\ln Z({\cal X}|y)$ remains
\be
\avi{E(x|y)}{\cal X, Y}
= -\frac{\partial}{\partial B} \avi{\ln Z ({\cal X}|y)}{\cal Y}
.
\label{spinglas}
\ee
For $y$--independent normalization
$Z({\cal X}|y)$ = $Z({\cal X})$
both approaches are equivalent
and $\avi{\ln Z ({\cal X}|y)}{\cal Y}$ = 
$\ln Z ({\cal X})$.
In general, however, the expectations $E_{ann}$ and $E_{quen}$ 
are different. 
Also, despite of the equality $(\ref{cond-prod})$,
the exponential families defined by
varying the parameters $\beta$ or $B$ (or $B(y)$)
are not the same.
The conditional temperatures or fields $B(y)$
do not influence the distribution $p(y)$,
while the joint temperature $1/\beta$ does.

In practice, for example,
it may take some time after changing temperature or an external field
until a stationary distribution $p(x,y)$ is reached.
Assume the dynamic of $y$ being much slower than that of $x$
(e.g.\ lower energy barriers for $x$ and higher energy barriers for $y$).
In a magnetic substance $x$ may stand for fast adapting local spins
and $y$ for very slowly moving impurities.
Then changing the physical temperature
will on short time scales (or low temperatures) 
only change the distribution of the fast adapting spins $x$,
while the slow impurities $y$ remain quenched.
Then the relevant physical temperature 
is the conditioned or quenched field $B(x)$.
On a much longer time scale 
(or at high enough temperatures, i.e., in an `annealed system')
also the impurities will approach an equilibrium distribution.
Then the physical temperature is a joint or annealed field $\beta$.
In addition, ensemble averages 
for variables which reach a stationary distribution
on short enough time scales 
are often measured as time averages
under a \mbox{(quasi--)}\-ergodic dynamic.
In contrast, averages over slow variables can in practice not
be obtained as time averages
and must be realized as ensemble averages.

Eq.(\ref{spinglas}) requires
the calculation of a partition sum 
$Z({\cal Y},x)$
for every $x$.
This is possible for a small number
of different $x$ values
or if the $x$--dependence can be calculated analytically
and the average over ${\cal X}$ be performed.
In general, however, Eq.(\ref{spinglas})  is 
rather difficult to solve.
One possibility to proceed is using the identity
\be
\ln Z = \lim_{n\rightarrow 0} \frac{Z^n-1}{n},
\ee
which is verified by expanding
\be
Z^n = e^{n\ln Z}=1+n\ln Z+\sum_{k=2}\frac{n^k}{k!}(\ln Z)^k.
\ee
Typically, the average over $Z^n$ for integer $n$
is easier to perform than over $\ln Z$.
Because $Z^n$  describes a system with $n$ 
independent `replicas' of the same system
this approach is known under the name replica method
\cite{Mezard-Parisi-Virasoro-1987,Fisher-Hertz-1991,Hertz-Krogh-Palmer-1991,Parisi-1992}.
Performing the average over $x$, however,
results usually in a coupling between the different replicas.
Also one has to be careful, 
because calculating $Z^n$ for integer $n$
does not uniquely determine the analytic continuation to
$n\rightarrow 0$.
For non--interacting disordered systems
a supersymmetric approach
(expressing the two--point correlation function 
 as a product of two gaussian integrals,
 one over commuting variables and another over
 anticommuting (Grassmann) variables)
 can avoid such difficulties of the replica approach
(\cite{Efetov-1983,Mudry-Chamon-Wen-1996}).

%For $B(x) = B = b = \beta$ the difference $E_c - E$ is calculated as
%\be
%E_c-E 
%= -\frac{\partial}{\partial \beta} \left(
%   \avi{\ln Z ({\cal Y}|x)}{\cal X} + \ln Z ({\cal Y})\right)
%\ee
%\be
%=
%-\frac{\partial}{\partial \beta} 
%  \avi{\ln Z ({\cal Y}|x)-\ln Z ({\cal Y})}{\cal X} 
%=
%-\frac{\partial}{\partial \beta} 
%  \avi{\ln \frac{Z ({\cal Y}|x)}{Z ({\cal Y})}}{\cal X} 
%=
%\frac{\partial}{\partial \beta} 
%  \avi{\ln {Z ({x})}}{\cal X} 
%\ee

\section{Bayesian decision theory}
\label{Bayesian}

\subsection{Basic definitions}

\subsubsection{The basic model}
\label{model}

Consider the following scenario\cite{Wolpert-1995}.
We assume that we can prepare a specific situation $x\in{\cal X}$
and measure outcome $y\in{\cal Y}$.
Furthermore, we assume that 
the probability $p(y|x,{h})$ of outcome $y$ 
is determined by $x$ and additional variables ${h}$
which we cannot observe directly. 
These additional variables ${h}$ will be called collectively
`state of Nature'.
Furthermore, we assume all knowledge $f$ 
we have accumulated about Nature in the past 
has been combined in form of a probability density
$p({h}|f)$ over the possible states of Nature ${h}\in {\cal {H}}$. 
The aim of learning is to update our knowledge about Nature 
$p({h}|f)\rightarrow p({h}|f^\prime(D,f))$
as more data $D$ arrive under the assumption that the underlying
`true state of Nature' ${h}_N$ producing the data does not change.

\begin{figure}
\begin{center}
\setlength{\unitlength}{1mm}
%\hspace{2cm}
%\linethickness{1pt}
\begin{picture}(40,40)
\put(0,0){\framebox(40,40)[]{}}
% top row: empty
% second row: x_1 ... x_n  q
\put(10,30){\makebox(0,0){$x_1$}}
\put(20,30){\makebox(0,0){$\cdots$}} 
\put(30,30){\makebox(0,0){$x_k$}}
\put(35,30){\makebox(0,0){$\cdots$}} 
% arrows to third row
\put(9.5,27){\vector(0,-1){4.5}}  
\put(29.5,27){\vector(0,-1){4.5}}  
% third row: y_1 ... y_q
\put(10,20){\makebox(0,0){$y_1$}}
\put(20,20){\makebox(0,0){$\cdots$}} 
\put(30,20){\makebox(0,0){$y_k$}}
\put(35,20){\makebox(0,0){$\cdots$}} 
% arrows from third row
\put(17,13){\vector(-1,1){4.5}}  
\put(23,13){\vector(1,1){4.5}}   
% fourth row: {h}
\put(20,10){\makebox(0,0){${h}$}}
\put(27,10){\vector(-1,0){5}}  
\put(30,10){\makebox(0,0){$f$}}
\end{picture}
\end{center}
\caption{
Graphical representation of a probabilistic model
factorizing according to
$p(x,y,{h}|f)$ = 
$p({h}|f)$
$\prod_{k} p(x_k) p(y_k|x_k,{h})$ 
.
}
\label{graph-model}
\end{figure}
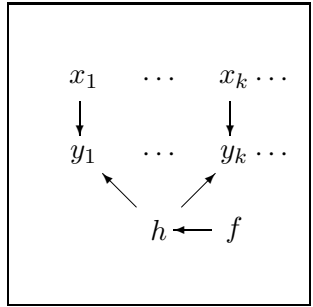

Hence, to define the basic model formally 
we split the random variables of interest  
into the two groups of 
\begin{itemize}
\item[1.]
{\it hidden} (not directly measurable) variables 
${h}\in {\cal {H}}$ ({\it model states}, 
{\it possible state of Nature}), assuming 
the {\it true state of Nature} ${h}_N$ is in ${\cal {H}}$,
and of 
\item[2.]
{\it visible} (directly measurable) variables 
consisting of (potential) {\it data} $\{x, y\}$  
and {\it state of knowledge} $f$,
where 
\begin{itemize}
\item[a.]
the vector $x\in {\cal X}$ collects all {\it independent variables}
(independent of ${h}$, may also be called 
questions, measurement devices, conditions/situations of measurement,
measured quantities, observables),
\item[b.]
the vector $y\in {\cal Y}$ encompasses 
all {\it dependent variables} (depending on ${h}$, may also be called  
answers, measurement results, responses, observed values), and
\item[c.]
the state of knowledge $f\in {\cal F}$ includes all 
{\it determining variables} 
(determining ${h}$, i.e., parameterizing the probability $p({h})$ 
of the ${h}$--producing process
and describing thus the situation under study).
\end{itemize}
\end{itemize}
Thus, a state of Nature or model state ${h}$ is described
by specifying its data generation densities 
($x$--conditional $y$--likelihoods of ${h}$)
$p(y|x,{h})$.
All together, the joint probability of the basic model 
factorizes according to 
\begin{equation}
p(x,y,{h}|f) 
=
p({h}|f)
\,p(x|{h}) 
\,p(y|x,{h}) 
= 
p({h}|f)
\,p(x) 
\,p(y|x,{h}) 
%\,p({h}|f).
.
\label{factorization-model}
\end{equation}
The variables $x$, $y$ may be vectors
of i.i.d.\ sampled (vector) variables. 
Repeated independent sampling under constant ${h}$,
i.e.,  
\be
p(x) = \prod_i^n p(x_i)
,\quad
p(y|x,{h})
= \prod_i p(y_i|x_i,{h}) 
,
\ee
where $x_i$ can contain (components of) $x_{0}$,
gives for the example 
of a discrete set of $x_i\in {\cal X}$ 
\begin{equation}
p(x,y,{h}) 
= 
p({h}|f)
\,\prod_{i} p(x_i) 
\,p(y_i|x_i,{h}) 
.
\label{factorization-model2}
\end{equation}
Fig.\ref{graph-model} 
shows 
a graphical representation of that model
as a directed acyclic 
graph \cite{Pearl-1988,Lauritzen-1996,Jensen-1996,Ripley-1996}.

\subsubsection{Learning: predictive and posterior density}
\label{Learningpp}

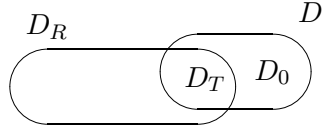
\begin{figure}
\begin{center}
\setlength{\unitlength}{1mm}
%\hspace{2cm}
%\linethickness{1pt}
\begin{picture}(55,25)
%\put(0,0){\framebox(55,25)[]{}}
\put(20,8){\oval(30,10)}
\put(35,10){\oval(20,10)}
\put(31,9){\makebox(0,0){$D_T$}}
\put(40,10){\makebox(0,0){$D_0$}}
\put(10,16){\makebox(0,0){$D_R$}}
\put(45,18){\makebox(0,0){$D$}}
\end{picture}
\end{center}
\caption{
Typical relation between relevant data $D_R$,
training data $D_T\subset D_R$, 
prior data $D_0$
and 
available data $D$ = $D_T\cup D_0$.
}
\label{data-relations}
\end{figure}

By {\it learning} we mean the change in state of knowledge 
$f\rightarrow f^\prime(D,f)$
as new data $D$ arrive.
We will distinguish potentially interesting and 
actually known data:
\begin{itemize}
\item[1.]
{\it relevant} or {\it test} data 
$D_{R}$ = $(x_{\!{}_R},y_{\!{}_R})$, 
$x_{\!{}_R}\in {\cal X}_R$, $y_{\!{}_R}\in {\cal Y}_R$
correspond to
potential (future) application situations of interest,
being thus the data we are actually interested in
and which we want to predict,
and 
\item[2.]
{\it available} 
%or {\it measured} 
data $D$ 
contribute to our {\it state of knowledge} about Nature.
For practical purposes those may be further divided in
\begin{itemize}
\item[a.] {\it training data} 
$D_{T}$ 
= $\{D_{T,i} | 1\le i\le n\}$ = $\{(x_i,y_i) | 1\le i\le n\}$
= $(x_{\!{}_T},y_{\!{}_T})$,
being an empirical sample of $D_R$, i.e.,
a finite number of pairs $(x_i,y_i)$
drawn i.i.d.\ according to $p(x_i)p(y_i|x_i,{h}_N)$
for relevant $x_i\in X_R$ under the true state of Nature ${h}_N$,
and 
\item[b.] {\it prior data} $D_0$ = $\{(x_{0},y_{0}),f,S\}$
%= $\{(q_j,y_{q_j}) | j\in J \}$
collecting all other available knowledge
(a priori information)
not contained in the training data.
Prior data can appear as
\begin{itemize}
\item[i.] {\it measured prior},
corresponding to measured data 
$(x_{0},y_{0})$ not considered as training data, 
as
\item[ii.] {\it generative prior} $f$, ({\it preparation control})
%{\It generative prior} $f$, 
corresponding to knowledge about
the (probabilistic) preparation process which generates 
the true state of Nature ${h}_N$,
or as
\item[iii.] {\it structural prior} $S$ ({\it model control})
refering to all knowledge concerning the specified 
dependency structure of the model variables.
\end{itemize}
\end{itemize}
\end{itemize}
Fig.\ref{data-relations}
summarizes the relations between the different
data types.

Being interested in the relevant data
the aim of learning is to find the 
{\it predictive density}
$p(y_{\!{}_R}|x_{\!{}_R},f^\prime (D,f))$,
or more shortly,
$p(y_{\!{}_R}|x_{\!{}_R},D)$,
after receiving training and prior data. 
Inserting the hidden variables ${h}$ the predictive density becomes
\be
p(y_{\!{}_R}|x_{\!{}_R},f^\prime)
p(y_{\!{}_R}|x_{\!{}_R},D)
= \int \!d{h}\, p({h}|f(D)) 
p(y_{\!{}_R}|x_{\!{}_R},{h})
.
\label{predictive}
\ee
Thus, the space ${\cal F}$ 
of possible states of knowledge
is the convex hull of the the space ${\cal {H}}$ 
of possible states of Nature.
The essential ingredient to be calculated in Eq.(\ref{predictive})
is the {\it posterior density} 
$p({h}|f^\prime(D,f))$, or more shortly $p({h}|D)$,
which can be obtained by obtained by 
inverting  the model (\ref{factorization-model2})
using Bayes' rule
\be
p({h}|D) = 
\frac{p(y_{\!{}_D}|x_{\!{}_D},{h}) p({h})}{p(y_{\!{}_D}|x_{\!{}_D})}
.
\ee

\subsubsection{The risk functional}
\label{risk}

Next we consider a set of possible actions 
$a(x)\in {\cal A}$
from which we can choose in situation $x$
before having seen $y$.
The action $a(x)$ can for example
be the value we expect for $y$ (regression)
or a complete density $p(y|x,a)$ (density estimation)
we expect for $y$ under $x$.
Furthermore, assume 
we suffer loss $l(x,y,a(x))$ if $y$ appears after we have chosen $a(x)$.
Common loss functions are for example
log-loss 
$l(x,y,a)$ = $-\ln p(y|x,a)$ for density estimation
or  mean square loss $(y-a(x))^2$, 
absolute loss $|y-a(x)|$, or $\delta (y-a(x))$ for regression
\cite{Berger-1980}.

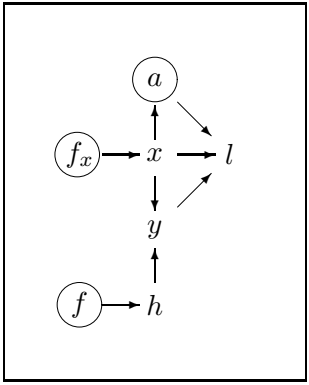
\begin{figure}
\begin{center}
\setlength{\unitlength}{1mm}
%\hspace{2cm}
%\linethickness{1pt}
\begin{picture}(40,50)
\put(0,0){\framebox(40,50)[]{}}
% top row: a
\put(20,40){\makebox(0,0){$a$}}
\put(20,40){\circle{6}}
% arrow to/from first row
\put(20,32){\vector(0,1){4.5}}  
\put(23,37){\vector(1,-1){4.5}}   
% second row: f_x -> x -> l
\put(10,30){\makebox(0,0){$f_x$}}
\put(9.7,30){\circle{6}}
\put(13,30){\vector(1,0){5}}  
\put(20,30){\makebox(0,0){$x$}} 
\put(23,30){\vector(1,0){5}}  
\put(30,30){\makebox(0,0){$l$}}
% arrows to/from third row
\put(20,27){\vector(0,-1){4.5}}  
%\put(23,27){\vector(1,-1){4.5}}   
%\put(30,23){\vector(0,1){4.5}}  
\put(23,23){\vector(1,1){4.5}}   
% third row:  y 
\put(20,20){\makebox(0,0){$y$}} 
% arrows from fourth row
\put(20,13){\vector(0,1){4.5}}  
% fourth row: {h}
\put(10,10){\makebox(0,0){$f$}}
\put(10,10){\circle{6}}
\put(13,10){\vector(1,0){5}}  
\put(20,10){\makebox(0,0){${h}$}}
%\put(27,10){\vector(-1,0){5}}  
%\put(30,10){\makebox(0,0){$f$}}
%\put(30,10){\circle{6}}
\end{picture}
\end{center}
\caption{The setting of Bayesian decision theory in a graphical model.
Circles indicate known variables.
In this figure the variable $f_x$ determining the probability 
$p(x)$ is shown explicitly.
$f_x$ is implicit in the other figures.}
\label{graph-decision}
\end{figure}

The decision model we consider has a graphical representation
shown in Fig.\ref{graph-decision},
where actions and loss are deterministic variables
\be
p(l|x,y,a) = \delta(l-l(x,y,a))
,
\ee
and
\be
p(a|x) = \delta(a-a(x))
.
\ee

Decision theory aims in minimizing a functional 
of the loss posterior $p(l|a,f)$.
The most common functional chosen to be minimized 
is the {\it expected risk} (expected loss)
\bea
r(a,f)
&=& \int \!dl\; l(x,y,a)\; p(l|a,f) 
\nonumber\\
&=& \ind{x}\ind{y}  p(x)p(y|x,f)\,l(x,y,a)
\nonumber\\
&=& \ind{{h}}\!\!\ind{x}\!\!\ind{y} p(x)\,p(y|x,{h})\,p({h}|f)\,l(x,y,a)
%=\avi{l(x,y,h)}{ {\cal X},{\cal Y},{\cal {H}}}.
.
\label{expected-risk}
\eea
%The expected risk is the expectation of the loss function
%over ${\cal X}_R$ and ${\cal Y}$ under the 
%predictive density.
%\begin{equation}
%p(y|x,f) 
%= \int \!d{h} \, p({h}|f) \, p(y|x,{h}) 
%= \avi{p(y|x,{h})}{{h}}
%%= p(y|x,{h}) 
%.
%\end{equation}
%The set of possible predictive densities,
%i.e., of possible states of knowledge $f\in {\cal F}$
%is the convex hull of the {\it pure} states ${h}\in{\cal {H}}$.

\subsection{Interpretation of priors}
\subsubsection{Measured and factorial priors}
\label{Measured-priors}

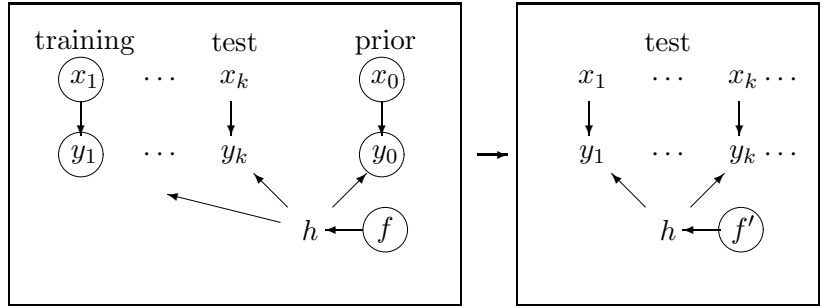
\begin{figure}
\begin{center}
\setlength{\unitlength}{1mm}
%\hspace{2cm}
%\linethickness{1pt}
\begin{picture}(60,40)
\put(0,0){\framebox(60,40)[]{}}
% top row: text
\put(10,35){\makebox(0,0){training}} 
\put(30,35){\makebox(0,0){test}} 
\put(50,35){\makebox(0,0){prior}} 
% second row: x_1 ... x_n  q
\put(10,30){\makebox(0,0){$x_1$}}
\put(9.5,30){\circle{6}}
\put(20,30){\makebox(0,0){$\cdots$}} 
\put(30,30){\makebox(0,0){$x_k$}}
%\put(29.5,30){\circle{6}}
\put(50,30){\makebox(0,0){$x_{0}$}}
\put(49.5,30){\circle{6}}
% arrows to third row
\put(9.5,27){\vector(0,-1){4.5}}  
\put(29.5,27){\vector(0,-1){4.5}}  
\put(49.5,27){\vector(0,-1){4.5}}  
% third row: y_1 ... y_q
\put(10,20){\makebox(0,0){$y_1$}}
\put(9.5,20){\circle{6}}
\put(20,20){\makebox(0,0){$\cdots$}} 
\put(30,20){\makebox(0,0){$y_k$}}
%\put(29.5,20){\circle{6}}
\put(50,20){\makebox(0,0){$y_{0}$}}
\put(49.5,20){\circle{6}}
% arrows from third row
\put(37,13){\vector(-1,1){4.5}}  
\put(43,13){\vector(1,1){4.5}}   
\put(36,11){\vector(-4,1){15}}  
% fourth row: {h}
\put(40,10){\makebox(0,0){${h}$}}
\put(47,10){\vector(-1,0){5}}  
\put(50,10){\makebox(0,0){$f$}}
\put(50,10){\circle{6}}
\end{picture}
\begin{picture}(5,40)
\put(1,20){\vector(1,0){4}}  
\end{picture}
%$\rightarrow$
%
\begin{picture}(40,40)
\put(0,0){\framebox(40,40)[]{}}
% top row: empty
\put(20,35){\makebox(0,0){test}} 
% second row: x_1 ... x_n  q
\put(10,30){\makebox(0,0){$x_1$}}
\put(20,30){\makebox(0,0){$\cdots$}} 
\put(30,30){\makebox(0,0){$x_k$}}
\put(35,30){\makebox(0,0){$\cdots$}} 
% arrows to third row
\put(9.5,27){\vector(0,-1){4.5}}  
\put(29.5,27){\vector(0,-1){4.5}}  
% third row: y_1 ... y_q
\put(10,20){\makebox(0,0){$y_1$}}
\put(20,20){\makebox(0,0){$\cdots$}} 
\put(30,20){\makebox(0,0){$y_k$}}
\put(35,20){\makebox(0,0){$\cdots$}} 
% arrows from third row
\put(17,13){\vector(-1,1){4.5}}  
\put(23,13){\vector(1,1){4.5}}   
% fourth row: {h}
\put(20,10){\makebox(0,0){${h}$}}
\put(27,10){\vector(-1,0){5}}  
\put(30,10){\makebox(0,0){$f^\prime$}}
\put(29.7,10){\circle{6}}
\end{picture}
\end{center}
\caption{
Graphical representation of a model 
with measured and with generative prior
$p(x,y,{h}|f)$ = 
$p({h}|f)$
$p(x_{0}) p(y_{0}|x_{0},{h})$
%$\times$ 
$\prod_{k=1} p(x_k) p(y_k|x_k,{h})$ 
.
Hereby $x_{0}$, $y_{0}$ may also factorize
into independent components.
%The model includes 
%a priori information in the form of measured data which is treated
%analogously to training data.
Circles indicate measured variables,
i.e., the available data $D$.
The predictive density for relevant or test data $D_R$ reads
$p(y_{\!{}_R}|x_{\!{}_R},f^\prime (D,f) )$ =
$\int \!d{h}\,p(y_{\!{}_R}|x_{\!{}_R},{h} )$
$p({h}|D)$
and requires calculation of the posterior density 
$p({h}|D) \propto p(y_{\!{}_D}|x_{\!{}_D},{h})p({h})$.
The figure on the right shows the situation after learning
where
data $D$ has been used to obtain the new state of knowledge
$f^\prime (f,D)$.
}
\label{graph-model-simple}
\end{figure}

The state of knowledge 
before evidence has been received for data $D$
appears as visible variable.
This visible variable, however, must also be based on 
some information which can be considered being data.
One may wish, therefore, 
to express a state of knowledge by giving explicitly 
the data it is based on. 
To do this, one has to include prior data 
in form of a measured prior $(x_{0},y_{0})$.
Such a measured prior represents a situation where
value $y_0$ have been found for ${h}$
as result of a (probabilistic) measurement 
of property $x_0$ (e.g., smoothness).
Thus, model (\ref{factorization-model2}) becomes
\bea
p(x,y,{h}) 
&=& 
p({h}|f)
p(x_0) \,p(y_0|x_0,{h}) 
\,\prod_{i=1} p(x_i) \,p(y_i|x_i,{h}) 
\nonumber\\&=& 
p({h}|f)
\,\prod_{i=0} p(x_i) 
\,p(y_i|x_i,{h}) 
.
\label{factorization-model3}
\eea
Its graphical representation is shown in Fig.\ref{graph-model-simple}.

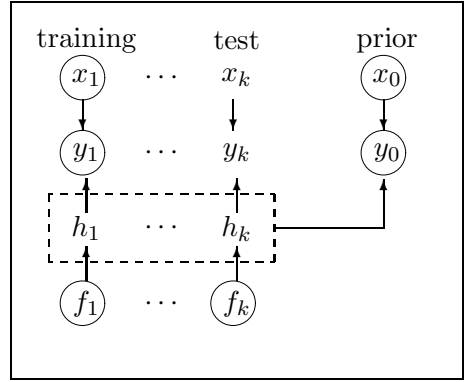
\begin{figure}
\begin{center}
\setlength{\unitlength}{1mm}
%\hspace{2cm}
%\linethickness{1pt}
\begin{picture}(60,50)
\put(0,0){\framebox(60,50)[]{}}
% top row: text
\put(10,45){\makebox(0,0){training}} 
\put(30,45){\makebox(0,0){test}} 
\put(50,45){\makebox(0,0){prior}} 
% second row: x_1 ... x_n  q
\put(10,40){\makebox(0,0){$x_1$}}
\put(9.5,40){\circle{6}}
\put(20,40){\makebox(0,0){$\cdots$}} 
\put(30,40){\makebox(0,0){$x_k$}}
%\put(29.5,40){\circle{6}}
\put(50,40){\makebox(0,0){$x_{0}$}}
\put(49.5,40){\circle{6}}
% arrows to third row
\put(9.5,37){\vector(0,-1){4}}  
\put(29.5,37){\vector(0,-1){4}}  
\put(49.5,37){\vector(0,-1){4}}  
% third row: y_1 ... y_q
\put(10,30){\makebox(0,0){$y_1$}}
\put(9.5,30){\circle{6}}
\put(20,30){\makebox(0,0){$\cdots$}} 
\put(30,30){\makebox(0,0){$y_k$}}
%\put(29.5,30){\circle{6}}
\put(50,30){\makebox(0,0){$y_{0}$}}
\put(49.5,30){\circle{6}}
% arrows from third row
\put(10,22){\vector(0,1){4.5}}  
\put(30,22){\vector(0,1){4.5}}  
%big arrow
\put(5,15.5){\dashbox{}(30,9)[br]{}}
%\put(10,27){\line(0,-1){3}}  
%\put(20,27){\line(0,-1){3}}  
%\put(30,27){\line(0,-1){3}}  
\put(35,20){\line(1,0){14.5}}
\put(49.5,20){\vector(0,1){6.5}} 
% fourth row: {h}
\put(10,20){\makebox(0,0){${h}_1$}}
\put(20,20){\makebox(0,0){$\cdots$}} 
\put(30,20){\makebox(0,0){${h}_k$}}
% arrows from fifth row
\put(10,13){\vector(0,1){4.5}}  
\put(30,13){\vector(0,1){4.5}}   
% fifth row: f
\put(10,10){\makebox(0,0){$f_1$}}
\put(9.5,10){\circle{6}}
\put(20,10){\makebox(0,0){$\cdots$}} 
\put(30,10){\makebox(0,0){$f_k$}}
\put(29.5,10){\circle{6}}
\end{picture}
\end{center}
\caption{
Graphical representation of 
a factorial prior
(with respect to relevant or test data)
$p(x,y,{h}|f({\rm factorial}))$ = 
$p(x_{0})p(y_{0}|x_{0},{h})$
$\prod_{k=1} p({h}_k|f_k) p(x_k) p(y_k|x_k,{h}_k)$ 
.
Without $y_{0}$ depending on ${h}_k$
no generalization is possible from training data $D_{l\ne k}$ 
to test data $D_{k}$.
}
\label{factorial-prior}
\end{figure}

Even, however, if measured priors are included 
the variable $f$ (characterizing now a lower level generative prior)
remains in the model. 
The question therefore arises what kind of $f$ should be chosen as 
a natural starting point of learning.
Consider therefore (as generative prior) 
a {\it factorial prior},
for which $p({h}|f)$ factorizes 
with respect to relevant data,
\be
p({h}|f) = \prod_{k=1} p({h}_k|f_k) 
.
\ee
The model,
shown in Fig.\ref{factorial-prior}, becomes
\be
p(x,y,{h}|f({\rm factorial})) =
p(x_{0} p(y_{0}|x_{0},{h})
\prod_{k=1} p({h}_k|f_k) p(x_k) p(y_k|x_k,{h}_k) 
.
\ee
Without prior term $p(y_{0}|x_{0},{h})$
this corresponds
to a diagram where only variables with the same index $i$ are
connected, 
which means that receiving data for $x_i$
would not allow any generalization 
to relevant data $D_{j\ne i}$
\be
p(y_j|x_j,f(D_{i\ne j})) =
p(y_j|x_j,f({\rm factorial}))
.
\ee
A factorial prior is therefore a natural choice 
for a formal starting point of learning
as it contains no information which allows generalization
from training to non--training data.
Under a factorial prior
it is therefore essential 
for generalization 
that the value $y_{0}$ of prior $x_{0}$
depends on all ${h}_i$.

Interestingly, the asymptotic end points of learning, i.e.,
the pure model states ${h}$,
also represent factorial states,
\be
p(y|x,{h}) = \prod_i p(y_i|x_i,{h}),
\ee
because the variables $(x_i,y_i)$ for different $i$
are independent conditioned on ${h}$.
Hence, in this picture learning starts and ends asymptotically
in a factorial state of knowledge.
The distance from a factorial state could be measured by
the ($x$--averaged) {\it mutual information}
\cite{Langton-1992,Deco-Obradovic-1996}
\be
M(f) = \int \!dx \int \!dy\, p(y|x,f)\frac{p(y|x,f)}{\prod_i p(y_i|x_i,f)}
.
\ee
Thus, starting at zero the mutual information
would be expected to increase at the beginning of learning
and to approach finally zero again asymptotically.
The final factorial state could again 
be starting point of learning under a new, finer model ${{\cal {H}}}^\prime$.

\subsubsection{Gaussian likelihoods}
%\subsubsection{Gaussian and non--gaussian likelihoods}
\label{Gaussian-data}

A model state ${h}$ is a shorthand notation for a parameter vector $\xi$
specifying the data generating densities $p(y|x,{h})$ = $p(y|x,\xi)$.
In this paper we mainly consider gaussian regression problems 
for which relevant and training
data are produced by gaussians with mean specified by ${h}$
and ${h}$-- and $x$--independent variance.
For example, for one--dimensional $y$
\be
p(y|x,{h}) = \frac{1}{\sqrt{2\pi \sigma^2}} e^{-\frac{(y-{h}(x))^2}{2\sigma^2}}
\label{regression}
.
\ee
Hence, in this case 
${h}$ is parameterized by the regression function $\xi (x)$ = ${h}(x)$.
In general density estimation problems
one use a parameter function 
$\xi(x,y)$ = $p(y|x,{h})$
under the additional
normalization constraint 
$\ind{y} \xi(y,x)$ 
= $\ind{y} p(y|x,{h})$ 
= $1$, $\forall x, {h}$.
Thus, in this case the $y$--likelihood of ${h}$
is not gaussian in $\xi$.

The integration 
$\ind{{h}}$ to obtain the predictive density stands for the 
integration $\int \prod_m d\xi_m$
over all components $\xi_m$ of parameter vector $\xi$.
In case ${h}$ is parameterized 
by a regression function 
$\xi_m\rightarrow {h}(x)$
as in Eq.(\ref{regression}) 
the integral reads
\be
\int \prod_m d\xi_m
\rightarrow
\int \prod_x d{h}(x).
\ee
As far as well defined, this becomes for continuous $x$ variable
a functional integral.
Similarly, for general density estimation problems
\be
\int \prod_m d\xi_m
\rightarrow
\int \prod_x \delta(\ind{y^\prime} p(y^\prime|x,{h})-1)
\prod_y dp(y|x,{h}).
\ee

Prior data have to depend on all ${h}(x)$ to allow for generalization.
Analogously to the gaussian $y$--likelihood of (\ref{gauss-prior}), 
we will in this paper mainly consider
gaussian prior information, e.g., for one--dimensional regression problems
\be
p(y_0|x_0,{h}) = (2\pi)^{-\frac{d}{2}} (\det K)^{\frac{1}{2}}
e^{-\frac{1}{2} \mel{{h}-y_0}{K(x_0)}{{h}-y_0}}
,
\label{gauss-prior}
\ee
for symmetric, positive (semi--)definite $K$ and
\be
\mel{{h}-y_0}{K(x_0)}{{h}-y_0}
=
\sum_{x=1}^d \sum_{x^\prime=1}^d
%\int \!dx\,dx^\prime\, 
({h}(x)-y_0(x))
K(x,x^\prime ) ({h}(x^\prime )-y_0(x^\prime ))
.
\ee
An infinite normalization factor
in Eq.(\ref{gauss-prior}) appearing for $d\rightarrow \infty$
will cancel when calculating
expectations.          % under $p(y_0|x_0,{h})$.
We will use therefore 
the convenient integral notation $\sum_x \Delta_x\rightarrow \int \!dx$
in the following.
The variable $x_0$ 
determines the kind of measurement 
for which $y_0$  is regarded as output.
In the gaussian case (\ref{gauss-prior})
$x_0$ defines 
the operator $K$ and thus the covariance $K^{-1}$.
To measure smoothness, for example, 
one may choose the laplacian $K$ = $\Delta$. 
The choice $K(x)$ = $|x><x|$ 
yields the standard gaussian training or test data
of Eq.({regression}).

\subsubsection{Generative priors}
\label{Generative-priors}

\begin{figure}
\begin{center}
\setlength{\unitlength}{1mm}
%\hspace{2cm}
%\linethickness{1pt}
\begin{picture}(40,50)
\put(0,0){\framebox(40,50)[]{}}
% top row: text
\put(10,45){\makebox(0,0){training}} 
\put(30,45){\makebox(0,0){test}} 
% second row: x_1 ... x_n  q
\put(10,40){\makebox(0,0){$x_1$}}
\put(9.5,40){\circle{6}}
\put(20,40){\makebox(0,0){$\cdots$}} 
\put(30,40){\makebox(0,0){$x_k$}}
%\put(29.5,40){\circle{6}}
% arrows to third row
\put(9.5,37){\vector(0,-1){4.5}}  
\put(29.5,37){\vector(0,-1){4.5}}  
% third row: y_1 ... y_q
\put(10,30){\makebox(0,0){$y_1$}}
\put(9.5,30){\circle{6}}
\put(20,30){\makebox(0,0){$\cdots$}} 
\put(30,30){\makebox(0,0){$y_k$}}
%\put(29.5,30){\circle{6}}
% arrows from third row
\put(17,23){\vector(-1,1){4.5}}  
\put(23,23){\vector(1,1){4.5}}   
% fourth row: {h}
\put(20,20){\makebox(0,0){${h}$}}
% arrows from fifth row
\put(13,13){\vector(1,1){4.5}}   
\put(27,13){\vector(-1,1){4.5}}  
% fifth  row: $f$
\put(10,10){\makebox(0,0){$x_{0}$}} %{$q$}}
\put(9.5,10){\circle{6}}
\put(30,10){\makebox(0,0){$y_{0}$}}
\put(29.5,10){\circle{6}}
\put(5,5){\dashbox{}(30,10)[br]{$f\;$}}
\put(20,7){\makebox(0,0){prior}} 
\end{picture}
\end{center}
\caption{
Graphical representation of a generative prior model 
$p(x,y,{h}|D_0)$ 
=      
% $p(x_{0}, y_{0})$
$p({h}|x_{0}, y_{0})$
$\prod_{k=1} p(x_k) p(y_k|x_k,{h})$.
Here, a priori information enters as posterior density
of a ${h}$--generating process.
The variables $\theta_{{h}}$ = $(x_{0}, y_{0})$ are
parameters of the state generating process.
For gaussians, for example, they may determine
mean and covariance structure.
The posterior density of ${h}$ under data $D$ becomes
$p({h}|D) \propto p(y_{\!{}_T}|x_{\!{}_T},{h})
p({h}|D_0)$.
}
\label{graph-model-simple-generative}
\end{figure}
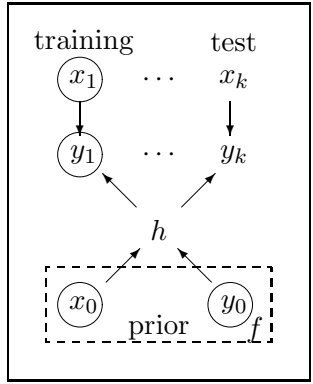

If the process generating ${h}_N$ is under explicit
control it is natural to model prior data 
as generative prior.
For example, consider a situation where ${h}$ is produced 
by a gaussian density
\be
p(f_0|x_0,y^0) 
%= p(f_0|y_0,K) 
= (2\pi)^{-\frac{d}{2}} (\det K)^{\frac{1}{2}}
e^{-\frac{1}{2} \mel{{h}-y_0}{K(x_0)}{{h}-y_0}}
,
\label{gauss-generation}
\ee
with mean $y_0$ and covariance $K^{-1}$ under control.
Then the corresponding generative prior model
is shown in 
Fig.\ref{graph-model-simple-generative}.

\subsubsection{More complex models}
\label{Complex-models}

The basic model of Section (\ref{model})
is completely general. It can however be useful
to implement additional structure.
{\it Input noise}, for example, corresponds to a decomposition
\be
p(y|x,{h}) = \int\!d\theta_x p(y|\theta_x,{h}) p(\theta_x|x)
.
\ee
For a finite number of discrete $\theta_x$
and gaussian components
$p(y|\theta_x,{h})$
this constitutes for $p(y|x,{h})$
a mixture of gaussians with varying mean.
Similarly, {\it output noise}
\be
p(y|x,{h}) = \int\!d\theta_y p(y|\theta_y)p(\theta_y|x,{h}) 
,
\ee
may be used to construct gaussian mixtures
with varying covariances.
Analogously, {\it generation noise} 
\be
p({h}|f) 
= \int\!d\theta_{{h}} p({h}|\theta_{{h}}) p(\theta_{{h}}|f)
,
\ee
can be used to obtain for example a mixture of gaussians
with varying mean and covariance.

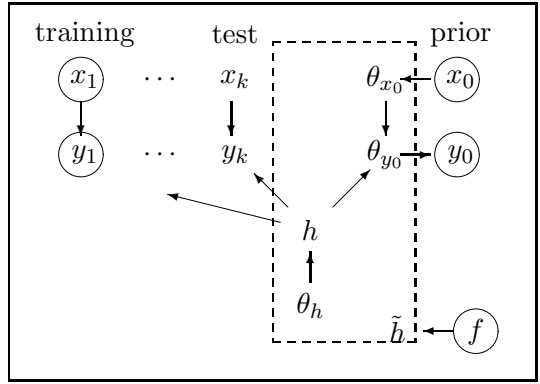
\begin{figure}
\begin{center}
\setlength{\unitlength}{1mm}
%\hspace{2cm}
%\linethickness{1pt}
\begin{picture}(70,50)
\put(0,0){\framebox(70,50)[]{}}
% top row: 
\put(10,46){\makebox(0,0){training}} 
\put(30,46){\makebox(0,0){test}} 
\put(60,46){\makebox(0,0){prior}} 
% second row: x_1 ... x_n \xi_q q
\put(10,40){\makebox(0,0){$x_1$}}
\put(9.5,40){\circle{6}}
\put(20,40){\makebox(0,0){$\cdots$}} 
\put(30,40){\makebox(0,0){$x_k$}}
%\put(29.5,40){\circle{6}}
\put(50,40){\makebox(0,0){$\theta_{x_0}$}}
\put(56,40){\vector(-1,0){4}}  
\put(60,40){\makebox(0,0){$x_{0}$}}  
\put(59.5,40){\circle{6}}
% arrows to third row
\put(9.5,37){\vector(0,-1){4.5}}  
\put(29.5,37){\vector(0,-1){4.5}}  
\put(50,37){\vector(0,-1){4.5}}  
% third row: y_1 ... theta_y_q -> y_q
\put(10,30){\makebox(0,0){$y_1$}}
\put(9.5,30){\circle{6}}
\put(20,30){\makebox(0,0){$\cdots$}} 
\put(30,30){\makebox(0,0){$y_k$}}
%\put(29.5,30){\circle{6}}
\put(50,30){\makebox(0,0){$\theta_{y_0}$}}
\put(52,30){\vector(1,0){4}}  
\put(60,30){\makebox(0,0){$y_{0}$}}    
\put(59.5,30){\circle{6}}
% arrows from third row
\put(37,23){\vector(-1,1){4.5}}  
\put(43,23){\vector(1,1){5}}   
\put(36,21){\vector(-4,1){15}}  
% fourth row: {h}
\put(40,20){\makebox(0,0){${h}$}}
% arrows from fifth row
\put(40,13){\vector(0,1){4}}  
% fifth  row: ${h}$
\put(40,10){\makebox(0,0){$\theta_{{h}}$}}
\put(35,5){\dashbox{}(19,40)[br]{$\tilde {h}$ }}
% f
\put(59,6.5){\vector(-1,0){4}}  
\put(62,6.5){\makebox(0,0){$f$}}
\put(62,6.5){\circle{6}}
\end{picture}
\end{center}
\caption{
Graphical representation of a model 
including besides ${h}$ additional hidden variables
$p(x,y,\tilde {h})$ 
= 
$p(x,y,{h},\theta_{x_0},\theta_{y_0},\theta_{{h}})$ 
= 
$p(\theta_{{h}})p({h}|\theta_{{h}})$
$p(x_0)p(y_0|\theta_{x_0})$
$p(\theta_{y_0}|\theta_{x_0},{h})p(\theta_{x_0}|x_0)$
$\prod_{k=1} p(x_k) p(y_k|x_k,{h})$.
The additional hidden variables
$\theta$ = $(\theta_{x_0},\theta_{y_0},\theta_{{h}})$
parameterize
input, output, and generation noise.
The posterior density of ${h}$ requires marginalization
(integration) over $\theta$ and becomes
$p({h}|D) \propto 
\int \!d\theta_{x_0}\,d\theta_{y_0}\,d\theta_{{h}}$
$p(y_{\!{}_T}|x_{\!{}_T},{h})$ 
$p(y_0|\theta_{y_0})$ 
$p(\theta_{y_0}|\theta_{x_0},{h})$ 
$p(\theta_{x_0}|x_0)$
$p({h}|\theta_{f_0})$ $p(\theta_{f_0})$
.
}
\label{graph-model-complete-xi}
\end{figure}

The variables $\theta$ = $(\theta_x, \theta_y, \theta_{{h}})$
represent additional hidden variables of the model.
Restricting for convenience the variables denoted by ${h}$ 
to that hidden variables
which determine the relevant data (relevant state of Nature),
an (extended/complete) state of nature $\tilde {h}$
is given by specifying $\tilde {h}$ = $({h},\theta)$.
A graphical representation of a model 
with additional noise variables
$\theta$ = $(\theta_x, \theta_y, \theta_{{h}})$
is shown in Fig.\ref{graph-model-complete-xi}.
%\be
%p(y|x,h^\prime)
%=
%\int \!d\vartheta\,
%p(\vartheta|h) 
%p(y|x,h,\vartheta )
%.
%\ee

\subsubsection{Structural priors and a posteriori control}
\label{structural-priors}

%The role of probabilistic models in learning is to provide rules
%inferring from training data to relevant data, 
%i.e., to obtain a predictive density as function of training data.
%Those rules represent the equivalent of prior data.
We have up to now discussed how prior data can enter a model
in the form of empirically measured data 
or as generative prior controlling the preparation of $f$. 
A third possibility consists in the direct specification of the
necessary dependency relations between
relevant and training data within the model.

The empirical validity a a priori information
implemented by a specific model is based the possibility
to control the required dependency structure 
between relevant and training data (model control).
As those dependencies have to be controled at the time
of testing (which is usually after having received training data)
the validity of a priori information
can be ensured by {\it a posteriori control}.
As a trivial example consider a bound $y_{\!{}_R}<c$
which can be enforced by ignoring values larger than $c$ 
at the time of testing.

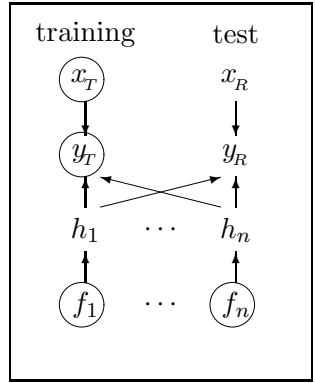
\begin{figure}
\begin{center}
\setlength{\unitlength}{1mm}
%\hspace{2cm}
%\linethickness{1pt}
\begin{picture}(40,50)
\put(0,0){\framebox(40,50)[]{}}
% top row: 
\put(10,46){\makebox(0,0){training}} 
\put(30,46){\makebox(0,0){test}} 
% second row: x_T ... x_R
\put(10,40){\makebox(0,0){$x_{\!{}_T}$}}
\put(9.5,40){\circle{6}}
\put(30,40){\makebox(0,0){$x_{\!{}_R}$}}  
% arrows to third row
\put(10,37){\vector(0,-1){4.5}}  
\put(30,37){\vector(0,-1){4.5}}  
% third row: y_T ... y_R
\put(10,30){\makebox(0,0){$y_{\!{}_T}$}}
\put(9.5,30){\circle{6}}
\put(30,30){\makebox(0,0){$y_{\!{}_R}$}}  
% arrows from third row
\put(10,23){\vector(0,1){4.0}}  
\put(12,23){\vector(4,1){16.0}}  
\put(30,23){\vector(0,1){4.0}}  
\put(28,23){\vector(-4,1){16.0}}  
% fourth row: {h}_1 ... {h}_k
\put(10,20){\makebox(0,0){${h}_1$}}
\put(20,20){\makebox(0,0){$\cdots$}}
\put(30,20){\makebox(0,0){${h}_n$}}
% arrows from fifth row
\put(10,13){\vector(0,1){4.0}}  
\put(30,13){\vector(0,1){4.0}}  
% fifth row: f_1 ... f_k
\put(10,10){\makebox(0,0){$f_1$}}
\put(9.5,10){\circle{6}}
\put(20,10){\makebox(0,0){$\cdots$}}
\put(30,10){\makebox(0,0){$f_n$}}
\put(29.5,10){\circle{6}}
\end{picture}
\end{center}
\caption{
Graphical representation of a structural prior model
$p(x,y,{h})$ = 
$p(x_{\!{}_T})$ $p(x_{\!{}_R})$ 
$p(y_{\!{}_T}|x_{\!{}_T},{h})$
$p(y_{\!{}_R}|x_{\!{}_R},{h})$
$\prod_n p({h}_n|f_n)$
with posterior
$p({h}|D)$
$\propto p({h})p(y_{\!{}_T}|x_{\!{}_T},{h})$
.}
\label{graph-model-posterior-control}
\end{figure}

Thus, 
%in our context 
it is useful to note that
a device producing $y$ with probability density $p(y|x,h)$ 
has a `passive' interpretation as {\it measurement device}
or an `active' interpretation as {\it control device}.
In the passive interpretation
the device is said to measure
the property $x$ of Nature $h$,
in the active interpretation
$y$ is said to be produced or controled by preparing $x$,
but modified by unknown parameters $h$.
The active interpretation is evident
if the probability density of $y$ is determined
by a function $g$ of $x$ and $y$ only
\be
p(y|x,h) = g(x,y)
,
\ee
which in the extreme case may even be deterministic
\be
p(y|x,h) = \delta ( y - g(x) )
.
\ee
Next, consider a control device 
\be
p(y|x,h) = g(x,y,\xi)
\ee
depending
on some parameters $\xi$
which are {\it unknown but
fixed for all $x$}.
Thus, the parameters $\xi$
represent the unknown state of Nature $h$
and the control device can be said to be a 
(indirect) measurement device for $\xi$.
Such a situation is shown in Fig.\ref{graph-model-posterior-control}.
Because hereby prior information is implemented
by choosing a model with specific structure 
this may be called a structural prior.

The number of parameters $\xi$ may be very large.
Consider, for example, a control device
producing $y$ according to a gaussian density 
with known variance and mean similar but probably not equal to $t(x)$.
Assuming now the true but unknown mean $\xi(x)$ = $h(x)$
is an empirical realization
of a gaussian process with mean $t(x)$
and covariance $K^{-1}$
would result in an
quadratic error term 
$(1/2)<h-t|K|h-t>$.
Thus, in this interpretation the template function $t(x)$
characterizes an average control device.
We have already discussed that
an approximative control device $t(x)$ is often provided
by human experts. For example they may specify
what constituents define an image to be a face.
Looking for $h(x)$ in the neighborhood of $t(x)$
approximate control devices $t(x)$
can than be refined by training data.

Summarizing, a measurement device can be seen as control device
with unknown but constant parameters
with a priori information specifying the amount
of control or knowledge we have about that device.

\subsection{Quantum Mechanics}
\subsubsection{Density operators}

The formalism can also be applied
to quantum mechanical problems
and can be used to solve
inverse quantum mechanical problems, i.e.,
problems where empirical, (e.g., scattering) data
are used to determine the Hamiltonian of a system.
In quantum mechanics a system is specified 
%by giving directly data production probabilities 
%$p(y|x,{h})$ but 
by a density operator $\rho$ describing the state of the system.
An observable $x$ is represented by a 
hermitian operator ${\bf O}_x$\footnote{
We used ${\bf O}_x$ instead of ${\bf x}$
to denote the operator representing a general observable,
because ${\bf x}$ is used in quantum mechanics to denote
the multiplication operator in coordinate space.}
and its eigenvalues are the possible measurements results $y$.
As a measurement in quantum mechanics changes $\rho$,
repeated measurement under the same $\rho$
requires a new preparation of $\rho$ before
each measurement.

In particular, the probability of measuring value $y$
for observable $x$ represented
by an hermitian operator
${\bf O}_x$ 
under density operator 
$\rho$
is given by
\be
p(y|x,{h})
= p(y|{\bf O}_x,\rho)
={\rm Tr} ( {\bf P}_y \rho)
= \sum_j^{n_y} <y_j|\, \rho \,|y_j>
,
\ee
where 
and
\be
{\bf P}_y
=\sum_j^{n_y} |y_j><y_j|
\ee 
is the projector 
( with $|y_j><y_j|$ denoting the dyadic product )
into the subspace
of (orthonormalized) eigenfunctions 
$y_j$ of the 
(hermitian) operator
${\bf O}_x$
with eigenvalue $y$
\be
{\bf O}_x|y_j> = y |y_j>,
\quad
\scp{y_j}{y^\prime_i} = \delta_{ji}\delta (y-y^\prime)
.
\ee
The indices distinguish $n_y$ different 
eigenvectors with equal eigenvalues $y$.
For nondegenerate eigenvalues $n_y = 1$.
The density operator characterizing the system
is a hermitian, positive definite operator
with eigenfunction decomposition
(the sum to be replaced by an integral for non discrete cases)
\be
\rho = \sum_i p(i) \, |\varphi_i><\varphi_i|
\label{density-operator}
\ee
with $0\le p_i\le 1$ and
$\sum_i p (i) = 1$
and orthonormal
$\scp{\varphi_i}{\varphi_j} = \delta_{ij}$.
In case the $\varphi_i$ are the eigenstates $y_i$ of ${\bf O}_x$
the $p(i)$ become $p(y|{\bf O}_x,\rho)$.
A density operator 
which is a projector,
\be
\rho^2 = \rho = |\varphi><\varphi|
,
\ee
consists of only one term and
is called a pure (quantum mechanical) state.
For a pure state 
\[
p(y|x,{h})
%= p(y|{\bf O}_x,\rho)
%={\rm Tr} ( {\bf P}_y \rho),
={\rm Tr} \Big( \sum_j^{n_y} |y_j><y_j|\varphi><\varphi| \Big)
\]\be
=\sum_j^{n_y} \scp{\varphi}{y_j} \scp{y_j}{\varphi}
=\sum_j^{n_y} |\scp{y_j}{\varphi}|^2
=\sum_j^{n_y} |\varphi (y_j)|^2.
\ee
Pure states already show all
(non classical) quantum phenomena
while statistical mixtures with $\rho^2\ne \rho$
just add a classical (non quantum mechanical) averaging
to quantum mechanical systems.

Consider a system with $\rho (t_0)$
prepared at time $t_0$.
To calculate the density operator $\rho (t)$ of that system
at a later time $t>t_0$ of measurement
it is necessary to study the time dependence of the system.
The time development of a quantum mechanical system
is given by a unitary evolution operator ${\bf U}$ 
(unitary means ${\bf U}^{-1}$ = ${\bf U}^\dagger$
with ${\bf U}^\dagger$ denoting the hermitian conjugate of ${\bf U}$)
\be
|\varphi (t)> = {\bf U}(t,t_0) |\varphi (t_0)>, 
\ee
leading for density operators to
\be
\rho (t) = {\bf U}(t,t_0) \rho (t_0) {\bf U}^\dagger (t,t_0)
.\ee
The evolution operator
is usually expressed by a Hamiltonian operator ${\bf H}$
\be
U(t,t_0) = e^{-i(t-t_0){\bf H} }
,
\ee
for time--independent Hamiltonian ${\bf H}$
(and disregarding a factor $\hbar$)
or (at least formally) by
\be
U(t,t_0) = T e^{-i\int_{t_0}^t {\bf H}(t) \,dt},
\ee
for time--dependent Hamiltonian ${\bf H}(t)$
with $T$ denoting the time--ordering operator
(see  for example
\cite{Itzykson-Zuber-1985,Negele-Orland-1988,Zinn-Justin-1989}
).

%\begin{exmp}
\subsubsection{Quantum statistics}
\label{Quant-statistics}

Stationary density operators
\be
\rho = \sum_{ij} p_{ij} |E_{ij}\!\!><\!\!E_{ij}|
\ee 
with $\rho (t)$ = $\rho (t_0)$
are built from (orthonormalized) eigenstates $|E_{ij}\!\!>$ of the Hamiltonian
\be
{\bf H} \, |E_{ij}\!\!> = E_i \, |E_{ij}\!\!>
,
\ee
the index $j$ distinguishing eigenstates
with equal eigenvalue.
For example,
a canonical ensemble at temperature $1/\beta$ 
for a Hamiltonian
${\bf H}(\xi)$
depending on parameters $\xi$
is given by the density operator
\cite{Balian-1991}
\be
\rho ({\bf H}(\xi))
=\frac{ e^{-\beta{\bf H(\xi)}}}
{Z_\xi}
=\frac{ e^{-\beta{\bf H(\xi)}}}
{{\rm Tr} \, e^{-\beta{\bf H(\xi)}}}
%{\sum_y \sum_{j}^{n_y}\mel{y_j}{e^{-\beta{\bf H(\xi)}}}{y_j}},
=\frac{ \sum_i  
e^{-\beta{E_i(\xi)}}
        \sum_j^{n_{E_i}}  
|E_{ij}\!\!><\!\!E_{ij}|
}
{
%{\rm Tr} \,
%e^{-\beta{\bf H(\xi)}}
 \sum_i  \sum_j^{n_{E_i}} e^{-\beta{E_i(\xi)}}
}
\label{canonical-ensemble}
\ee
yielding
\be
p(y|x,{h}) = p(y|{\bf O}_x,\rho({\bf H}(\xi)))
=\frac{\sum_j^{n_y}\mel{y_j}{e^{-\beta{\bf H(\xi)}}}{y_j}}
{\sum_y \sum_{j}^{n_y}\mel{y_j}{e^{-\beta{\bf H(\xi)}}}{y_j}}
.
\label{canonical-likelihood}
\ee
Here $\rho$ has been expressed by eigenfunctions of ${\bf H}$
by expanding the exponential and
inserting the eigenfunction expansion 
\be
{\bf H} = \sum_i \sum_j^{n_{E_i}} E_i |E_{ij}><E_{ij}|.
\ee
(Replace the sum by an integral 
if ${\bf H}$ has a continuous spectrum.)
%\end{exmp}

A Bayesian approach now uses
the likelihood (\ref{canonical-likelihood})
to update a given prior density $p(\xi)$
to a new posterior density $p(\xi|D)$
after new data $D$ became available.
Because the measurement of a quantum mechanical system
changes $\rho$, repeated data for the same $\rho$
requires the repeated preparation of the canonical ensemble 
before each measurement.
If learning about $\xi$ is intended the canonical ensembles 
must hereby correspond to fixed (but unknown) $\xi$. 
%In the case of a system at finite temperatures
This can simply mean
waiting long enough between two measurements 
until a given system with fixed but unknown hamiltonian
(described by $\xi$)
is again in thermal equilibrium.

%\begin{exmp}
\subsubsection{Quantum mechanical scattering}
\label{Scattering}

As a second example we prepare a pure state $|z(t_0)\!\!>$ at time $t_0$
with ${\bf z} |z\!\!> = z |z\!\!>$.
This corresponds to the density operator at time $t$
\be
\rho (t)
= |z(t)\!\!><\!\!z(t)|
= {\bf U}(t,t_0) |z(t_0)\!\!><\!\!z(t_0)| {\bf U}^\dagger (t,t_0) 
.
\ee
Measuring then at time $t$ a non degenerated eigenvalue $y$
of observable ${\bf y}$ has probability
\bea
p(y|x,{h}) 
&=&p(y|{\bf O}_x,\rho(t)) 
=|z(t,y)|^2
=<y(t)|z(t)\!\!><\!\!z(t)|y(t)>
\nonumber\\
&=&
| \scp{y(t)}{ z(t) } |^2
=| \scp{y(t)}{ {\bf U}(t,t_0)z(t_0) } |^2
.
\eea
In scattering theory one takes the limit
$t\rightarrow \infty$
and
$t_0\rightarrow -\infty$
and assumes the asymptotic states 
$\lim_{t\rightarrow \infty}y(t)$
and 
$\lim_{t_0\rightarrow -\infty}z(t_0)$
converge (in the weak sense)
to (`free´) states 
$y^0$ or $z^0$, respectively, fulfilling
\be
{\bf H}_{in} |z^0_j> = z^0 |z^0_j>,
\quad
{\bf H}_{out} |y^0_j> = y^0 |y^0_j>
.
\ee
(We will skip the degeneration index $j$
assuming there exist non--degenerate (i.e., unique) states
$|z^0>$ and  $<y^0|$.
In case the eigenvalues $z^0$ or $y^0$ are degenerated
the states can be made unique by measuring additional
commuting observables commuting also with 
${\bf H}_{in}$  or ${\bf H}_{out}$, respectively. 
For non-unique states, see below.)
One obtains
\be
p(y|x,{h}) 
=\!\lim_{t\rightarrow\infty} p(y|{\bf O}_x,\rho(t)) 
=\!\!\lim_{t\rightarrow\infty} 
|\scp{y(t)}{z(t)}|^2
=\!\!\lim_{{t\rightarrow\infty \atop t_0\rightarrow-\infty}} 
  |\scp{y(t)}{{\bf U}(t,t_0)z(t_0)}|^2
,
\ee
or, inserting
the free in (initial, preparation) states $|z^0\!\!>$ 
and out (final, measured) states $<\!y^0|$,
\be
\lim_{{t\rightarrow\infty \atop t_0\rightarrow-\infty}} 
p(y|{\bf O}_x, \rho(t,z(t_0))) 
%=\lim_{{t\rightarrow\infty \atop t_0\rightarrow-\infty}} 
%  |\scp{y^0(0)}{{\bf U}^\dagger_{out}(t,0) 
%                {\bf U}(t,t_0) 
%                {\bf U}_{in}(t_0,0)  z^0(0) }|^2
=  \Big| \Big\langle y^0(0) \Big| {\bf S} z^0(0) \Big\rangle \Big|^2
\ee
which defines 
the scattering operator 
\be
{\bf S} = \lim_{{t\rightarrow\infty \atop t_0\rightarrow-\infty}} 
                {\bf U}^\dagger_{out}(t,0) 
                {\bf U}(t,t_0) 
                {\bf U}_{in}(t_0,0)
,
\ee
with (for time--independent ${\bf H}_{in}$, ${\bf H}_{out}$)
\be
{\bf U}_{in}(t,t_0) = e^{-i (t-t_0) {\bf H}_{in} },
\quad
{\bf U}_{out}(t,t_0) = e^{-i (t-t_0) {\bf H}_{out} }
.
\ee
Often only a partial measurement is performed which does not allow
to identify a unique final state $y^0$ (or $y$).
Then there is a set $A$ of $y^0$ which can not be 
distinguished by the measurement.
%resolved 
Also, often the preparation 
is not a pure state but a mixture of states 
$z^0 \in B$ 
(or $z \in B$, 
possibly also with varying preparation observables ${\bf z}$
or ${\bf H}_{in}$)
with  probability 
$p(z^0)$.
In that case
one has to sum over out states in $A$ 
and average over in states in $B$
\[
\sum_{y\in A} \sum_{{h}\in B}
p(y|x,{h})p({h}(z^0))
\]
\be
=\sum_{y^0\in A} \sum_{z^0\in B}
p(z^0) p(y|{\bf O}_x,\rho) 
=
\sum_{y^0\in A} \sum_{z^0\in B}
p(z^0)
\Big| \Big\langle y^0(0) \Big| {\bf S}(\xi ) z^0(0) \Big\rangle \Big|^2
.
\label{scatt-eq}
\ee
Eq.\ref{scatt-eq}
links quantum mechanics 
to the Bayesian framework
and allows to determine a system
Hamiltonian ${\bf H}(\xi)$
from scattering experiments
given a prior $p(\xi)$ over parameters $\xi$.
%\end{exmp}

\subsubsection{Disordered systems}

A general density operator $\rho$ may be constructed
as a mixture of components $\rho (\bf H (\xi))$ 
which are stationary with respect
different ${\bf H}(\xi)$.
Consider, for example, a situation where the preparation of a system
with stationary density with respect to a constant (but possibly unknown)
${\bf H} (\xi)$ 
is not possible before each measurement.
Assuming that at least a constant (but possibly unknown) mixture 
can be prepared
density operators representing
states of Nature $\rho ({h})$
have the form
\be
\rho ({h})
= \int \!d{\vartheta}\; p(\vartheta|{h}) \, \rho({\bf H}(\vartheta))
,
\label{disorder-rho}
\ee
so that (see Section \ref{Complex-models})
\be
p(y|x,{h}) =
\int \!d{\vartheta}\; p(\vartheta|{h}) \;
p(y|x, {h}, \vartheta)
=
\int \!d{\vartheta}\; p(\vartheta|{h}) \;
p(y|{\bf O}_x,\rho({\bf H}(\vartheta))
.
\ee
There are many cases where it is easier to specify  a system 
in a conditional (disordered) form (\ref{disorder-rho})
than to give directly the joint density $p(y,\vartheta|x,{h})$.
However, as discussed in Section \ref{disorder},
technical complications 
arise because Eq.(\ref{disorder-rho}) requires
calculation of possibly $\vartheta$--dependent
normalizations $Z_\vartheta$ 
of $\rho({\bf H}(\vartheta))$
for all $\vartheta$.
In such situations where the likelihood
is defined as a  mixture of conditional densities
$p(y|x, {h}, \vartheta)$ and we want to 
emphazise the need to deal with $\vartheta$--dependent $Z_\vartheta$  
we will also speak of a {\it disordered system}.
Note however,
that written in its eigenbasis the density operator for disordered systems
takes again the form of Eq.(\ref{density-operator}), and vice versa
a general $\rho$ 
of that form is a mixture of components 
$|\varphi_i><\varphi_i|$.
If such a formulation has been found, however,
one already has $Z_i$ = Tr$(|\varphi_i><\varphi_i|)$ = $1$
for orthonormalized $\varphi_i$.

We remark that 
most problems studied in textbooks
of quantum mechanics (or, analogously, of quantum field theory)
assume a given $\rho$
and aim in calculating 
$p(y|{\bf O}_x,\rho)$.
Hereby $\rho$ can, for example, be 
a pure (quantum mechanical) stationary state (bound state problems),
a pure (quantum mechanical) non--stationary state 
(scattering with completely determined initial state),
a stationary mixture (e.g., a system at finite temperature),
a mixture of conditional $\rho (\vartheta)$ (disordered system),
or, equivalently but differently specified,
a general non--stationary mixture 
(e.g., scattering with not completely observed initial states).
For such problems with fixed $\rho$
no learning can occur.
To allow learning a space of possible
$\rho$ together with a prior density
$p(\rho)$ must be specified
which is updated
to obtain a posterior
$p(\rho|D)$
after new data $D$ have been received.
The following table shows possible forms of density operators
with $\varphi_i({\bf H})$ denoting orthonormalized eigenfunctions of ${\bf H}$
\begin{center}
\begin{tabular}{|c|c|}
\hline
&$\rho$\\
\hline
stationary pure state & $|\varphi ({\bf H}) ><\varphi ({\bf H})|$ \\
general pure state & $|\varphi><\varphi|$ \\
stationary mixture state & $\sum_i p(i|{\bf H}) 
\;|\varphi_i ({\bf H}) ><\varphi_i ({\bf H}) |$ \\
disordered mixture state & $
\int\!d\vartheta\sum_i p(\vartheta) p(i|\vartheta) 
\;|\varphi_i(\vartheta)><\varphi_i (\vartheta)|$ \\
general mixture state & $\sum_i p(i) 
\;|\varphi_i><\varphi_i|$ \\
\hline
\end{tabular}
\end{center}

%\begin{itemize}
%\item[1.]
%Quantum field theory:
%$\rho$ pure state 
%looking for $p(y|x,\rho)$ 
%\item[2.]
%Finite temperature:
%$\rho$ = $e^{-\beta {\bf H}}/Z$
%\item[3.]
%Disordered systems:
%$p(\rho)$
%\item[4.]
%Inverse field theory
%$p(\rho|D)$
%\end{itemize}

\subsubsection{Path integrals}

The path integral approach provides an alternative
to the operator formalism for
quantum mechanics 
or quantum field theory, respectively
\cite{Glimm-Jaffe-1987,Rivers-1987,Zinn-Justin-1989,Montvay-Muenster-1994}.
For example,
a density operator of a canonical ensemble (\ref{canonical-ensemble}) 
can be expressed as a path or functional integral
\be
<y|\,\rho\,|y> =  \frac{1}{Z} \int \!d\pi \int \!d\phi \,
\phi_0(y) \phi^*_\beta(y)
e^{\int_0^\beta d\tau \int dx^3 
\left(
      i\pi \frac{\partial \phi}{\partial \tau}
      -H(\pi,\phi) 
\right)
}
,
\ee
with
\be
Z = \int \!d\pi \int_{\phi_0=\phi_\beta}
\!\! %\!\!\!\!\!\!\!\!\!\!\!\!
 d\phi
\; %\quad
e^{\int_0^\beta d\tau \int dx^3 
\left(
      i\pi \frac{\partial \phi}{\partial \tau}
      -H(\pi,\phi) 
\right)
}
.
\ee
Hereby 
$H(\pi,\phi)$ 
is a classical function depending on classical fields
$\phi$ and $\pi$,
and $\int \!d\pi\,d\phi$ denotes a functional integral
over functions $\pi(x,\tau)$ and $\phi (x,\tau)$.
The function $H$ corresponds to a Hamiltonian 
\be
{\bf H} = \int\! dx\, H(\hat \pi,\hat \phi)
,
\ee
expressed in terms of
field operators $\hat \phi (x,\tau)$
and their canonical conjugates $\hat \pi (x,\tau)$.
For $H$ which are quadratic in $\pi$ the $\pi$--integral
is gaussian and can be performed analytically.
For more details see for example \cite{Kapusta-1989}.

Calculating the functional integral in saddle point approximation
yields the classical field equations.
Such a saddle point approximation can be combined with the saddle
point approximation for the ${h}$--integral
which will be discussed in Section \ref{spa} \cite{Uhlig-1999}.

\subsection{Bayes' rule for complete data} 
\subsubsection{Posterior probabilities and likelihoods} 
\label{postandlike}

Typically, model states ${h}$ are defined
by giving their data generating probabilities or likelihoods
$p(D|{h})$.
The posterior probabilities 
$p({h}|f)$  = $p({h}|D)$ we are interested in
are related to the likelihoods $p(D|{h})$
by Bayes' rule
\be
p({h}|f) = p({h}|D)
%=p({h}|D_T,D_0)
= \frac{p(D|{h}) p({h})}{p(D)}
= \frac{p(D|{h}) p({h})}{\ind{{h}}p(D|{h}) p({h})}
\label{bayesinv}
.\ee
Restricting this Bayesian inversion for the moment to training data
one finds
\be
p({h}|D)
= \frac{p(D_T|{h}) p({h}|D_0)}{\ind{{h}}p(D_T|{h}) p({h}|D_0)}
= \frac{\prod_i p(y_i|x_i,{h})p({h}|D_0)}
       {\ind{{h}} \prod_i p(y_i|x_i,{h}) p({h}|D_0)}
,
\ee
where the training data term 
has been written as product over all training data
$D_T = {(x_{\!{}_T},y_{\!{}_T})}$
= $\{(x_i,y_i) | 1\le i\le n\}$
($x_{\!{}_T}$, $y_{\!{}_T}$ denoting the vectors of $x_i$, $y_i$),
\be
\frac{p(D_T|{h})}{p(x_{\!{}_T})}
=  p(y_{\!{}_T}|x_{\!{}_T},{h})
= \prod_i p(y_i|x_i,{h})
.
\ee
Hence the effective distribution
under the posterior state of knowledge $f$
becomes the quotient of two correlation functions
\begin{equation}
p(y|x,f)  
%= \avi{p(y|x,{h})}{{h}} 
= \frac{\avi{p(y|x,{h})  \prod_i p(y_i|x_i,{h})}{P}}
       {\avi{\prod_i p(y_i|x_i,{h})}{P} },
\end{equation}
where
$\avi{g({h})}{P}$ denotes the prior average
$\ind{{h}} p({h}|D_0) \, g({h})$.

In the following prior data $D_0$ will be treated
analogously to training data $D_T$.
We also assume enough prior data so that all information
leading to nonuniform $p({h}|D)$ is contained in $D$.
Then,
$p({h})$ is
uniform (possibly improper, i.e., non-normalizable)
and thus ${h}$--independent, so that
$p({h},D)$ 
= $p(y_{\!{}_D}|x_{\!{}_D},{h}) p({h})p(x_{\!{}_D})$
 $\propto p(y_{\!{}_D}|x_{\!{}_D},{h}) p(x_{\!{}_D})$.
We call such data $D=D_T\cup D_0$ {\it complete}.

The (training and prior data) generating
probability or (complete) likelihood $p(D|{h})$ 
is related to the posterior probability
according to Eq.(\ref{bayesinv})
\be
p({h}|D)
=p({h}|y_{\!{}_D},x_{\!{}_D})
=\frac{p(y_{\!{}_D}|x_{\!{}_D},{h})p({h}|x_{\!{}_D})}{p(D|x_{\!{}_D})}
=\frac{p(y_{\!{}_D}|x_{\!{}_D},{h})p({h}|x_{\!{}_D})}{\ind{{h}} p(y_{\!{}_D}|x_{\!{}_D},{h})p({h}|x_{\!{}_D})}
,
\ee
which becomes for complete data or uniform 
$p({h}|x_{\!{}_D})$ 
\be
%p({h}|f) = 
p({h}|D)
%=\frac{p(D|{h})p({h})}{p(D)}
=\frac{p(y_{\!{}_D}|x_{\!{}_D},{h})}{\ind{{h}} p(y_{\!{}_D}|x_{\!{}_D},{h})}
%\propto p(D_T,D_0|{h}) 
\propto p(y_{\!{}_D}|x_{\!{}_D},{h}).
\ee

\subsubsection{Posterior and likelihood energies}

Let us now introduce 
%log--probability, 
posterior energy $E({h}|f)$ , posterior temperature $1/\beta_{\!{}_P}$,
with corresponding free energy $F({\cal F}_0|f)$
and partition sum $Z({\cal F}_0|f)$ 
to write the posterior
\begin{equation}
p({h}|f) 
=p({h}|D) 
%= e^{L({h})}
% = e^{-\beta_{\!{}_P} E({h})+c(f,\beta_{\!{}_P} )}
= \frac{Z ({h}|D)}{Z ({\cal {H}}|D)}
= e^{-\beta_{\!{}_P} \big(E({h}|D)-F({\cal F}_0|D)\big)}
,
\label{posterior}
\end{equation}
with
\be
Z({h}|D) = e^{-\beta_{\!{}_P} E({h}|D)},
\quad
Z({\cal {H}}|D) = \int\! d{h} \, e^{-\beta_{\!{}_P} E({h}|D)} 
= e^{-\beta_{\!{}_P} F({\cal {H}}|D)}.
\ee 

Analogously, the $x_i$--conditional 
training likelihood $p(y_i|x_i,{h})$
becomes in terms of (conditional) training likelihood energy
$E(y_i|x_i,{h})$
and (conditional) training likelihood temperature $1/\beta_T$ 
\be
p(y_i|x_i,{h}) 
= e^{-\beta_T \sum_i \big(E(y_i|x_i,{h})-F({\cal Y}|x_i,{h})\big)}
= \frac{e^{-\beta_T \sum_i E(y_i|x_i,{h})}}{\prod_i Z({\cal Y}|x_i,{h})}
,
\ee
with free energy 
\be
F({\cal Y}|x_i,{h})
= -\frac{1}{\beta_{\!{}_L}} Z({\cal Y}|x_i,{h}),
\ee
and normalization factor $Z({\cal Y}|x_i,{h})$
over responses $y_i\in {\cal Y}$ 
for given ${h}$ and $x_i$
\be
Z({\cal Y}|x_i,{h})
= \int_{\cal Y} dy_i\, e^{-\beta_{\!{}_L} E(y_i|x_i,{h})}
.
\ee

The complete conditional likelihood $p(y_{\!{}_D}|x_{\!{}_D},{h})$
%(training and prior data) generating probability or (complete) likelihood 
reads in terms of (complete, conditional) likelihood energy $E(y_{\!{}_D}|x_{\!{}_D},{h})$
and (complete, conditional) likelihood temperature $1/\beta_{\!{}_L}$
\begin{equation}
p(y_{\!{}_D}|x_{\!{}_D},{h})
%= p({h}|D,D_0) 
%= p(D_T|{h}) \, p(D_0|{h},D_T)
%= \frac{ Z_L({h}|D) Z_0({h}|D_0)}{Z({\cal {H}}|D)Z({\cal {H}}|D_0)}
%= e^{L_T + L_P}
= e^{-\beta_{\!{}_L} (E(y_{\!{}_D}|x_{\!{}_D},{h})-F({\cal Y}|x_{\!{}_D},{h}))}
= \frac{e^{-\beta_{\!{}_L} E(Y|x_{\!{}_D}{h})}}{Z({\cal Y}|{h})}
,
\end{equation}
with $E(y_{\!{}_D}|x_{\!{}_D},{h}) = E_T + E_0$ and 
$E_T = E_T(y_{\!{}_T}|x_{\!{}_T},{h})$, $E_0= E(y_0|x_0,{h},D_T)$,
free energy
\be
F({\cal Y}|x_{\!{}_D},{h})
= -\frac{1}{\beta_{\!{}_L}} Z({\cal Y}|x_{\!{}_D},{h})
,
\ee
and normalization factor over data $y_{\!{}_D}$ for given state ${h}$
\be
Z({\cal Y}|x_{\!{}_D},{h})
= \int_{\cal Y} dy_{\!{}_D}\, e^{-\beta_{\!{}_L} E(y_{\!{}_D}|x_{\!{}_D},{h})}
.
\ee
%$L_P = \ln p(D_0|{h})$,
%$L_T = \ln p(D_T|{h})$
%and $Z({\cal D})$
%integrated over $D\in {\cal D}$.
% \propto  \sum_i E (y_i|x_i,{h})$.
%$p({h}|D_0) \propto e^{-\beta E_0({h}|D_0)} $,
%$p(x|y,{h}) \propto e^{-\beta E_D(y|x,{h})}$,

%\begin{equation}
%p_0 ({h})=
%\frac{Z_{0,{h}}}{Z_{0,{h}}} =
%e^{- \beta E_0 (L) + c_0}
%\prod_x \delta (\int \!\!dy e^{L(x,y)} -1 ) 
%%e^{- \!\frac{1}{2}\int\! dx dy dx^\prime dy^\prime  
%%     L(x,y) {\bf O}(x,y;x^\prime,y^\prime ) L(x^\prime ,y^\prime)+c}.
%.
%\label{priorP}
%\end{equation}
%with
%$c_0$ = $-\ln Z_{0,{h}}$, the partition sum
%$Z_{0,{h}} = \int \! d{h}\, Z_{0,{h}}$.

\subsection{Saddle point approximation}
\label{spa}

\subsubsection{Maximum a posteriori approximation}
\label{map}

An exact analytical solution 
of the full integral
in $r(a,f)$ is most times not possible
and approximations have to be used.
We also remark
that for functions ${h}(x)$ 
of continuous variables $x$,
like in field theory in physics,
the integral $\ind{{h}}$ is a functional (or path) integral.
Such integrals have typically to be defined by perturbation theory, 
starting from a well defined, e.g.\ gaussian, case.
Alternatively, the integral can also be discretized, 
like it is done in lattice field theory \cite{Montvay-Muenster-1994}
and like we will do for numerical calculations.

All approximations can only calculate 
a finite number of solvable terms.
A solvable term can correspond
to a solvable infinite sum or, e.g., a gaussian, integral. 
Low temperature approximations restrict
the evaluation to the most important terms
(thus they replace integration by maximization),
high temperature approximations
start from a case where all terms are equal
(e.g., a cumulant expansion starts with the mean),
while Monte Carlo integration 
evaluates a random sample of terms
(so that under certain conditions 
the sampled sum converges
to the true result).
We will discuss in the following the maximum posterior approximation,
which is a special variant of a saddle point approximation
\cite{De-Bruijn-1981,Bleistein-Handelsman-1986}.
%It is
%a low temperature type approximation, valid for small $1/\beta_{\!{}_P}$.
%(The neighborhood near the terms with maximal contribution
%is, however, approximated using a moment expansion, 
%i.e., a high temperature type approximation, 
%but with $\beta_{\!{}_P}$ instead of $1/\beta_{\!{}_P}$ 
%playing the role of temperature.)

The ${h}$--integral within the risk $r(a,f)$ (\ref{expected-risk})
involves two ${h}$--dependent factors
\be
\ind{{h}} p({h}|f)p(y|x,{h}) = 
\ind{{h}}
e^{-\beta_{\!{}_P} \left(E({h}|f)-F({\cal {H}})\right)}
e^{-\beta_{x,{h}} \left( E(y|x,{h})-F({\cal Y}|x,{h})\right)}
\ee
with 
posterior temperature $1/\beta_{\!{}_P}$
and
posterior energy $E({h}|f)$  = $E({h}|D)$, 
which will in the following be written simply $E({h})$.
For a Taylor expansion of the energy 
$E({h})$ with respect to ${h}$
around a minimum ${h}^*$ of $E({h})$,
the first order terms vanish 
and the Hessian $H$, i.e., the matrix of second derivatives 
is positive definite.
Hence,
\be
E({h}) 
=
e^{ \scp{{h}-{h}^*}{\nabla} } 
E({h}^*)
\ee
\[
=
E({h}^*)
%+\frac{({h}-{h}^*)^2}{2} 
%\frac{\partial^2 E({h})}{\partial ({h})^2}\big|_{{h}={h}^*} 
%              {\frac{\partial^2 E({h})}{\partial ({h})^2}\big|_{{h}={h}^*}}
+\frac{1}{2}\mel{\Delta {h}}{H}{\Delta {h}}
+\sum_{n=3}^\infty
 \frac{1}{n!} 
 \scp{\Delta {h}}{\nabla}^n 
 E({h})\big|_{{h}={h}^*} 
\,\,, 
\]
with $H$ the positive definite Hessian of $E({h})$ at ${h}^*$
\be
H ({h})=\frac{\partial^2 E({h})}
{\partial {h}(x,y)\partial {{h}}(x^\prime,y^\prime)}\Bigg|_{{h}={h}^*},
\ee
and $\nabla$ acting on $E({h})$
but not on 
$\Delta {h} = {h}-{h}^*$.

Now assume $p(y|x,{h})$ is slowly varying at
the stationary point 
(i.e., it has a $\beta_{\!{}_P}$--independent Taylor expansion
at the stationary point at ${h}^*$)
and approximate it by
its value $p(y|x,{h}^*)$ at ${h}^*$.
%(More precisely, that 
%for the following gaussian approximation
%to be valid it is sufficient to assume that $p(y|x,{h})$ 
%has a Taylor expansion around ${h}^*$.
%See \cite{De-Bruijn-1981,Bleistein-Handelsman-1986}.)
Then the second order term results in a gaussian 
with mean ${h}^*$
and the Hessian of $E({h})$ at ${h}^*$
as inverse covariance matrix $H$. 
Diagonalizing the positive definite $H$ 
by an orthogonal transformation $O$
\be
H = O^T D O,
\ee
changing the integration variables
from ($q$--dimensional) ${h}$ to $g$ according to
\be
\Delta {h} = \sqrt{\beta_{\!{}_P}}(\sqrt{D} O)^{-1} g
\ee
with Jacobian
\be
J=
\det\left(\frac{\partial {h}(x)}{\partial g(y)}\right)
=\beta_{\!{}_P}^{-\frac{q}{2}} \det H^{-\frac{1}{2}}
\label{jacobian}
\ee
the $q$--dimensional gaussian integral can be performed analytically
\be
\ind{{h}} e^{-\frac{\beta_{\!{}_P}}{2}\mel{\Delta {h}}{H}{\Delta {h}}
}
= J {\pi}^{\frac{q}{2}}
= (\det H)^{-\frac{1}{2}} 
  \left(\frac{\pi}{\beta_{\!{}_P}}\right)^{\frac{q}{2}}.
\label{exactgauss}
\ee
Thus in case we restrict to the contribution of only the lowest
minimum of $E^*$, 
i.e., assuming other local minima of $E$ to be 
sufficiently larger than $E^*$
for a given $\beta_{\!{}_P}$, we have
for $\beta_{\!{}_P}$ large enough
(and no other stationary points contribute)
\be
\ind{{h}}
p({h}|f)\,p(y|x,{h}) 
\approx
%=
(\det H)^{-\frac{1}{2}} \left(\frac{\pi}{\beta_{\!{}_P}}\right)^{\frac{q}{2}}
e^{\beta_{\!{}_P} F({\cal {H}})}
e^{-\beta_{\!{}_P} E({h}^*)}
p(y|x,{h}^*)
%+O(\frac{1}{\beta_{\!{}_P}})
.
\label{saddle}
\ee

The factors beside $p(y|x,{h}^*)$
are ${h}$, $x$, $y$ and $a$--independent
and can therefore be skipped from the risk
in case only one minimum of $E({h})$ contributes
to the ${h}$--integral.
Moreover, evaluating
$e^{-\beta_{\!{}_P} F({\cal {H}})}
=Z({\cal {H}})
= \ind{{h}} e^{-\beta_{\!{}_P} E({h}|f)}$
also by saddle point approximation
one finds for $\beta_{\!{}_P}$ large enough
\be
\frac{1}{Z({\cal {H}})} = e^{\beta_{\!{}_P} F({\cal {H}})}
= \ind{{h}} e^{-\beta_{\!{}_P} E({h}|f)}
\approx
(\det H)^{\frac{1}{2}} \left(\frac{\pi}{\beta_{\!{}_P}}\right)^{-\frac{q}{2}}
e^{\beta_{\!{}_P} E({h}^*)}
%+O(\frac{1}{\beta_{\!{}_P}})
%+{\cal O}(\frac{1}{\beta_{\!{}_P}})
\ee
so the factors in Eq.(\ref{saddle}) cancel
yielding
\be
\ind{{h}}
p({h}|f)\,p(y|x,{h}) 
=
%\approx
p(y|x,{h}^*)
+{\cal O}(\frac{1}{\beta_{\!{}_P}}).
\label{saddlepoint}
\ee
(For the justification of the symbol 
${\cal O}(\frac{1}{\beta_{\!{}_P}})$ see Section \ref{loopex}.)

A maximum posterior approximation
of the ${h}$ integral in $r(a,f)$ 
minimizes instead of the
expected risk $r(a,f)$ of (\ref{expected-risk})
the risk 
\be
r(a,{h}^*) = \ind{x}\ind{y} p(x) p(y|x,{h}^*)l(x,y,a)
\ee
for
\be
{h}^* = 
{\rm argmax}_{{h}\in{\cal {H}}} p({h}|f) 
= {\rm argmin}_{{h}\in{\cal {H}}} E({h}|f), 
\ee
which assumes slowly varying $p(y|x,{h})$ at ${h}^*$.
Hence, integration is replaced by minimization of $E({h}|f)$
(maximization of $p({h}|f)$).
%When minimizing $p({h}|f)$ 

%If a maximum posterior approximation
%leads to a stationarity equation with
%several maxima their contributions can be added.
%This, however, requires calculation of volume factors, i.e.,
%the determinant of a Hessian or covariances for gaussians.
For multiple minima of $E({h})$
which are 
well enough separated the volume factor
$|\det H|^{-\frac{1}{2}}$ can be used as weighting factor 
for the contributions of different minima.
Well enough separated,
means that the gaussian approximations around the different maxima do
not considerably overlap.
Note however, that in this case the (nontrivial) 
calculation of the Hessian is required.
If this is not possible, the ratio 
$\det H_1 / \det H_2$ = $\det H_1 H_2^{-1}$ 
of the determinant of two Hessians $H_1$, $H_2$ 
can be approximated by expanding the logarithm
according to
\be
\ln(1+x) = \sum_{n=1}^\infty \frac{(-1)^{n-1}}{n}x^n
\ee
around the identity matrix $I$
in $\Delta = H_1 H_2^{-1}-I$
\be
\det H_1 H_2^{-1}
= e^{{\rm Tr} \ln (I+\Delta)}
.
\ee
A graphical algorithm for expansion of a determinant
is given by the polymer expansion 
\cite{Montvay-Muenster-1994}.

For the case of minima with overlapping
contributions see
\cite{Berry-1966,Miller-1970,Connor-Marcus-1971}.
%In as far as the resulting approximation
%gives a predictive distribution
%$p(y|x,f)$ is not in ${\cal {H}}$
%The following Appendix \ref{Approximation-problems}
%shows, that there cases 
%a second minimization step has to be performed.

\subsubsection{Complete maximum likelihood approximation}

In this Section the saddle point approximation will  
be discussed 
in the limit of small (conditional) likelihood temperatures $1/\beta_{\!{}_L}$
instead of posterior temperatures $1/\beta_{\!{}_P}$.
It should be stressed, that in contrast to a (conditional)
maximum likelihood approximation which  maximizes the likelihood
$p(y_{\!{}_T}|x_{\!{}_T},{h})$ = $\prod_i p(y_i|x_i,{h})$
for training data 
we consider here the complete (conditional) likelihood 
$p(y_{\!{}_D}|x_{\!{}_D},{h})$ = $p(y_{\!{}_T}\cup y_{\!{}_P}|x_{\!{}_D},{h})$
containing training and prior data.
The two approximations are equivalent only for uniform $p({h}|D_0)$.

To be a bit more general, we also include
continuous hidden variables $\theta$ 
and discrete hidden variables $i$.
%as defined in Appendix \ref{model}.
In the following the integral over 
$\theta$ will be treated in saddle point approximation
analogous to the ${h}$--integral
while the (finite) sum over $i$ will be treated exactly.
Then, one finds for complete data, i.e., uniform, possibly improper, $p({h})$
for the predictive density
\bea
p(y|x,D) 
&=&
\ind{{h}} p({h}|D)\,p(y|x,{h})
\\&=&
\frac{\ind{{h}} p(y_{\!{}_D}|x_{\!{}_D},{h}) \,p(y|x,{h})}
     {\ind{{h}} p(y_{\!{}_D}|x_{\!{}_D},{h})}
\\&=&
\frac{\ind{{h}} \ind{\theta} 
\sum_i p(y_{\!{}_D},\theta,i|x_{\!{}_D},{h})p(y|x,{h})}
{\ind{{h}} \ind{\theta} \sum_i p(y_{\!{}_D},\theta,i|x_{\!{}_D},{h})}
\\&=&
\frac{1}{Z_L}{\ind{{h}} \ind{\theta} 
\sum_i p(y_{\!{}_D},\theta,i|x_{\!{}_D},{h})p(y|x,{h})}
\eea
introducing 
$Z_L$ = $\ind{{h}} \ind{\theta} \sum_i p(y_{\!{}_D},\theta,i|x_{\!{}_D},{h})$.
Decomposing 
$\theta$ = $(\theta_x,\theta_y)$
and
$i$ = $(i_x,i_y)$
%Under the assumption
%\be
%p({h},\theta,i|x_{\!{}_D}) = p({h}) p(\theta,i|x_{\!{}_D}), 
%\,\,{\rm i.e.,}\quad
%p(\theta,i|x_{\!{}_D},{h}) = p(\theta,i|x_{\!{}_D}), 
%\ee
in an input and an output noise component %and generation noise
(see Section \ref{Complex-models}),
this yields
\bea
p(y|x,D) &=&
\frac{1}{Z_L}
\ind{{h}} \ind{\theta} \sum_i 
p(\theta_x,i_x|x_{\!{}_D}) p(y_{\!{}_D}|\theta_y,i_y)
\nonumber\\&&\times
p(\theta_y,i_y|x_{\!{}_D},{h},\theta_x,i_x)
p(y|x,{h})
,
\eea 
The likelihood may be written in terms of 
(the complete, $\theta_x$--, $i_x$--conditional) 
likelihood energy 
%$E(y_{\!{}_D}|x_{\!{}_D},{h},\theta,i)$
$E(\theta_y,i_y|\theta_x,i_x,{h})$
and corresponding
%(complete, $x_{\!{}_D}$--, $\theta$--, $i$--conditional) 
likelihood temperature $1/\beta_{\!{}_L}$ 
\be
p(\theta_y,i_y|x_{\!{}_D},{h},\theta_x,i_x) =
%\frac{1}{Z_L}{\ind{{h}} \ind{\theta} \sum_i p(\theta,i|x_{\!{}_D})
e^{-\beta_{\!{}_L} \big( 
%E(y_{\!{}_D}|x_{\!{}_D},{h},\theta,i) 
E(\theta_y,i_y|\theta_x,i_x,{h})
    - F({\Theta}_y ,{\cal I}_y|\theta_x,i_x,{h}) \big)} 
%p(y|x,{h})}
,
\ee
with
$F({\Theta}_y ,{\cal I}_y|\theta_x,i_x,{h})$ the free energy
corresponding to normalization 
over $\theta_y$ and $i_y$.
Assuming small enough likelihood temperature $1/\beta_{\!{}_L}$ 
to calculate the ${h}$ and $\theta$ integrals
%in numerator and denominator $Z_L$ 
in saddle point approximation
the contributions from numerator and denominator $Z_L$ cancel 
yielding, like Eq.(\ref{saddlepoint})
\be
\ind{h} p({h}|D) \, p(y|x,{h})
\approx p(y|x,{h}^*).
\ee
If
the normalization
$Z({\Theta}_y ,{\cal I}_y|\theta_x,i_x,{h})$
over $\theta_y$ and $i_y$
%$Z({\cal Y}_D|x_{\!{}_D},{h},\theta,i)$
is
$i_x$--, $\theta_x$--, ${h}$--independent 
then ${h}^* = {h}^*(\theta^*)$
with
\bea
{h}^* &=& 
{\rm argmax}_{{h}}\sum_i 
p(\theta_x^*,i_x|x_{\!{}_D}) p(y_{\!{}_D}|\theta_y^*,i_y)
e^{-\beta_{\!{}_L} E(\theta_y^*,i_y|\theta_x^*,i_x,{h})
%E(y_{\!{}_D}|x_{\!{}_D},{h},\theta^*,i)
},
%{\rm argmax}_{\theta, {h}\in{\cal {H}}} p({h}|f) 
\\
\theta^* &=& {\rm argmax}_{\theta}
\sum_i 
p(\theta_x,i_x|x_{\!{}_D}) p(y_{\!{}_D}|\theta_y,i_y)
e^{-\beta_{\!{}_L} E(\theta_y,i_y|\theta_x,i_x,{h}^*)
%E(y_{\!{}_D}|x_{\!{}_D},{h}^*,\theta,i)
},
\eea
which are 
solutions of the coupled stationary equations
\bea
0&=&
  \frac{\partial }{\partial {h}}
\sum_i 
p(\theta_x,i_x|x_{\!{}_D}) p(y_{\!{}_D}|\theta_y,i_y)
e^{-\beta_{\!{}_L} E(\theta_y,i_y|\theta_x,i_x,{h})}
%\sum_i e^{-\beta_{\!{}_L} E(y_{\!{}_D}|x_{\!{}_D},{h},\theta,i)}
,\\
0&=&
  \frac{\partial }{\partial \theta}
\sum_i 
p(\theta_x,i_x|x_{\!{}_D}) p(y_{\!{}_D}|\theta_y,i_y)
e^{-\beta_{\!{}_L} E(\theta_y,i_y|\theta_x,i_x,{h})}
%\sum_i e^{-\beta_{\!{}_L} E(y_{\!{}_D}|x_{\!{}_D},{h},\theta,i)}
.
\eea
If 
$i_x$--, $\theta_x$--, or ${h}$--dependent, 
normalization terms 
$Z({\Theta}_y ,{\cal I}_y|\theta_x,i_x,{h})$
%$Z({\cal Y}_D|x_{\!{}_D},{h},\theta,i)$
also appear in the stationarity equations.

In the case of $n$ training data with likelihood energy 
being the sum
\be
\sum_j^n E_j(y_j|x_j,{h}) 
= n \frac{1}{n} \sum_j^n E_j(y_j|x_j,{h}) 
= n \, \avi{E_j}{{\rm training}}
\ee
one can choose
\be
\beta_{\!{}_L} = n 
,
\ee
provided additional prior information 
ensures that the expectation 
$(1/n) \sum_i E_i$
=
$n \avi{E_j}{{\rm training}}$
converges for large $n$.
In that case the saddle point approximation
is a large $n$ approximation.

\subsubsection{Loop expansion}
\label{loopex}

One can go beyond a maximum posterior approximation
and include  higher order contributions.
We will discuss in the following the expansion in posterior temperatures.
For that purpose,
we will write a full expectation over $p=p(x|y,{h})$
as a sum of gaussian expectations, i.e.,
symbolically,
\be
%\avi{p\,\frac{e^{-\beta_{\!{}_P} E}}{Z}}{{\rm uniform}} =
\avi{\,p\,}{{\rm full}}
=a^* \avi{\,p \,e^{R}}{{\rm gauss}}
=a^* \avi{\,p \,\left( 1+R+\frac{R^2}{2}+\cdots \right)}{{\rm gauss}}
%=a^* \avi{\,p \,\sum_{n=0}^\infty \frac{1}{m}R^m}{{\rm gauss}}
,
\ee
where 
$a^*$ = 
 $e^{-\beta_{\!{}_P} E({h}^*)} (\pi^{\frac{q}{2}}/(JZ)$
is a maximum a posteriori result.
The expansion of the exponential $e^R$ 
corresponds to a moment expansion
analogous to Eqs.(\ref{multidim},\ref{non-zero}).
This expectation can be expressed in terms of moments
of ${h}$ by expanding $p$ and the remaining term $R$ around ${h}^*$.
This will result in an expansion in powers of $1/\beta_{\!{}_P}$.
Expanding also $Z$ in the denominator
common prefactors cancel
\be
\avi{\,p\,}{{\rm full}}
=
\frac{\avi{\,p \,\left( 1+R+\frac{R^2}{2}+\cdots \right)}{{\rm gauss}}}
 {\avi{ 1+R+\frac{R^2}{2}+\cdots }{{\rm gauss}}}
.\ee

Hence, we begin by expanding $p(y|x,{h})$ around a stationary point ${h}^*$
\be
p(y|x,{h}) = 
e^{\scp{\Delta {h}}{\nabla}} 
 p(y|x,{h})\big|_{{h}={h}^*}.
=
\sum_{n=0}^\infty \frac{1}{n!} 
 \scp{\Delta {h}}{\nabla}^n 
 p(y|x,{h})\big|_{{h}={h}^*}.
\ee
Changing for $H = O^T D O$ the integration variables
from ${h}$ to $g$ 
with
$
\Delta {h}$ = ${h}-{h}^*$ = $\sqrt{\beta_{\!{}_P}}(\sqrt{D} O)^{-1} g
$
and Jacobian $J$ of Eq.(\ref{jacobian}),
we find
\[
\ind{{h}} p({h}|f)\,p(y|x,{h})
\]
\[
=
\frac{1}{Z}
\ind{{h}}
\left(
%\sum_{m=0}^\infty \frac{1}{m!} 
% \scp{\Delta {h}}{\nabla}^m 
% p(y|x,{h}^*)
 e^{\scp{\Delta {h}}{\nabla}} 
 p(y|x,{h}^*)
\right)     %\big|_{{h}^*}     %{{h}={h}^*}
e^{-\beta_{\!{}_P}
\left(
 e^{\scp{\Delta {h}}{\nabla}}
 E({h}^*)
\right)}       %\big|_{{h}={h}^*}}.
\]
\[
=\frac{1}{Z}
e^{-\beta_{\!{}_P} E({h}^*)}
\ind{{h}}
\left(
\sum_{m=0}^\infty \frac{1}{m!} 
 \scp{\Delta {h}}{\nabla}^m 
 p(y|x,{h}^*)
\right)
\]
\[
\times
e^{-\beta_{\!{}_P}
\left(
\frac{1}{2} \mel{\Delta {h}}{H}{\Delta {h}}+
\sum_{n=3}^\infty
 \frac{1}{n!} 
 \scp{\Delta {h}}{\nabla}^n 
 E({h}^*)
\right)
}  
\]
%\[
%=\frac{1}{Z}
%e^{-\beta_{\!{}_P} E({h}^*)}
%\ind{{h}}
%\left(
%\sum_{m=0}^\infty \frac{1}{m!} 
% \scp{\Delta {h}}{\nabla}^m 
% p(y|x,{h}^*)
%\right)     %\big|_{{h}^*}     %{{h}={h}^*}
%e^{-\beta_{\!{}_P}\sum_{n=2}^\infty
% \frac{1}{n!} 
% \scp{\Delta {h}}{\nabla}^n 
% E({h}^*)}       %\big|_{{h}={h}^*}}.
%\]
%\[
%=
%\frac{1}{J\,Z}
%e^{-\beta_{\!{}_P} E({h}^*)}
%\ind{g}
%e^{-\frac{\scp{g}{g}}{2}}
%\left(
%\sum_{m=0}^\infty 
% \frac{1}{m!\beta_{\!{}_P}^{\frac{m}{2}}} 
% \scp{(\sqrt{D} O)^{-1} g}{\nabla}^m 
% p(y|x,{h}^*)
%\right)     %\big|_{{h}^*}     %{{h}={h}^*}
%\]
%\[
%\times
%e^{%-\frac{\scp{g}{g}}{2}
% -\sum_{n=3}^\infty
% \frac{1}{n!\beta_{\!{}_P}^{\frac{n}{2}-1}} 
% \scp{(\sqrt{D} O)^{-1} g}{\nabla}^n 
% E({h}^*)}       %\big|_{{h}={h}^*}}.
%\]
\[
=
\frac{1}{Z}
e^{-\beta_{\!{}_P} E({h}^*)}
\ind{{h}}
e^{ -\frac{\beta_{\!{}_P}}{2} \mel{\Delta {h}}{H}{\Delta {h}}}
\left( \sum_{m=0}^\infty \frac{1}{m!} 
 \scp{\Delta {h}}{\nabla}^m 
 p(y|x,{h}^*)
\right)     %\big|_{{h}^*}     %{{h}={h}^*}
\]
\[
\times
\sum_{k=0}^\infty \frac{(-\beta_{\!{}_P})^k}{k!}
\left(
 \sum_{n=3}^\infty
 \frac{1}{n!} 
 \scp{\Delta {h}}{\nabla}^n 
 E({h}^*)    
\right)^k
\]
\[
=
\frac{1}{J\,Z}
e^{-\beta_{\!{}_P} E({h}^*)}
\ind{g}
e^{-\frac{\scp{g}{g}}{2}}
\left(
\sum_{m=0}^\infty 
 \frac{1}{m!\beta_{\!{}_P}^{\frac{m}{2}}} 
 \scp{(\sqrt{D} O)^{-1} g}{\nabla}^m 
 p(y|x,{h}^*)
\right)     %\big|_{{h}^*}     %{{h}={h}^*}
\]
%\be
\[
\times
\sum_{k=0}^\infty \frac{1}{k!}
\left(
 -\sum_{n=3}^\infty
 \frac{1}{n!\beta_{\!{}_P}^{\frac{n}{2}-1}} 
 \scp{(\sqrt{D} O)^{-1} g}{\nabla}^n 
 E({h}^*)    
\right)^k
,
%\label{loop}
%\ee
\]
\[
=
\frac{1}{J\,Z}
e^{-\beta_{\!{}_P} E({h}^*)}
\ind{g}
e^{-\frac{\scp{g}{g}}{2}}
\sum_{m=0}^\infty 
\sum_{k=0}^\infty 
\sum_{n_1=3}^\infty \cdots  \sum_{n_k=3}^\infty
 \frac{(-1)^k}{k!m! \prod_l^k n_l!}
\frac{1}{\beta_{\!{}_P}^{\frac{m}{2}+ \sum_l^k \frac{n_l}{2}-k}} 
\]
\be
\times
\left(  \scp{(\sqrt{D} O)^{-1} g}{\nabla}^m p(y|x,{h}^*) \right)     
\prod_l^k
\left(  \scp{(\sqrt{D} O)^{-1} g}{\nabla}^{n_l} E({h}^*) \right)
,
\label{loop}
\ee
with normalization factor
\be
{J\, Z}
=
e^{-\beta_{\!{}_P} F({\cal {H}})}
\beta_{\!{}_P}^\frac{q}{2} (\det H)^{\frac{1}{2}}
%= 1+{\cal O}(\left(\frac{1}{\beta_{\!{}_P}}\right)
.
\ee
The normalization integral $Z$
can be treated analogously in saddle point approximation
leading in first order to the cancellation of the prefactor.
The individual terms are moments of a multidimensional gaussian
and can be evaluated using Wick's theorem (\ref{Wick}).
Because the gaussian has mean zero odd moments vanish
and Eq.(\ref{loop}) is an expansion in 
 $1/\beta_{\!{}_P}$, also known as loop expansion. 
Only if not expanding around a saddle point linear terms 
would survive.
Higher order terms are usually represented as Feynman diagrams
\cite{Itzykson-Zuber-1985,Glimm-Jaffe-1987,Pokorski-1987,Rivers-1987,Negele-Orland-1988,Zinn-Justin-1989,Itzykson-Drouffe-1989,Le_Bellac-1991,Binney-Dowrick-Fisher-Newman-1992,Montvay-Muenster-1994}.
Surprisingly, the presence of the denominator 
$e^{-\beta_{\!{}_P} F({\cal {H}})}=Z({\cal {H}})$
leads to a simplification.
It can be shown that it cancels exactly the so called vacuum diagrams. 
Further simplifications arise
for expanding the cumulant generating functional
$W=\ln Z$ where only connected diagrams contribute
or its Legendre transform 
$\Gamma$ 
%= $- W + \scp{J}{\Delta {h}}$
where only amputated and one--particle irreducible diagrams
have to be considered.
$\Gamma$ is the generating function of so called (proper) vertex functions.
Equations
connecting moments, cumulants, and vertex functions of different order
can be obtained
(e.g., Ward--identities, 
equations of motion or Dyson--Schwinger equations).
Solving Eq.(\ref{selfenergy}), for example,
corresponds to summing up infinite
subclasses of diagrams.
The name loop expansion stems from the fact that
expansion of $Z$ 
(which does not contain $p(y|x,{h})$)
or $\Gamma$
in powers of $\beta_{\!{}_P}^{-1}$ 
is equivalent to an expansion in the number of loops
of the corresponding Feynman diagrams.

\subsubsection{Stationarity equations}

A specific state ${h}$ is given by 
a parameter vector  $\xi$
determining $p(y|x,{h})$, 
$\forall x\in {\cal X}$,$\forall y\in {\cal Y}$.
Thus, the stationarity equation
to be solved for an a--posteriori approximation
is of the form
\be
\frac{\partial E({h})}{\partial \xi_m}=0,\quad \forall m
,
\ee
where $E({h})$ is an exponent to be minimized.

In the present paper we use 
equally weighted quadratic error terms for training data
assuming gaussian $p(y|x,{h})$
according (\ref{regression})
with equal variance (and therefore equal normalization) for all $x$
and mean specified by a regression function ${h}(x)$.
Then the stationarity equation is obtained by
setting the functional derivatives
\be
\frac{\partial E({h})}{\partial {h}(x)}=0,\quad \forall x\in {\cal X}_R.
\ee

In the more general case, like in density estimation
where one wants to allow for non--gaussian densities $p(y|{h})$
(then also the data terms $\ln p(y_i|{h})$
in the error functional become non--quadratic), arbitrary
$p(y|x,{h})$ can be used, as long as they fulfil
the normalization conditions 
\be
\frac{\partial E({h})}{\partial p(y|x,{h})}=0,\quad 
\forall x\in {\cal X}_R,y\in {\cal Y}
\quad{\rm with}
\ind{y} p(y|x,{h}) = 1, \forall x\in {\cal X}_R,{h}\in{\cal {H}}.
\ee
In that case the posterior probability can be maximized 
with respect to $p(y|x,{h})$
using standard techniques
of constraint optimization 
\cite{Fletcher-1987,Bazaraa-Sherali-Shetty-1993,Bertsekas-1995}.
The normalization conditions 
%$Z({\cal Y}|x,{h})$
%$\ind{y} p(y|x,{h})$ = $1$, $\forall x\in {\cal X}$ 
can be implemented by $\delta$--functions
%is hereby equivalent to the division by a normalization factor 
\begin{equation}
p({h}|f) \propto 
e^{-\beta \left( \sum_i E(y_i|x_i,{h}) + E({h}|D_0) \right) }
\,\delta ( \int \! dy \, e^{-\beta E(y|x,{h})+c(x)}-1)
%= \frac{1}{2 \pi i} 
%\int_{-i \infty}^{i \infty} \!d\Lambda (x)\, 
%e^{\Lambda (x)  
%\left( 1- \int \! dy \, e^{-\beta E(y|x,{h})} \right) }
,
\label{delta}
\end{equation}
where
$p({h}|D_0) \propto e^{-\beta E({h}|D_0)} $,
$p(x|y,{h})\propto e^{-\beta E(y|x,{h})}$,
$c(x$) = $-\ln Z({\cal Y}|x,{h})$ = $(1/\beta) F({\cal Y}|x,{h})$,
${h}$--independent factors have been skipped,
and we have written $\beta_{\!{}_L}=\beta$.
Using the 
Fourier representation of the $\delta$--function 
\begin{equation}
\delta (x)
= \frac{1}{2 \pi} \int_{-\infty}^{\infty} \!dk\, e^{i k x }
=  \frac{1}{2 \pi i} \int_{-i\infty}^{i\infty} \!dk\, e^{- k x },
\label{fourierdelta}
\end{equation}
%gives
%\begin{equation}
%%\delta ( \int \! dy \, e^{-\beta E(y|x,{h})}-1)
%%=
% \frac{1}{2 \pi i} 
%\int_{-i \infty}^{i \infty} \!d\Lambda (x)\, 
%e^{\Lambda (x)  
%\left( 1- \int \! dy \, e^{-\beta E(y|x,{h})} \right) }. 
%\end{equation}
and performing the $\Lambda$ integral
by a saddle point approximation (which is exact for the delta function)
yields
%shows $\Lambda^*$ as Lagrange parameter determined by
%the normalization constraint
\begin{equation}
p({h}|f) 
%= e^{-\beta E_D+ (c-c_0)} p_0 ({h}) 
\propto
e^{-\beta \left( \sum_i E(y_i|x_i,{h}) + E({h}|D_0) \right) 
    +\int \!\!dx\, \Lambda^* (x) 
      \left( 1-\ind{y} e^{-\beta E(y|x,{h})+c(x)} \right) }.
 \label{lambda}
\end{equation}
Here the Lagrange parameter function $\Lambda^*(x)$ 
denotes the stationary value 
of $\Lambda (x)$,
i.e., a solution of the saddle point condition
$\partial P({h}|f)/\partial \Lambda (x)$ = $0$.
It is easy to see that this stationarity condition
which $\Lambda$ has to fulfil at the end of the optimization procedure
is equivalent to the normalization constraint.
There exist many standard iteration procedures
to perform the maximization of (\ref{lambda}).

Alternatively,
the normalization constraint can be implemented
by writing
%including the normalization factor $Z({\cal Y}|x,{h})$
%in the parameterization
\be
p(y|x,{h}) = \frac{g(x,y,{h})}{Z({\cal Y}|x,{h})}
\ee
and solve for a stationary point $g(x,y,{h}^*)$
\be
\frac{\partial E(\frac{g(x,y,{h})}{Z({\cal Y}|x,{h})})}
{\partial g(x,y,{h})}=0,\quad 
\forall x\in {\cal X}_R,y\in {\cal Y}.
\ee
This is equivalent to the
insertion of the $\delta$--functions in Eq.(\ref{delta})
and using always the
stationary value $\Lambda^*(x)$ during iteration.

\subsection{Approximation problems}
\label{Approximation-problems}

In the most common kinds of  problems the hypothesis space 
is identified with the model space
${\cal A} = {\cal {H}}$.
Let an {\it approximation problem} be defined
as a situation with ${\cal A} = {\cal {H}}$ and log--loss 
\be
b(x) l(x,y,a \!=\! {h}) +c(x,y)= -\ln p(y|x,{h}) 
=\beta E(y|x,{h})
\label{log-loss}
\ee
with $a$--independent coefficients $c(x,y)$ and $b(x)>0$.
Notice that the probability
$p(y|x,{h})$ is that governing the production of
test data and is in principle not related to the posterior $p({h}|D)$
or likelihood probabilities $p(y_{\!{}_D}|x_{\!{}_D},{h})$.
Often, however, training data are also assumed to be produced 
according to the same $p(y|x,{h})$.
For log--loss it is easy to see,
by using Jensen's inequality, that
\be
a^* = {\rm argmin}_{a\in{\cal A}} r(a,{h}^*)
    = {h}^* 
%    = {\rm argmin}_{{h}\in{\cal {H}}} p({h})
, 
\ee
and thus
\be
p(y|x,a^*) = p(y|x,{h}^*)
. 
\ee
Thus, for approximation problems the maximum posterior
approximation ${h}^*$ gives already a consistent solution
$a={h}^*$.
In this case the energy function of the posterior probability
$p({h}|f)$ to be minimized is also called {\it error} function,
and the maximum a posteriori approximation is equivalent
to empirical risk minimization \cite{Vapnik-1982,Vapnik-1995,Wahba-1990}.
In the typical case of gaussian states $p(y|x,{h})$
the log--loss is up to a constant
proportional to the squared error $(y-a(x))^2$.

\begin{figure}
\begin{center}
%\hspace{-2cm}
\begin{minipage}[b]{7.2cm}
%\begin{center}
\raisebox{-20mm}[110mm][0mm]{
\hspace{-3.5cm}
\includegraphics[scale = .53 ]{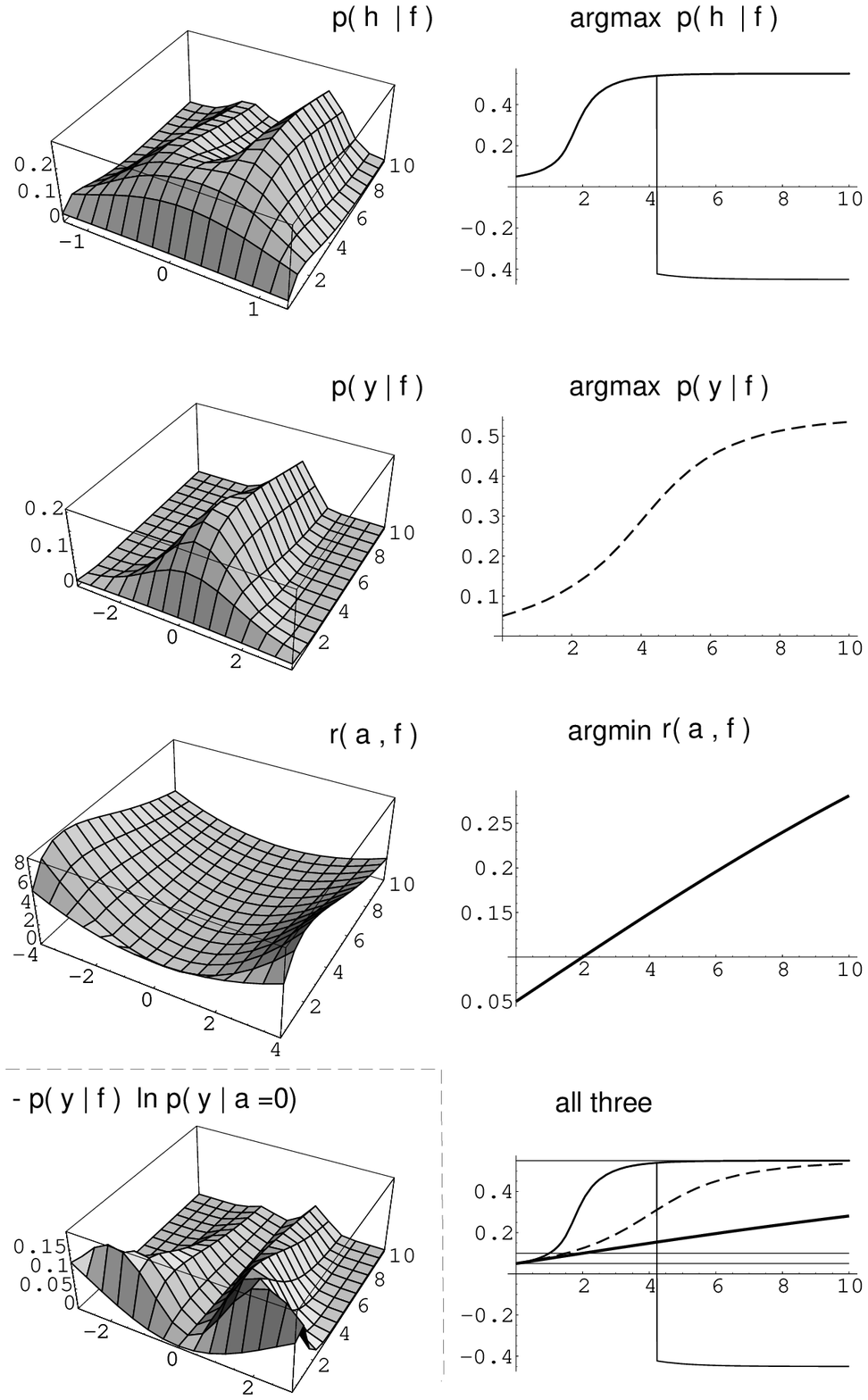}
}
%\end{center}
\caption{}
\label{map0.1}
\end{minipage}
\hspace{-1.8cm}
\begin{minipage}[b]{6cm}
%\begin{center}
\raisebox{-20mm}[110mm][0mm]{
%\hspace{-2cm}\includegraphics[scale = .53 ]{map.d0.5.ps}
\hspace{-2cm}\includegraphics[scale = .53 ]{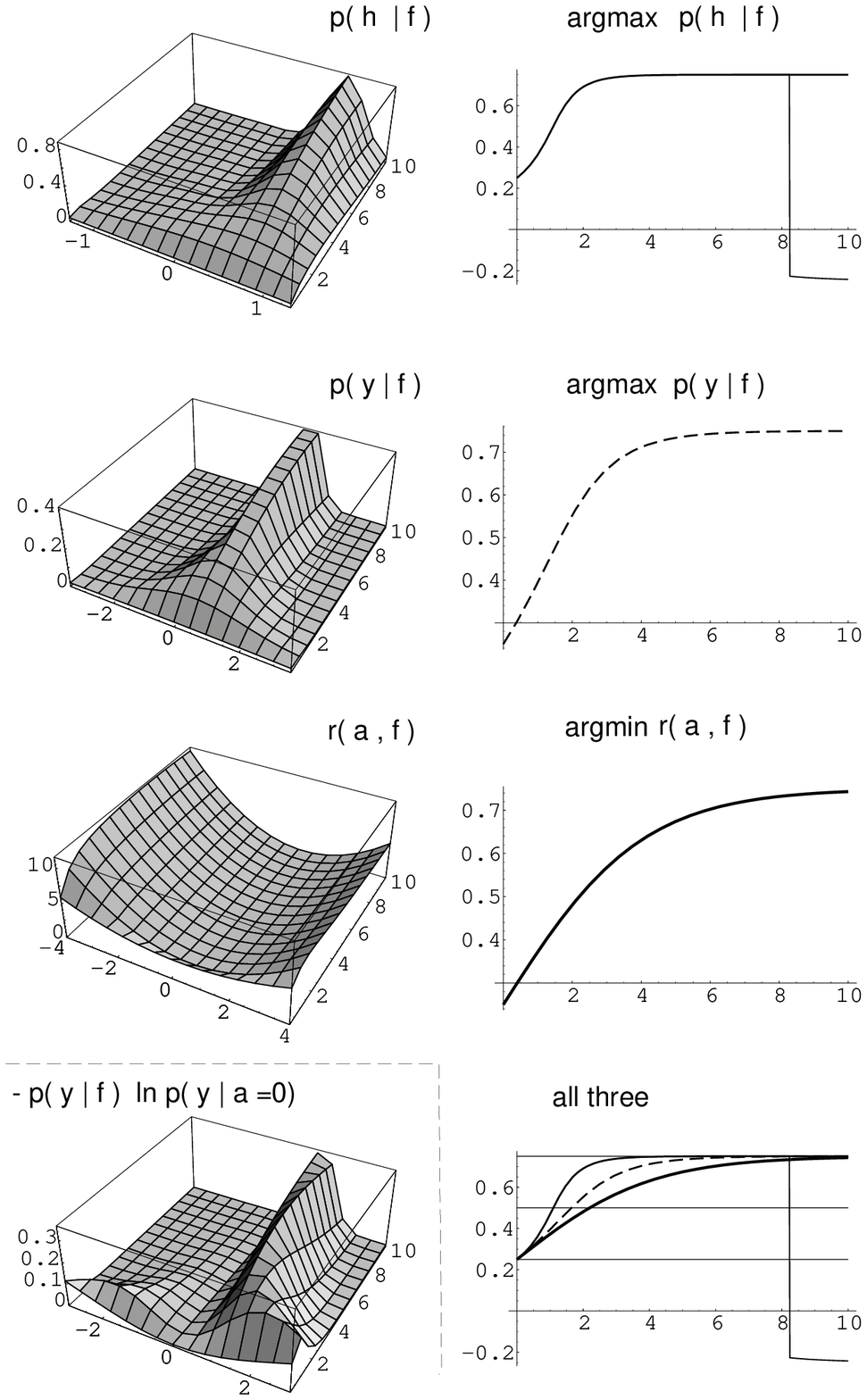}
}
%\end{center}
\caption{}
\label{map0.5}
\end{minipage}
\end{center}
{\footnotesize
The figures compare 
the maximum posterior approximation (top row) with
the mode of the posterior (second row)
and a full Bayesian risk minimization (third row)
for the model defined by Eq.(\ref{mod1},\ref{mod2},\ref{mod3}).
(Fig.\ref{map0.1} for $d=0.1$ and Fig.\ref{map0.5} for $d=0.5$.)
Row 1. Left: $p({h}|f)$ (unnormalized) for 
$-1.3\le {h}\le 1.3$ and $0\le \beta\le 10$.
Right: argmax${}_{{h}\in {\cal {H}}} p({h}|f)$ for 
$0\le \beta\le 10$. 
The second local maximum which appears for
larger $\beta$ (low temperature) is also shown.
Row 2. Left: $p(y|f)$ (unnormalized) for 
$-3.3\le y\le 3.3$ and $0\le \beta\le 10$.
Right: argmax${}_{y\in {\cal Y}} p(y|f)$ for 
$0\le \beta\le 10$. 
Row 3: Left: $r(a,f)$ for 
$-4\le a\le 4$ and $0\le \beta\le 10$.
Right: argmin${}_{a\in {\cal A}} r(a,f)$ for 
$0\le \beta\le 10$. 
Row 3: Left: The 
contributions of $-3\le y\le 3$ to the full risk
for $a=0$ and $0\le \beta\le 10$.
Right: All three approximations within the same diagram.
}
\end{figure}

A similar result holds for the full
Bayesian approach.
Choosing actions $a$ from the space ${\cal A}$ = ${\cal F}$
of states of knowledge $f$
instead from the model space ${\cal {H}}$ of states of Nature ${h}$
(${\cal F}$ is the convex hull of ${\cal {H}}$), 
and using log--loss (\ref{log-loss})
%\be
%a(x) l(x,y,a \!=\! f) +b(x,y)= -\ln p(y|x,f) 
%%=-\beta E(y|x,f)
%\ee
gives 
%analogously
\be
a^* = {\rm argmin}_{a\in{\cal A}} r(a,f)
    = f 
%    = {\rm argmin}_{{h}\in{\cal {H}}} p({h})
, 
\ee
and thus
\be
p(y|x,a^*) = p(y|x,f)
.
\ee
%Notice, however, that a log--loss is of quite unusual form
%for multimodal $p(y|x,f)$.
Other choices are 
the squared error $(y-a(x))^2$
for which the mean of $y$ over
the predictive distribution $p(y|x,f)$ is optimal,
absolute error loss $|y-a(x)|$ which leads to the median,
and zero--one loss $1-\delta_{y,a(x)}$ resulting in the mode
to be optimal.
For gaussian distributions mean, median, and mode coincide.

For non--approximation problems
calculating the maximum posterior solution ${h}^*$ 
(or the state of knowledge $f$, respectively)
has to be followed by the second step 
of finding the optimal $a^*$ 
which  minimizes
$r(a,{h}^*)$   (or $r(a,f)$)
for given  %maximum posterior solution 
${h}^*$ (or $f$) \cite{Lemm-1996}.
For squared error but non--gaussian $p(y|x,{h}^*)$, for example,
the second minimization step would
require the calculation of the mean of $y$
under $p(y|x,{h}^*)$.

%Formulations as approximation problem are very common.
There are typical non--approximation situations.
Complexity costs, for example, which penalize a complex $a$ 
(e.g.\ due to computational resources or to facilitate comprehensibility)
are typical for non--approximation problems.
Only when there are reasons to believe that
also the real state of Nature ${h}$ fulfils such complexity constraints
this results in an approximation problem.
This means, one has to distinguish between complexity costs
and knowledge about true simplicity in nature.

Related is the distinction between
a generative and a reconstruction model.
Consider the face detection task
of Example \ref{face-detecion} in Section \ref{Definitions}.
Hereby, a reconstruction model $p(y|x,{h})$
would specify a probability of face vs.\ non--face given an image,
while a generative model  $p(x|y,\tilde {h})$
specifies the probability of an image provided 
it represents a face or non--face.
The roles of $x$ and $y$ as independent and dependent variables are
herein exchanged.
Looking for an optimal $y=a(x)$
can be formulated as approximation problem for the reconstruction model.
However, the probability of representing a face
does depend on all non--faces which can appear.
Hence changing the set of non--faces the reconstruction model
has to be adapted.
The specification of the inverse generative model
is often easier.
Here the generative model for faces $p(x|y,\tilde {h})$
does not have to be changed if the class of non--faces changes.
Modeling physical processes of the mapping from objects to images
also falls in this class.
But a generative model does not define an approximation problem
for the function $a$ which returns an answer $y=a(x)$ for a given $x$
and not vice versa.
Then a maximum posteriori approximation requires a second minimization.

Finally, Figs.\ref{map0.1},\ref{map0.5}
show the relations between the optimal $a$ for a full Bayesian risk $r(a,f)$
and a maximum posterior approximation (or empirical risk minimization)
for a simple situation with 
only one $x$--value, i.e., ${\cal X}=\{x\}$,
so $x$ and $x_{\!{}_D}$ can be skipped from the notation.
Assumed are gaussian model states parameterized by their
mean ${h}\in {\cal {H}} = I\!\!\!\:R$
\be
p(y|{h}) = \sqrt{\frac{\beta}{2\pi}} e^{-\frac{\beta}{2} ({h}-y)^2}
,
\label{mod1}
\ee
real numbers
as possible actions $a\in {\cal A}={\cal {H}}= I\!\!\!\:R$
with log--loss
\be
l(y,a) = (y-a)^2 
=-\frac{2}{\beta}\ln p(y|{h}=a)+
\frac{1}{\beta}\ln \frac{\beta}{2\pi}
.
\label{mod2}
\ee
Data $y_{\!{}_D}$ of the form $d$ AND ($b$ OR $c$) 
with $b=1$, $c=-1$ have been represented by a
gaussian mixture with equally weighted components
\be
%p(f|{h}) = 
p(y_{\!{}_D}|f) =
\frac{\beta}{2\pi}
e^{-\beta ({h}-d)^2/2}
\left(
e^{-\frac{\beta}{2}({h}-1)^2} + e^{-\frac{\beta}{2}({h}+1)^2}
\right)
.
\label{mod3}
\ee
This means for complete data, i.e., uniform prior $p({h})$
\be
p({h}|f) = \frac{p(y_{\!{}_D}|{h})}{\ind{{h}} p(y_{\!{}_D}|{h})}
=
\frac{p(y_{\!{}_D}|{h})}{
\frac{1}{2}
\sqrt{\frac{\beta}{\pi}}
\left(
e^{-\frac{\beta}{4}(d-1)^2} + e^{-\frac{\beta}{4}(d+1)^2}
\right)
}
.
\ee
The expected risk reads
\be
r(a,f) = \ind{y} p(y|f)\, l(y,a)
,
\ee
with predictive distribution
\be
p(y|f) = \ind{{h}} p(y|{h})\, p({h}|f)
.
\ee
Shown are the $\beta$--dependency of 1.\
the maximum posterior solutions
${h}^*$ = argmax${}_{{h}\in {\cal {H}}} p({h}|f)$
%= argmax${}_{{h}\in {\cal {H}}} p(f|{h})$
which for the given approximation situation corresponds
to an empirical risk minimization
$a^*$ = argmin${}_{a\in {\cal A}} E(a)$
for error $E(a)=-\ln p(a|f)$,
2.\ the mode of the predictive distribution
argmax${}_{y\in {\cal Y}} p(y|f)$,
and 3.\
the full Bayesian solution
$a^*_{Bayes}$ = argmin${}_{a\in {\cal A}} r(a,f)$.
In Fig.\ref{map0.1} 
the `data' term $d=0.1$ is nearly in the middle
between the two alternative templates $b=1$ and $c=-1$,
while in Fig.\ref{map0.5} the `data' term $d=0.5$ 
is much nearer to $b=1$.
Compared to Fig.\ref{map0.1} 
the second local maximum of the posterior probability $p({h}|f)$
appears at larger $\beta$ (or lower temperature) in Fig.\ref{map0.5},
where also the  maximum posterior approximation is better.
Indeed it is well known from physics that 
mean field (maximum posterior) approximations break
down near phase transitions. 
On the other hand, using 
an adapted $\beta^\prime$ unequal to the true $\beta$
an improved solution can be obtained by a maximum posterior approximation.

\end{appendix}

\acknowledgements{The work was started during a research stay at 
the Massachusetts Institute of Technology.
The author was supported by a 
Postdoctoral Fellowship (Le 1014/1--1)
from the Deutsche Forschungsgemeinschaft and a NSF/CISE Postdoctoral
Fellowship. 
%He also wants to thank Tomaso Poggio, Federico Girosi,
%and the interesting time at M.I.T.
%Chris Williams, 
%Achim Weiguny and Joerg Uhlig for stimulating discussions.
He also wants to thank 
Federico Girosi, 
Gernot M\"unster, 
Tomaso Poggio, 
Joerg Uhlig,
Achim Weiguny,
and 
Christian Wieczerkowski,
%, and Chris Williams
for stimulating discussions.
}
%%%********** End of text entry ****************

%

\end{document}